%% file: 0-main.tex
\documentclass[pdflatex, sn-apa, 10pt]{sn-jnl}

\usepackage{hyperref}       
\usepackage{url}            
\usepackage{booktabs}       
\usepackage{amsfonts}       
\usepackage{nicefrac}       
\usepackage{microtype}      
\usepackage{graphicx}
\usepackage{caption}
\usepackage{rotating}       
\usepackage{multirow}
\usepackage{mwe}  
\usepackage{paralist}

\usepackage{pdflscape}
\usepackage{fancyhdr} 
\usepackage{booktabs}
\usepackage{adjustbox}

\usepackage{fonts-tlwg}

\graphicspath{ {./images/} }

\jyear{2023}%

\theoremstyle{thmstyleone}%
\theoremstyle{thmstyletwo}%

\theoremstyle{thmstylethree}%

\begin{document}

\title[THAI Speech Emotion Recognition (THAI-SER) corpus]{THAI Speech Emotion Recognition (THAI-SER) corpus}


\author[1]{\fnm{Jilamika} \sur{Wongpithayadisai}$^\dagger$}\email{jilamikaw\_pro@vistec.ac.th}

\author[1,2]{\fnm{Chompakorn} \sur{Chaksangchaichot}$^{\dagger,\ddagger}$}\email{6472014221@student.chula.ac.th}

\author[1]{\fnm{Soravitt} \sur{Sangnark}}\email{soravitts\_pro@vistec.ac.th}

\author[2]{\fnm{Patawee} \sur{Prakrankamanant}}\email{6671005621@student.chula.ac.th}

\author[2]{\fnm{Krit} \sur{Gangwanpongpun}}\email{6070108321@student.chula.ac.th}

\author[2]{\fnm{Siwa} \sur{Boonpunmongkol}}\email{siwa.b@alumni.chula.ac.th}

\author[3]{\fnm{Premmarin} \sur{Milindasuta}}\email{prem.milin@gmail.com}

\author[3]{\fnm{Dangkamon} \sur{Na-Pombejra}}\email{dangkamon2011@gmail.com}

\author[1]{\fnm{Sarana} \sur{Nutanong}}\email{snutanon@vistec.ac.th}

\author[2]{\fnm{Ekapol} \sur{Chuangsuwanich}}\email{ekapol.c@chula.ac.th}

\affil[1]{\orgdiv{Information Science and Technology}, \orgname{Vidyasirimedhi Institute of Science and Technology}, \orgaddress{\city{Rayong}, \postcode{21210}, \country{Thailand}}}

\affil[2]{\orgdiv{Department of Computer Engineering}, \orgname{Chulalongkorn University}, \orgaddress{\city{Bangkok}, \postcode{10330}, \country{Thailand}}}

\affil[3]{\orgdiv{Department of Dramatic Arts}, \orgname{Chulalongkorn University}, \orgaddress{\city{Bangkok}, \postcode{10330}, \country{Thailand}}}

\affil[$\dagger$]{\small These authors contributed equally as co-first authors}

\affil[$\ddagger$]{\small Corresponding Author}

\abstract{
We present the first sizeable corpus of Thai speech emotion recognition, THAI-SER, containing 41 hours and 36 minutes (27,854 utterances) from 100 recordings made in different recording environments: Zoom and two studio setups. The recordings contain both scripted and improvised sessions, acted by 200 professional actors (112 females and 88 males, aged 18 to 55) and were directed by professional directors. There are five primary emotions: neutral, angry, happy, sad, and frustrated, assigned to the actors when recording utterances. The utterances are annotated with an emotional category using crowdsourcing. To control the annotation process's quality, we also design an extensive filtering and quality control scheme to ensure that the majority agreement score remains above 0.71. We evaluate our annotated corpus using two metrics: inter-annotator reliability and human recognition accuracy. Inter-annotator reliability score was calculated using Krippendorff's alpha, where our corpus, after filtering, achieved an alpha score of 0.692, higher than a recommendation of 0.667. For human recognition accuracy, our corpus scored up to 0.772 post-filtering. We also provide the results of the model trained on the corpus evaluated on both in-corpus and cross-corpus setups. The corpus is publicly available under a Creative Commons BY-SA 4.0\footnote{On Github Release - \href{https://github.com/vistec-AI/dataset-releases/releases/tag/v1}{{https://github.com/vistec-AI/dataset-releases/releases/tag/v1}}
On HuggingFace - \href{https://huggingface.co/datasets/airesearch/thai-ser}{{https://huggingface.co/datasets/airesearch/thai-ser}}}, as well as our codes for the experiments\footnote{\href{https://github.com/tann9949/thaiser-experiments}{{https://github.com/tann9949/thaiser-experiments}}}.
}

\keywords{Speech Emotion Recognition, Thai Dataset}

\maketitle
\input{1-introduction}
\input{2-corpus_design}
\input{3-annotation}
\input{4-data-evaluation}
\input{6-experiments}
\input{7-discussion}
\input{8-conclusion}

\bibliography{sn-bibliography}
\end{document}

%% file: 1-introduction.tex
\section{Introduction}
\label{sec1}
\label{sec:introduction}

There has been a surge of interest in designing a system that allows a computer to infer human emotion from facial expressions, text, or speech.
Classifying emotion from a given speech input is called Speech Emotion Recognition (SER).
Unlike facial expression and text, speech modality is more challenging due to its nature of solely recognizing emotion via speech signal, which limits the most dominant features used by humans, such as semantic context and facial expression.
Several works showed the potential usage of SER in various domains.
\cite{callcenter} and \cite{mobileservice} proposed that SER could assist in observing interactions between agents and customers in call centers. 
In the smart home use case, an intelligent household robot could function better with the ability to recognize emotional information~\citep{robot}. 
SER can also be used in online learning to detect students' satisfaction in each class, aiding customization in learning approaches for each student~\citep{onlinelearning}. 
Despite all those use cases, most corpus are collected in Western languages as shown in \autoref{tab:speechemotioncorpus}.
Since the recognition of emotions in speech was partially dependent on linguistic and cultural variables \citep{langdiif}, most corpus in Western languages limits the application of SER in countries where linguistic and cultural values differ from Western, such as in Asia.
Therefore, there is a need for a non-Western SER corpus to contribute to the diversity and improvement of SER research.

\subsection{Existing SER Datasets and Design Considerations}
Our work focuses on developing the SER corpus in Thai.
Since Thai is a tonal language, the acoustic features of tonal languages (e.g., Thai, Mandarin) differ significantly from those of non-tonal languages (e.g., English, German) in conveying emotion \citep{doi:10.1177/0022022108321178, chong2015exploring}. 
This difference poses challenges for Speech Emotion Recognition (SER) systems in understanding emotions across language types. 
To the best of our knowledge, existing Thai speech corpora, such as the ORCHID-SPEECH CORPUS and the NECTEC-ATR Thai speech corpus \citep{Kasuriya_1}, primarily served for speech recognition rather than SER. 
Thus, there is a substantial need to develop a dedicated Thai SER corpus.

SER corpora are usually divided into three categories: \textbf{natural}, \textbf{acted}, and \textbf{elicited} \citep{Mehmet_2}, as shown in \autoref{tab:speechemotioncorpus}. 

\textbf{Natural corpus} is either the recording of real-life events or from different sources, such as a recordings from TV talk shows \citep{Grimm_3}, interviews \citep{Abrilan_4}, or a mixture of films and television plays \citep{Li_5}. 
While the natural corpus is closer to daily life environments and more natural in conversation, there are still concerns regarding privacy, copyright, and environmental control issues such as word usage, situation diversity, microphone position, or background noise that are challenging to handle.

\textbf{Acted corpus} was introduced to address the limitations of natural corpus, providing a more controlled alternative.
Several corpora collected using the acted approach include Emo-DB \citep{Emo-DB}, \textit{RAVDESS} \citep{RAVDESS}, \textit{CREMA-D} \citep{CREMA-D}, and \textit{EMOVO Corpus} \citep{EMOVO}.
An acted corpus was created in controlled environments, introducing more control over dialogue, recording environments, and background noise control. 
Nevertheless, such corpora lack naturalness due to the amount of control in the dialogue.

\textbf{Elicited corpus} finds a better trade-off between environmental control and naturalness and offers an alternative to find the middle ground between the two.
Elicited corpus was designed to mitigate the naturalness problem by reducing control over a dialogue script, allowing only brief actors to improvise in dyadic interactions with spontaneous situations. 
Several corpora such as \textit{MSP-IMPROV} \citep{MSP-IMPROV}, \textit{DEMoS} \citep{DEMoS}, and \textit{IEMOCAP} \citep{IEMOCAP} contained both elicited and acted subset. 
\cite{IEMOCAP} showed that incorporating improvisation could mitigate the naturalness problem,  which tended to contain more intense emotions and a wider lexicon than scripted recordings (acted corpus) \citep{CarlosBusso}.
However, despite more naturalness and diversity, the elicited corpus had a higher rate of overlapped speech than the scripted recordings, which introduces challenges in utterance segmentation.
Elicited corpus was usually recorded in the studio for quality and background noise control, except for \textit{SEWA} \citep{SEWA} that conducted the recording via video chat.

When curating SER data, there are two main approaches to how emotion was annotated: crowdsourcing and in-house.
The first approach, crowdsourcing, is where annotation was collected publicly through an online platform \citep{wangwebsite,amazonmechanicalturk}.
This approach allows for easier annotator recruitment, which makes the annotation process faster.
For instance, CREMA-D collected annotations from over 1,000 annotators, which would not have been feasible with an in-house approach. 
However, crowdsourcing suffers from data quality control due to its nature of sourcing from the public, as well as controlling annotator expertise due to its online sourcing nature.
The second approach, in-house annotation, was done by recruiting annotators, either freelancers or domain-specific experts, to annotate the corpus.
This allows for more fine-grained control over annotator expertise and easier data quality control.





\subsection{Proposed Thai SER Dataset}

To this end, we proposed \textbf{THAI-SER}, a large-scale speech emotion recognition dataset that bridges the gap between Western language-based corpus and South-East-Asia-based corpus.
Following \cite{IEMOCAP}, we collect our SER corpus containing both acted and elicited subsets.
The acted split --- a ``scripted'' session --- was collected by fixing sentences across all emotions, isolating the speech information (e.g., tonal, speaking rate, etc.) from potential contextual information.
Elicited split - or ``improvised session'' - was collected using a conventional method where we provide a situation for actors to act based on assigned emotion.
Our speech data was recorded using 200 actors who graduated from or are currently studying at the Department of Dramatic Arts.
We collected a total of five emotions, including four basic emotions (\textit{neutral}, \textit{angry}, \textit{happy}, and \textit{sad}) as well as \textit{frustration}, which plays an important role in call center applications.
THAI-SER was recorded in both studio settings, allowing fine-grained control over echoes, noises, and reverbs. 
It also included online meeting sessions via Zoom, which reflects a noisier environment.
The recordings are then post-processed and chunked into multiple utterances before being annotated using crowdsourcing platforms \citep{wangwebsite,hopewebsite}.

\subsection{Research Questions and Contributions}

In addition to the dataset, this work provides a standard benchmarking scheme for THAI-SER, following a standard k-fold cross-validation with speaker-independent settings similar to \cite{ser_kfold:journals/corr/abs-1802-05630}. 

Leveraging THAI-SER, we conducted a series of empirical studies guided by the following research questions:
\begin{compactenum}[\bf RQ1]
\item \emph{How reliably can emotional categories be perceived and annotated in Thai speech across different session types, speaker demographics, and recording environments?}
To answer this, in Section~\ref{sec:evaluation}, we analyzed inter-annotator agreement and human recognition accuracy across acting styles (scripted vs. improvised), emotional intensity, gender, age groups, and studio vs. Zoom recordings.
\item \emph{How does the choice of acting style (scripted-only, improvised-only, or both) affect model performance in speech emotion recognition?}
We trained models under each session configuration and evaluated their performance using speaker-independent k-fold cross-validation in Section~\ref{sec:acting}.  
\item \emph{To what extent do SER models trained on THAI-SER generalize across domains and datasets?}
We assessed robustness to distributional shifts by testing models on Zoom recordings (excluded from training) and conducted cross-corpus evaluations involving THAI-SER and existing SER datasets in Section~\ref{sec:cross}
\end{compactenum}

The contributions of our work are as follows.
\underline{First}, we introduce THAI-SER, a large-scale, high-quality Thai speech emotion recognition (SER) dataset comprising 41.6 hours of scripted and improvised speech from 200 professional actors, recorded across both studio and Zoom environments. The dataset includes five practically relevant emotions and is annotated through crowdsourcing with rigorous quality control measures, including pretests, gold utterances, consistency checks, and agreement-based filtering. 
\underline{Second}, we present a systematic analysis of human annotation reliability across session types, speaker demographics, and recording conditions, highlighting how perception varies with emotional intensity, actor profile, and recording setup. 
\underline{Third}, we establish a benchmark for Thai SER and conduct a series of experiments to assess model generalization across session types, environments, and datasets, including cross-corpus evaluations with IEMOCAP and other existing SER corpora.

%
%
%

%
%
%
%

\input{tables/intro/intro_ser-corpus}

%% file: tables/intro/intro_ser-corpus.tex
\begin{landscape}
\begin{table}[!ht]
\begin{adjustbox}{width=0.95\linewidth,center}
{
    \centering
    \begin{tabular}{ccccccccc}
    \toprule 

    \textbf{Corpus} 
    & \textbf{Language}                                                                                 
    & \textbf{Type}  
    & \textbf{Modality}  
    & \textbf{Environment} 
    & \textbf{\begin{tabular}[c]{@{}c@{}}Size \\ (utterances)\end{tabular}}          
    & \textbf{\begin{tabular}[c]{@{}c@{}}Subject \\ (speakers)\end{tabular}}   
    & \textbf{Emotion} 
    & \textbf{Annotation}    
    \\ \midrule

    \begin{tabular}[c]{@{}c@{}}
        Emo-DB 2005\\  \cite{Emo-DB}
        \end{tabular} 
    & German                                                   
    & Acted      
    & \begin{tabular}[c]{@{}c@{}}Audio \\ (1 mic)\end{tabular}   
    &  Studio
    & 500                   
    & 10  & \begin{tabular}[c]{@{}c@{}}Anger, Happiness, \\ Anxiety, Fear, \\ Boredom, Disgust, \\ and Neutral\end{tabular}            &    \begin{tabular}[c]{@{}c@{}} in-house \\ (20 annotators)\end{tabular} 
    \\ \midrule

    \begin{tabular}[c]{@{}c@{}}
        IEMOCAP 2007\\  \cite{IEMOCAP}
        \end{tabular}
    & English     
    & Acted \& Elicited 
    & \begin{tabular}[c]{@{}c@{}}Audio/Video/Motion \\ (2 mic/2 cam/8 motion cam) \end{tabular} 
    &  Studio
    & \begin{tabular}[c]{@{}c@{}}10,039 \\  (12 hours)\end{tabular}
    & 10     & \begin{tabular}[c]{@{}c@{}}Anger,  Happiness, \\ Sadness, Frustration, \\ and Neutral\end{tabular}    
    & \begin{tabular}[c]{@{}c@{}} in-house \\ (6 annotators)\end{tabular} 
    \\ \midrule

    \begin{tabular}[c]{@{}c@{}}
        EMOVO 2014\\  \cite{EMOVO}
        \end{tabular}
    & Italian                                                                                  
    & Acted
    & \begin{tabular}[c]{@{}c@{}}Audio \\ (2 mic)\end{tabular}
    &  Studio
    & 588
    & 6
    & \begin{tabular}[c]{@{}c@{}}Anger,  Surprising, \\ Sadness, Disgust, \\ Fear, Joy, \\ and Neutral\end{tabular}
    & \begin{tabular}[c]{@{}c@{}} in-house \\ (24 annotators)\end{tabular}
    \\ \midrule

    \begin{tabular}[c]{@{}c@{}}
        CREMA-D 2015\\  \cite{CREMA-D}
        \end{tabular}
    & English
    & Acted
    & \begin{tabular}[c]{@{}c@{}}Audio/Video \\ (1 mic/1 cam)\end{tabular}
    &  Studio
    & 7,442
    & 91   & \begin{tabular}[c]{@{}c@{}}Anger,  Happiness, \\ Sadness,  Disgust, \\ Fear, and Neutral\end{tabular}
    &  \begin{tabular}[c]{@{}c@{}} crowdsourcing \\ (2,443 annotators)\end{tabular}
    \\ \midrule

    \begin{tabular}[c]{@{}c@{}}
        MSP-IMPROV 2017\\  \cite{MSP-IMPROV}
        \end{tabular}
    & English
    & Elicited
    & \begin{tabular}[c]{@{}c@{}}Audio/Video \\ (1 mic*/1 cam*) \end{tabular}
    &  Studio
    & \begin{tabular}[c]{@{}c@{}}7,818 \\ (9 hours)\end{tabular}
    & 12
    & \begin{tabular}[c]{@{}c@{}}Anger,  Happiness,\\ Sadness,Disgust,\\ Fear, and Neutral\end{tabular}
    &  crowdsourcing                                                      
    \\ \midrule

    \begin{tabular}[c]{@{}c@{}}
        RAVDESS 2018\\  \cite{RAVDESS}
        \end{tabular}
    & English
    & Acted
    & \begin{tabular}[c]{@{}c@{}}Audio/Video \\ (1 mic/1 cam)\end{tabular}
    &  Studio
    & 7,356
    & 24  & \begin{tabular}[c]{@{}c@{}}Anger,  Happiness, \\ Sadness, Frustration, \\Fear, Disgust, \\ Surprising, and Calm\end{tabular} 
    &   \begin{tabular}[c]{@{}c@{}} in-house \\ (319 annotators)\end{tabular}   
    \\ \midrule


    \begin{tabular}[c]{@{}c@{}}
        DEMoS 2019\\  \cite{DEMoS}
        \end{tabular}
    & Italian
    & Elicited
    & \begin{tabular}[c]{@{}c@{}}Audio \\ (1 mic)\end{tabular}
    &  Studio
    & 9,697
    & 68
    & \begin{tabular}[c]{@{}c@{}}Anger,  Surprising,\\ Sadness, Disgust,\\ Happiness, and Guilt\end{tabular}
    & \begin{tabular}[c]{@{}c@{}} in-house \\ (86 annotators)\end{tabular}                      
    \\ \midrule

    \begin{tabular}[c]{@{}c@{}}
        SEWA 2019 \\   \cite{SEWA}
        \end{tabular}
    & \begin{tabular}[c]{@{}c@{}}
        Chinese, English, \\German, Greek, \\ Hungarian, \\ and Serbian
        \end{tabular} 
    & Elicited
    & \begin{tabular}[c]{@{}c@{}}Audio/Video \\ (1 mic*/1 cam*) \end{tabular}
    &  Video-chat
    & \begin{tabular}[c]{@{}c@{}}1,990 \\ (44 hours)\end{tabular} 
    & 398 
    & \begin{tabular}[c]{@{}c@{}}Continuous emotions \\ (valence and arousal)\end{tabular}
    & \begin{tabular}[c]{@{}c@{}} in-house \\ (30 annotators)\end{tabular}
    \\ \midrule

    \begin{tabular}[c]{@{}c@{}}%
        MSP-Podcast 2019\\ \cite{MSP-PODCAST}%
    \end{tabular} &
    English &
    Natural &
    Audio &
    Podcast &
    \begin{tabular}[c]{@{}c@{}}151\,654 \\ (237.93 hours)\end{tabular} &
    1\,409 &
    \begin{tabular}[c]{@{}c@{}}Anger, Happiness, \\ Sadness, Disgust, \\ Surprise, Fear, \\ Contempt, Neutral, Other\end{tabular} &
    crowdsourcing \\ \midrule

    \begin{tabular}[c]{@{}c@{}}%
        TESS 2020\\ \cite{TESS}%
    \end{tabular} &
    English &
    Acted &
    \begin{tabular}[c]{@{}c@{}}Audio \\ (1 mic)\end{tabular} &
    Studio &
    2\,800 &
    2 &
    \begin{tabular}[c]{@{}c@{}}Anger, Disgust, \\ Fear, Happiness, \\ Pleasant Surprise, \\ Sadness, Neutral\end{tabular} &
    in-house \\ \midrule

    \begin{tabular}[c]{@{}c@{}}
        LSSED 2021 \\ \cite{lssed}
        \end{tabular} 
      & English
      & Natural 
      & Audio 
      & Real-world 
      & \begin{tabular}[c]{@{}c@{}}147,025 \\ (206 hours)\end{tabular} 
      & 820 
      & \begin{tabular}[c]{@{}c@{}}Anger, Happiness, \\ Sadness, Disappointment, \\ Boredom, Disgust, \\ Excitement, Fear, \\ Surprise, Neutral, Other\end{tabular} 
      & \begin{tabular}[c]{@{}c@{}}crowdsourcing\end{tabular} \\ \midrule

    \textbf{\begin{tabular}[c]{@{}c@{}}
        THAI-SER \\ 2021 
        \end{tabular}}
    & \textbf{Thai}
    & \textbf{Acted \& Elicited}
    & \textbf{\begin{tabular}[c]{@{}c@{}}Audio \\ (2 mic per speaker,\\ 1 center mic) \end{tabular}}
    & \textbf{Studio/Video-chat}                          
    & \textbf{\begin{tabular}[c]{@{}c@{}}27,854 \\  (41.61 hours)\end{tabular}}  
    & \textbf{200}  & \textbf{\begin{tabular}[c]{@{}c@{}}Anger,  Happiness, \\ Sadness, Frustration, \\ and Neutral\end{tabular}}      
    & \textbf{\begin{tabular}[c]{@{}c@{}} crowdsourcing \\ (984 annotators)\end{tabular}} \\ 
    \bottomrule 
    \end{tabular}
}
\end{adjustbox}
\caption{\label{tab:speechemotioncorpus}Existing speech emotion corpus. In the annotation column, ``in-house'' means annotating at the lab/provided place; ``crowdsourcing'' means annotating from anywhere via an online platform. The * in modality column denotes per speaker unit.}
\end{table}
\thispagestyle{plain}
\end{landscape}

%% file: 2-corpus_design.tex

\section{Corpus Design and Collection Methodology}
\label{sec2}
\label{sec:design}

This section provides a detailed description of how the data was collected.
Generally, the THAI-SER dataset was recorded in multiple sessions, where each session contained both scripted sessions (acted subset) and improvised sessions (elicited subset).
Each session was recorded in specific environmental setups, such as in a studio or through online meetings.
This section will go through the details about how we sourced directors and actors, recording environments, tools used for the recordings, and how we set up improvised and scripted sessions.

\subsection{Director and Actors Sourcing}
Six professional directors, who are students, alumni, or professors of the Department of Dramatic Arts\footnote{The department was operated under Faculty of Arts, Chulalongkorn University}, were invited to direct the actors' performance in all recordings.

Regarding actors, we also recruited students and alumni from the Department of Dramatic Arts.
As shown in \autoref{tab:actorage}, we recruited 200 Thai native actors (112 females and 88 males, according to biological gender\footnote{We also recorded the actor profile, and we allowed actors to freely describe their sexuality. There are six actors who were not described as male or female. (LGBTQ:3, non-binary:1, gay:1 and queer:1).}), aged between 18 and 55 years old and averaged at 29 years old. 
%

\input{tables/corpus_design/actors_distribution}

\subsection{Recording environmental design}

We conducted a total of 100 recorded sessions from two different environments.
This includes 80 sessions recording in a controlled studio room and another 20 sessions via online meeting (which later we referred to as ``Zoom'' sessions).
Each session has its own unique pair of actors assigned, where directors were rotated based on their availability.
Thus, consider that there are a total of 100 sessions, and the total number of actors is 200.

\subsubsection{Studio recordings}
\label{subsubsec:studio_recordings}
The THAI-SER dataset was recorded in two recording environments: studio recording and online recordings via Zoom.
There are two unique rooms for studio recordings environments; both take place at the Faculty of Arts, Chulalongkorn University.
We named these two different room environments for studio recordings as follows:

\begin{itemize}
    \item \textbf{Studio A}: A controlled studio room with soundproof walls, measuring 4.75 $\times$ 3.45 $\times$ 3.2 (length, width, height) meters.
    \item \textbf{Studio B}: A standard room without soundproofing or noise control, the room size was 5.3 $\times$ 7 $\times$ 4.3 (length, width, height) meters.
\end{itemize}

\begin{figure}[!ht]
    \centering
    \begin{minipage}{0.45\textwidth}
        \centering
        \includegraphics[width=1.0\textwidth]{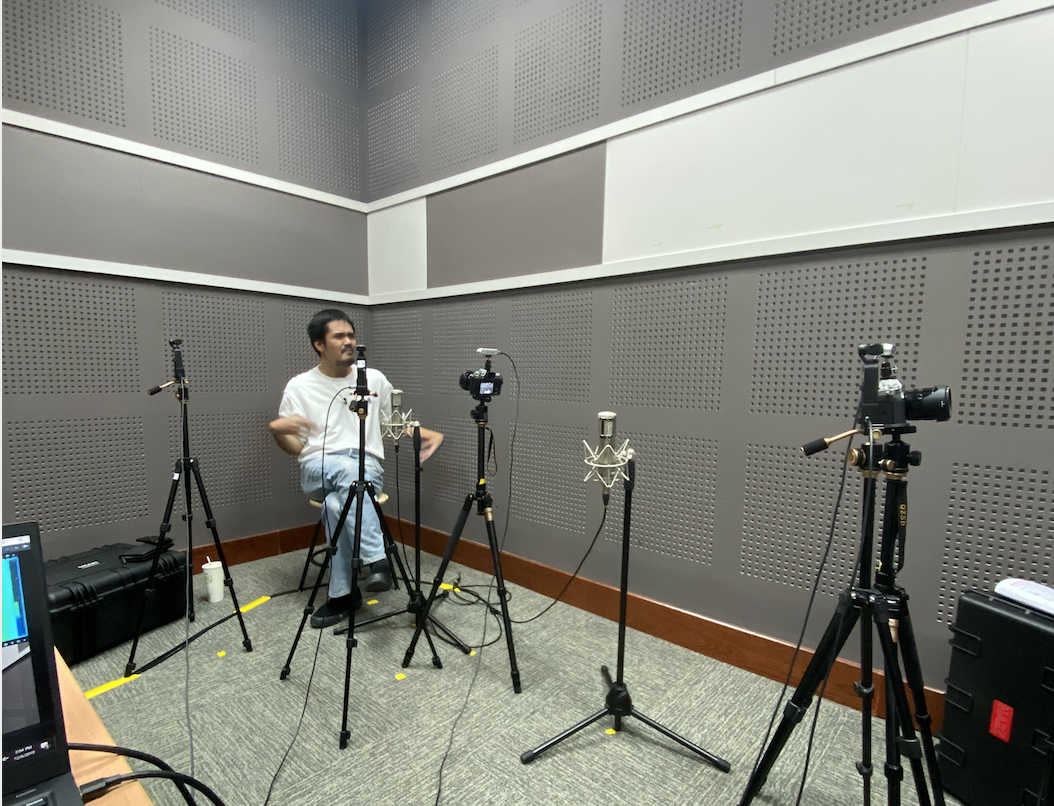} 
        \caption*{The recording of scripted session in studio A}
    \end{minipage}\hfill
    \begin{minipage}{0.45\textwidth}
        \centering
        \includegraphics[width=1.0\textwidth]{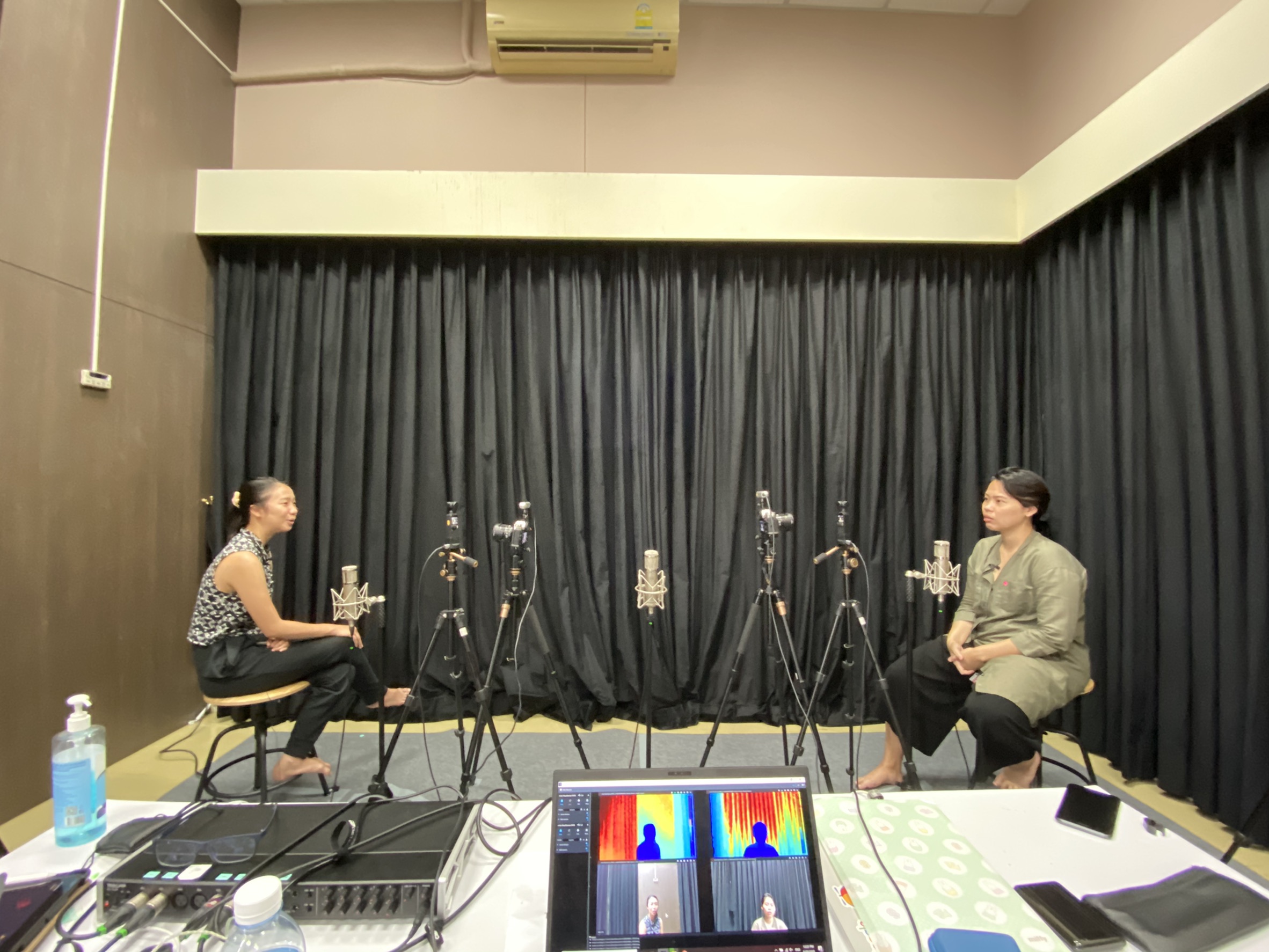} 
        \centering
        \caption*{The recording of improvised session in studio B}
    \end{minipage}
    \caption{\label{fig:studio}Studio recording environments and recording tools}
\end{figure}

\begin{figure}[!ht]
    \centering
    \includegraphics[width=0.8\textwidth]{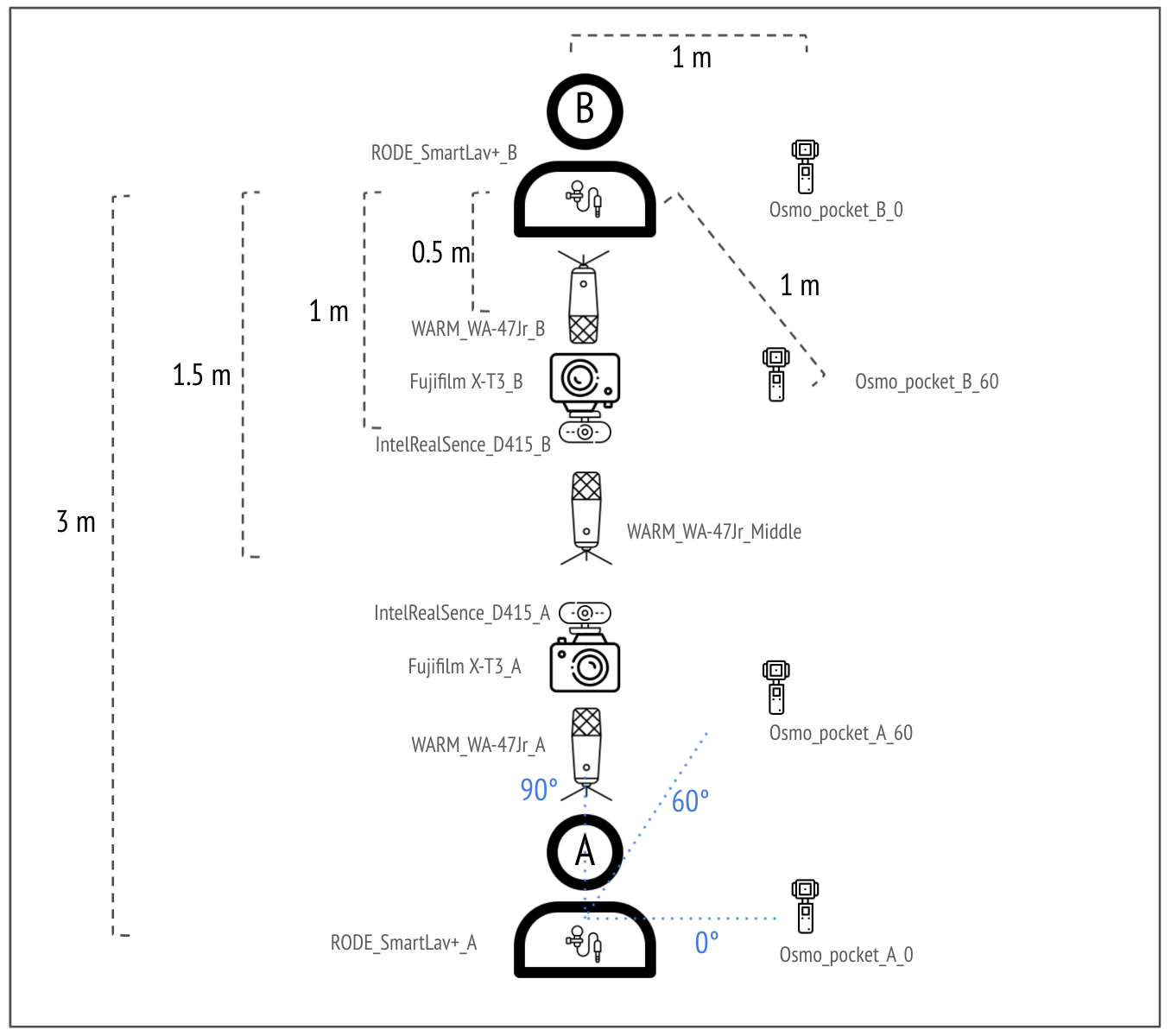}
    \caption{\label{fig:studiorecording}Studio setup: actors, microphones, and camera placements }
\end{figure}

\paragraph{Recording tools} 
We recorded the audio and video using several recording tools, which were placed at different angles and distances.
\autoref{fig:studiorecording} summarizes both microphone and camera setups of the studio recordings.

For each session, which contains two actors sitting 3 meters apart, there will be a total of five microphones.
Two of them are clipped to the actor's shirt collar around 0.2 meters away from the actor's mouth.
For these two clipped microphones, we use RODE Smart Lav+, which is an omnidirectional lavalier microphone.
In addition to the clipped microphones, we also have two cardioid condenser microphones (WARM WA-47Jr), placed 0.5 meters away from the actor.
These two condenser microphones were set to receive audio from the side of the actor only.
These condenser and clipped microphones were set with the objective to record the audio from each of the actors.
Finally, another WARM WA-47Jr cardioid condenser microphone was placed in the middle between the actors.
This middle condenser microphone was used only during an improvisation and was set as a figure-8 condenser to receive audios from both sides.

All audio recordings were recorded via the audio interface (TASCAM US-16x08) and GarageBand software on the MacBook Air.
The audio was recorded at a 44.1kHz sampling rate, 16 bits, and WAV format.

During the session, we also record the video using a total of four cameras for each actor.
Among these four cameras, one is a front-facing digital camera, which we use Fujifilm X-T3.
This front-facing camera also has a special camera-Intel RealSense D415-recording a depth of the actors' faces with 1280x720 resolutions at 30 fps.
On the side, there are two Osmo Pocket cameras recorded at 60 degrees and 0 degrees from the side.
The goal of these two cameras was to record the actors' faces from different angles.
These two Osmo cameras were recorded at a 4K resolution at 60 fps.
All cameras were adjusted individually based on the actors' height when they were sitting, ensuring that the actors' faces were properly captured.

\subsubsection{Zoom recordings}

For the online meeting environment, which we named ``Zoom sessions'', we recorded a total of 20 sessions in this environment.
Due to the COVID-19 pandemic, there is a rise in usage of online meeting platforms.
This raises a potential usage of SER in an online meeting scenario since online meetings usually have multiple noises that can't be found in studio recordings, such as video and audio compression loss, lagging, or varied item/room environments.
In this session, we use Zoom as a main online meeting platform while not controlling the specs of the cameras, microphones, and rooms.
For audio recording, we used a paid version of Zencastr, capturing at 48 kHz sample rate, 16 bits in WAV format, apart from using raw Zoom audio.
The file quality for the Zoom video was 1280×720 resolution at 30 fps in MP4 format.

\subsection{Recording process}

We exclusively worked with the five emotions: neutrality, anger, happiness, sadness, and frustration.
The emotions were selected based on the perceived usefulness in call center and robotics applications.
The details of each sentence and situation will be discussed in \autoref{subsec:recordingprocess}.

\label{subsec:recordingprocess}
The recordings were separated into two main subsets: scripted and improvised.
Initially, during the planning phase prior to all recordings, the directors were tasked with creating 3 sample sentences for scripted sessions and 15 improvised situations for the improvised sessions.
Then, for each session, the director will brief actors about the process, which will typically start with a scripted session followed by an improvised session.
It is important to note that the feelings and emotions of the actors must be authentically realistic throughout the recording session.
Therefore, within the recording session, we allowed actors and the director to have 15-30 minutes of break to prevent exhaustion building within actors, as it could potentially result in worse emotional expression in the session.
Each session took approximately 3 hours to record both scripted and improvised sessions.

\subsubsection{Scripted session}

\begin{figure}[!ht]
    \centering
    \includegraphics[width=1\textwidth]{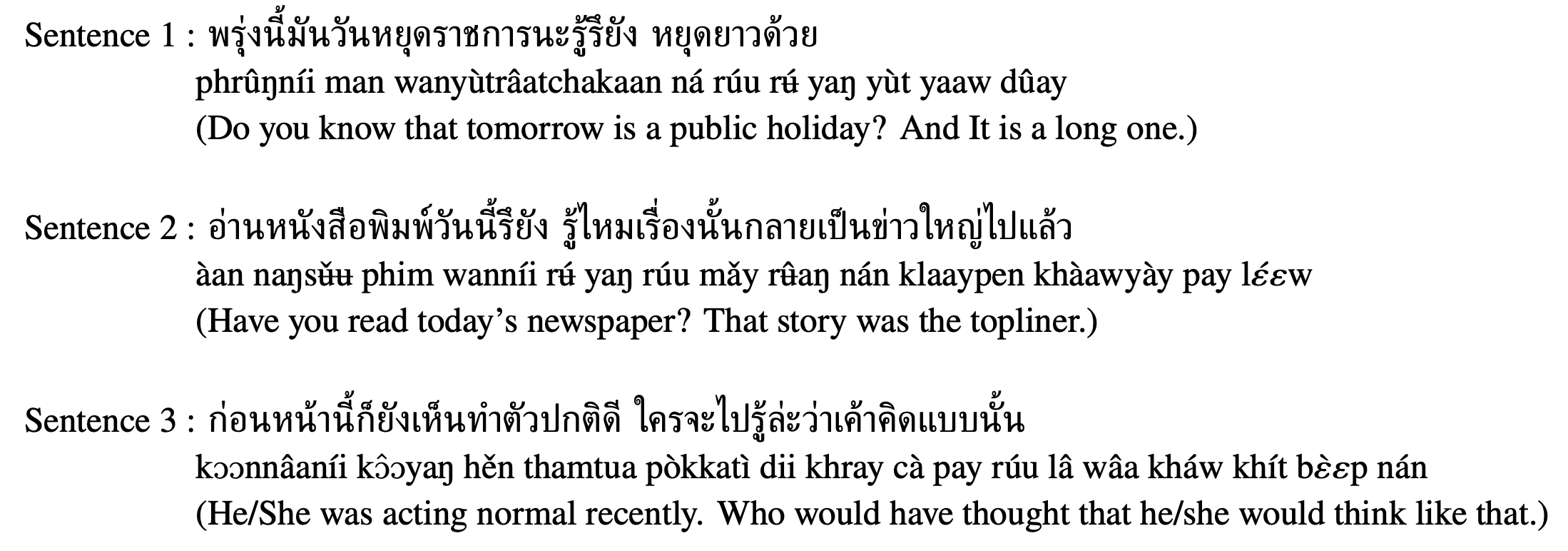}
    \caption{The three sentences from a scripted session. The first line is the sentence written in Thai, followed by its phonetic transcription using IPA in the second line, and finally, its translation is given in the third line.}
    \label{fig:script_example_sents}
\end{figure}

The scripted session was an acted speech where we asked the actors to speak the given sentence in a specific emotion.
However, to solely isolate the contextual and semantic information in the sentence, we designed the sentences to be compatible with any emotions, allowing actors to freely design their scenario according to the given sentences.
In addition, these sentences were designed to cover many Thai tones, consonants, and diphthongs.
\autoref{fig:script_example_sents} showed the sentences used in the scripted session.

\begin{figure}[hbt!]
    \centering
    \includegraphics[width=1\textwidth]{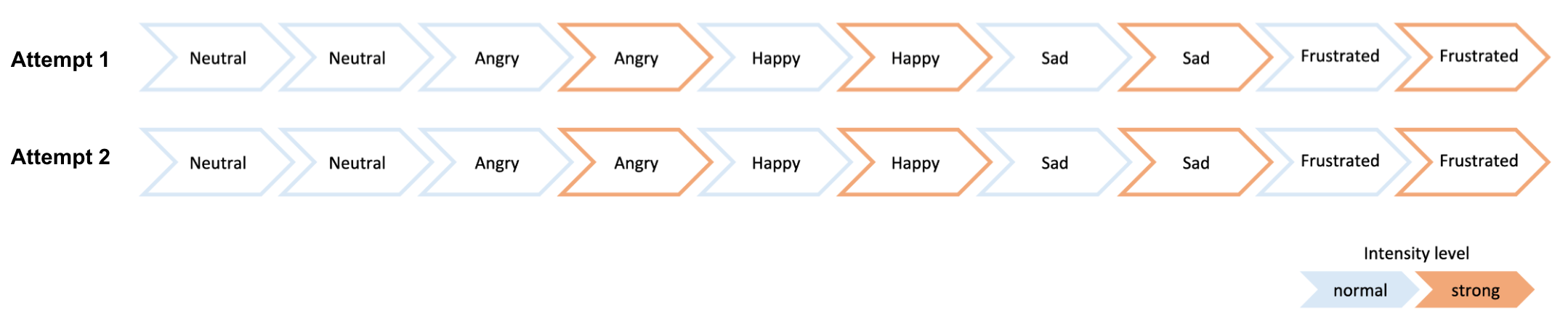}
    \caption{\label{fig:scriptorder}The order of the scripted recording. The process was repeated twice resulting in a total of 20 acts per sentence per actor: 5 emotions, 2 intensities, and 2 attempts.
    }
\end{figure}

To further ensure that actors can speak the sentences fluently, we asked the actors to practice with the sentences at least one day prior to the recording day.
During the script session recordings, we recorded the actors speaking this sentence twice with two different degrees of emotional intensity, resulting in 4 recordings per sentence.
Thus, for each session, there will be a total of $3\text{ sentences} \times 2 \text{ levels} \times 5 \text{ emotions} \times 2 \text{ recordings} = 60 \text{ recordings}$ script utterances in total.
We also design the order of emotions when recording the scripted session to ensure minimal emotion conflict when transitioning to different emotions.
The order of the recordings was illustrated in \autoref{fig:scriptorder}.

After finishing each sentence, the professional director reviewed the acting according to the following criteria:

\begin{enumerate}
    \item Accuracy of pronunciation and lexical.
    \item Emotions and intensity are expressed clearly through facial expression and voice.
\end{enumerate}

If the actors did not pass the criteria, they were asked to perform again.

\subsubsection{Improvised session}

Although scripted sessions represent the speech that decouples semantics and emotion, as described in \autoref{sec2}, they lack naturalness in the speech.
To tackle this problem, we also recorded the improvised session following \cite{IEMOCAP}.
To do so, the actors were tasked with performing over 15 different situations designed by the directors, where each actor would have their emotion assigned to their role.
\autoref{tab:situations_improvisation} shows the situations that were used in the dataset.
These situations were designed to have a clear objective, conflict, relationship, characteristics, and action.
Meanwhile, the characters' conditions in each situation must be performable, doable, and active enough.

To further increase our improvised situation variance, we designed three different batches of scenarios across different sessions as shown in \autoref{tab:recording_situation_env}.
The objective of this design is to prevent the trained model from being overfitted to specific keywords that might appear often in some situations.
For each scenario in each batch, we ensured that the scenarios covered all five distinct emotions. 
In addition, we also sometimes maintain the scenario while assigning a different emotion to each actor to prevent the trained model from overfitting emotions to keywords in each scenario.
For example, situation 1 of the first batch and situation 6 in the second batch were both conversations between the hotel's receptionist and the customer.
The difference between these two was the assigned emotions.

Unlike the scripted sessions, the actors were allowed to use any words they wished as long as the assigned emotions were properly conveyed during the period (3 minutes).
It usually takes a minute or so for the situation to develop into certain emotions such as sadness or anger.
At the end of each improvised situation, the director evaluated the acting by considering the emotional accuracy of both speech and facial expression.
If the director judged the acting unsatisfactory, the actors were asked to perform the improvised situation again.

In the released corpus, the actor with the odd-numbered \texttt{actor\_id} played the role A, and the even-numbered \texttt{actor\_id} played the role B. 

\input{tables/corpus_design/improvised_scenario}

\input{tables/corpus_design/improvised_room}

\subsection{Audio/Video Alignment and Segmentation}
\label{subsec:alignment}

As described in \autoref{subsubsec:studio_recordings}, there are multiple microphones and cameras that result in multiple audio and video sources that were recorded asynchronously.
Thus, it was necessary to synchronize these many source files before further processing.
We aligned the timestamp of audio and videos using a simple sliding window cross-correlation between video and audio signals.
The algorithm was summarized in Algorithm \ref{alg:convolve}.

First, we select the source audio with the highest audio quality as an anchor recording.
Then, cut the audio recordings which is typically the first 5 to 10 seconds since the audio start.
The cut anchor audio, \texttt{windowed\_anchor} in Algorithm \ref{alg:convolve}, is then used to apply sliding window cross-correlation with audio from other sources.
Note that we limit the start and end ranges for sliding-window cross-correlation to save our computation.
Finally, the timestamp that yielded the maximum cross-correlation value was selected as the \texttt{shift\_time} which was further used to align multiple audio files.

After the alignment process, the aligned recordings were then segmented into multiple utterances manually.
Note that when segmenting the audio, we omit all silent segments across all utterances.
After segmentation, we applied amplitude normalization across all audio files making sure that all utterances have the same average amplitude of -20 dBFS.
Finally, the utterances are then exported into both WAV and FLAC format.

\begin{algorithm}
\caption{Audio Alignment with Cross-correlation}
\label{alg:convolve}
\begin{algorithmic}[1]
\footnotesize
    \item[] \textbf{Input} : 
        \item[] \hspace{\algorithmicindent}\texttt{anchor\_audio} - Anchored audio file.
        \item[] \hspace{\algorithmicindent}\texttt{audio\_to\_align} - Audio from another source.
        \item[] \hspace{\algorithmicindent}\texttt{windows\_size} - Window size of the anchor audio.
        \item[] \hspace{\algorithmicindent}\texttt{start\_time} - Start time to cut anchor audio. This is usually the time that the first person start speaking.
        \item[] \hspace{\algorithmicindent}\texttt{time\_margin} - Maximum allowable time deviation to search for the alignment.
        
    \item[] \textbf{Output}: 
        \item[] \hspace{\algorithmicindent}\texttt{shift\_time} - Time lag between \texttt{anchor\_audio} and \texttt{audio\_to\_align}
        
    \item[] \textbf{Procedure}:
    
    \State \texttt{start\_range} $\gets$ \texttt{start\_time} $-$ \texttt{time\_margin}
    \State \texttt{end\_range} $\gets$ \texttt{start\_time} $+$ \texttt{time\_margin}
    \State \texttt{windowed\_anchor} := \texttt{anchor\_audio[start\_time : start\_time + window\_size]}
    \State \texttt{max\_conv\_value} := 0
    \State \texttt{max\_conv\_time} := 0
    
    \State {\textbf{FOR} \texttt{i} \textbf{FROM} \texttt{start\_range} \textbf{TO} \texttt{end\_range}:}
        \State \hspace{\algorithmicindent} \texttt{conv\_value} := \texttt{cross\_correlate(windowed\_anchor, audio\_to\_align[i:i + window\_size])}
        
            \State \hspace{\algorithmicindent} \textbf{IF} \texttt{conv\_value $\gt$ max\_conv\_value}:
                \State \hspace{\algorithmicindent}\hspace{\algorithmicindent} \texttt{max\_conv\_value} := \texttt{conv\_value}
            \State \hspace{\algorithmicindent} \texttt{max\_conv\_time} := \texttt{i}
            \State \texttt{shift\_time := ( max\_conv\_time + start\_range ) - start\_time}
    \State \textbf{return} \texttt{shift\_time}
\end{algorithmic}
\end{algorithm}


%% file: tables/corpus_design/actors_distribution.tex
\begin{table}[!ht]
\centering
\begin{tabular}{ccc}
\toprule 
\textbf{Gender} & \textbf{Age}   & \textbf{Count} \\ \hline
Male     & under 20       & 1             \\
(88)    & 20-29          & 46            \\ 
         & 30-39          & 31             \\ 
         & 40-49          & 9             \\ 
        & over 50        & 1              \\ \hline
Female   & under 20       & 3              \\ 
(112)         & 20-29          & 75            \\ 
         & 30-39          & 26             \\ 
         & 40-49          & 7            \\ 
         & over 50        & 1              \\ \hline
\textbf{TOTAL}    &               & \textbf{200}             \\ 

\midrule
\end{tabular}
\caption{\label{tab:actorage} Actor's gender and age distribution}
\end{table}

%% file: tables/corpus_design/improvised_scenario.tex
\begin{table}[!ht]
\begin{adjustbox}{width=\linewidth,center}{
    \centering
    \begin{tabular}{cccc}
    \toprule 
    \textbf{Batch} &\textbf{Situation} & \textbf{Actor A}   & \textbf{Actor B}  \\ \midrule 
    &1                 & \begin{tabular}[c]{@{}c@{}}\textbf{(Neutral)} \\ A hotel receptionist trying to explain \\ and service the customer\end{tabular}                   & \begin{tabular}[c]{@{}c@{}}\textbf{(Angry)} \\ An angry customer who is dissatisfied \\ with the hotel services\end{tabular}                               \\  \vspace{2mm}
    
    &2                 & \begin{tabular}[c]{@{}c@{}}\textbf{(Happy)} \\ A person excitingly talking with  \\ another person about his/her marriage plan\end{tabular}                      & \begin{tabular}[c]{@{}c@{}}\textbf{(Happy)}\\ A person happily talking with another person \\ and helping him/her plan the ceremony\end{tabular}               \\  \vspace{2mm}
    1&3                 & \begin{tabular}[c]{@{}c@{}}\textbf{(Sad)} \\ A patient feeling depressed\end{tabular}                                                              & \begin{tabular}[c]{@{}c@{}}\textbf{(Neutral)} \\ A doctor attempting to talk with A neutrally\end{tabular}                                   \\ \vspace{2mm}
    &4                 & \begin{tabular}[c]{@{}c@{}}\textbf{(Angry)} \\ A furious boss talking with the employee\end{tabular}                                               & \begin{tabular}[c]{@{}c@{}}\textbf{(Frustrated)} \\ A frustrated person attempting to argue \\ with his/her boss\end{tabular}                \\ \vspace{2mm}
    &5                 & \begin{tabular}[c]{@{}c@{}}\textbf{(Frustrated)} \\ A person frustratingly talk about \\ another person's action\end{tabular}                      & \begin{tabular}[c]{@{}c@{}}\textbf{(Sad)} \\ A person feeling guilty and sad about \\ his/her action\end{tabular}  \\ 
     \midrule
    &6                 & \begin{tabular}[c]{@{}c@{}}\textbf{(Happy)}\\ A happy hotel staff\end{tabular}                                                                    & \begin{tabular}[c]{@{}c@{}}\textbf{(Happy)}\\ A happy customer\end{tabular}                                                                    \\ \vspace{2mm}
    
    &7                 & \begin{tabular}[c]{@{}c@{}}\textbf{(Sad)} \\ A sad person who felt unsecured about \\ the upcoming marriage\end{tabular}                           & \begin{tabular}[c]{@{}c@{}}\textbf{(Frustrated)} \\ A person who is frustrated about \\ A's insecureness\end{tabular}              \\ \vspace{2mm}
    2&8                 & \begin{tabular}[c]{@{}c@{}}\textbf{(Frustrated)} \\ A frustrated patient\end{tabular}                                                              & \begin{tabular}[c]{@{}c@{}}\textbf{(Neutral)}\\ A Doctor talking with the patient\end{tabular}                                               \\ \vspace{2mm}
    &9                 & \begin{tabular}[c]{@{}c@{}}\textbf{(Neutral)} \\ A worker who is assigned to tell his/her \\ co-worker about the company's bad situation\end{tabular} & \begin{tabular}[c]{@{}c@{}}\textbf{(Sad)} \\ An employee feeling sad after listening\end{tabular}                                           \\ \vspace{2mm}
    &10                & \begin{tabular}[c]{@{}c@{}}\textbf{(Angry)} \\ A person raging about another person's behavior\end{tabular}                                        & \begin{tabular}[c]{@{}c@{}}\textbf{(Angry)}\\ A person who feels like being blamed \\ by the other\end{tabular}                         \\ 
    \midrule
    &11                & \begin{tabular}[c]{@{}c@{}}\textbf{(Frustrated)} \\ A director unsatisfied with a co-worker\end{tabular}                                              & \begin{tabular}[c]{@{}c@{}}\textbf{(Frustrated)} \\ A frustrated person who is trying their best \\ at the job\end{tabular}                           \\ \vspace{2mm}
    &12                & \begin{tabular}[c]{@{}c@{}}\textbf{(Happy)} \\ A person who gets a new job or promotion\end{tabular}                                               & \begin{tabular}[c]{@{}c@{}}\textbf{(Sad)} \\ A person who is desperated about his/her job\end{tabular}                                              \\ \vspace{2mm}
    3&13                & \begin{tabular}[c]{@{}c@{}}\textbf{(Neutral)} \\ A patient inquiring some information\end{tabular}                                                        & \begin{tabular}[c]{@{}c@{}}\textbf{(Happy)} \\ A happy doctor telling his/her patient \\ more information\end{tabular}                      \\ \vspace{2mm}
    &14                & \begin{tabular}[c]{@{}c@{}}\textbf{(Angry)} \\ A person who is upset with his/her work\end{tabular}                                                   & \begin{tabular}[c]{@{}c@{}}\textbf{(Neutral)} \\ A calm friend who listened \\ to A's problem\end{tabular}                      \\ 
    &15                & \begin{tabular}[c]{@{}c@{}}\textbf{(Sad)}\\ A person sadly telling another person \\ about his/her relationship\end{tabular}                                & \begin{tabular}[c]{@{}c@{}}\textbf{(Angry)} \\ A person who feels angry after listening to \\ A's bad relationship\end{tabular} \\
    \midrule 
    \end{tabular}
}
\end{adjustbox}
\caption{situations for the improvised sessions. Each actor pair is assigned to a batch of situations}
\label{tab:situations_improvisation}
\end{table}

%% file: tables/corpus_design/improvised_room.tex
\begin{table}[!ht]
\centering
\begin{tabular}{ccc}
\toprule 
Recording    & Environment & Situation batch \\
 \midrule 
Studio 1-18  & Studio: Room A        & 1               \\
Studio 19-20 & Studio: Room B        & 1               \\
Zoom 1-20    & Zoom                  & 1               \\
Studio 21-50 & Studio: Room B        & 2               \\
Studio 51-80 & Studio: Room B        & 3                \\
\midrule 
\end{tabular}
\caption{\label{tab:recording_situation_env} The mapping of the recording session, recording environment, and the situation batch.}
\end{table}

%% file: 3-annotation.tex

\section{Annotation}
\label{sec4}
\label{sec:annotation}

Once we obtained the processed utterances (in a format of both video and audio), we used two crowdsourcing platforms to annotate our dataset, including ``wang.in.th'' \citep{wangwebsite} and ``HOPE Annotation'' \citep{hopewebsite}.
The whole annotation process was done using audio modality only, as we want to solely rely on audio-based emotion for this dataset.
However, as mentioned in \autoref{sec:introduction}, quality controlling crowdsourcing annotations is challenging because annotators could be anyone with different backgrounds. 
This section will go through different approaches that we used to control data quality on the online annotation platform, how the annotator annotates the data through our online annotation platform, and the full processe we adopted for data quality control, respectively.

\subsection{Data Quality Control with Crowdsourcing}
\label{subsec:dqcwithcrowdsourcing}
\input{tables/annotation/crowdsource_qc}

Several works have proposed a method for screening low-quality data.
The overview of the crowdsourcing data quality control scheme was shown in \autoref{tab:annotation_method}.
Each method has its advantages, disadvantages, limitations, and suitability for different tasks. 
Choosing the proper method will result in higher quality data, and it may also save time and budget spent on annotating the data. 
In this work, we adopted three data quality control methods as follows: 

\begin{enumerate}

    \item \textbf{Majority Decision}: Collect 3 to 8 annotations per utterance and use the majority emotion.
    
    \item \textbf{Control Questions or Gold Standard Questions}: Randomly insert two gold standard questions (gold utterances) into each set of tasks assigned to the annotators. 
    These gold utterances, selected by expert studio directors, clearly express emotions that should be easily recognized by any attentive individual. 
    They serve as a quality control measure to detect whether annotators were paying attention or attempting to cheat.
    
    \item \textbf{Intrinsic matrix}: Add a duplicate utterance (consistency utterance) to each set of questions for consistency checking.
    
\end{enumerate}

In addition to the above methods, we also added a pretest assessment to control the quality of the annotator.

\subsection{Pretest assessment} 
\label{sec:pretest}

Before annotators begin working on the THAI-SER dataset, they are required to complete a pretest assessment.
The purpose of this pretest is twofold: to familiarize annotators with the task and to filter out low-performing candidates from contributing to the dataset.
The pretest includes a short tutorial video that demonstrates examples of each emotional category. 
This ensures a shared understanding and consistent interpretation of emotion labels across all annotators.
Following the video, annotators are given ten sample utterances as a screening task to assess their comprehension of the annotation process.
To further ensure engagement, one of the questions is a trick question that can only be answered correctly if the annotator has watched the video carefully. 
Specifically, at a random point in the tutorial, an animal sound is inserted, and the trick question asks which animal sound they heard.
To pass the pretest, annotators must correctly answer at least five of the standard questions as well as the trick question.
Out of 1,759 unique applicants, only 984 annotators (56\%) successfully passed the assessment.

\subsection{Annotation task}
\label{subsec:annotation_task}

To maximize the efficiency of the annotation, we conduct the annotation process in parallel to the recordings.
The annotations were divided into several batches named ``projects'', which were later added to the crowdsourcing platform.
Once the annotator passed the pretest assessment, they were subjected to choosing a ``project'' on the crowdsourcing platform.
Each project consists of around 600-800 tasks, and each task consists of 10 utterances to annotate.
For each task, the annotators were given a total of 10 utterances to annotate, where each utterance is one of the following:

\begin{itemize}
    \item \textbf{Regular utterance}: These are the utterances that need to be annotated.
    \item \textbf{Gold utterance}: Those that were explicit in their emotions and easy to identify, as guaranteed by the directors. This type of utterance was used to assess annotators' reliability.
    \item \textbf{Consistency utterance}: The duplicated utterance for the purpose of examining the consistency of the annotation. The consistency utterance was chosen from the regular utterance.
\end{itemize}

Within the crowdsourcing platform, the task was separated into two pages: the first page (4 regular utterances and 1 gold utterance) and the second page (3 regular utterances, 1 gold utterance, and 1 consistency utterance).

\begin{figure}[!ht]
    \centering
    \includegraphics[width=0.7\textwidth]{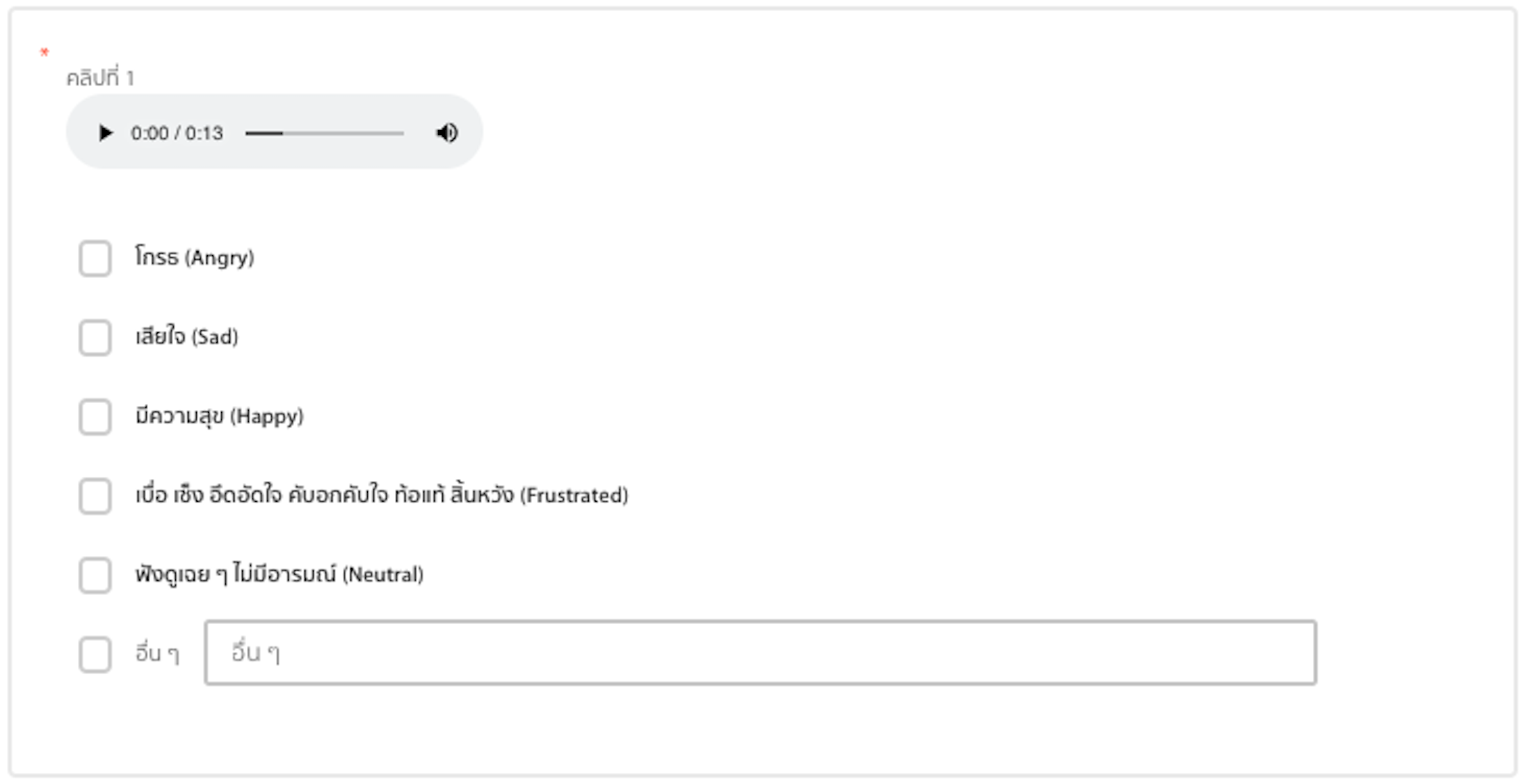}
    \caption{\label{wang_task}A user interface of annotating an utterance.}
\end{figure}

\autoref{wang_task} presented an annotation interface for each utterance, providing six multiple choices: angry, happy, neutral, sad, frustrated, and others. 
Annotators could choose ‘others’ when the five emotions were not appropriate or report any technical problems (e.g., no audio, audio error, etc.). 
We allowed annotators to select more than one choice when they felt more than one emotion in the utterance. 
After completing each task, the annotators pressed a submit button before proceeding to the next task, or they could decide to leave. 
We would not collect any completed annotations anyhow if the annotators did not complete all the utterances of the task by pressing the ``submit'' button.

For each annotation project, we restricted annotators to not do more than 30 tasks.
This was to avoid annotators' fatigue and situation memorization, which led to bias in perception. 
Since we only had 15 improvised situations, certain keywords might bias the more perceptive annotators if they realized what situation the utterance might be from.

\subsection{Annotation Quality Control Process}

\begin{figure}[!ht]
    \centering
    \includegraphics[width=0.9\textwidth]{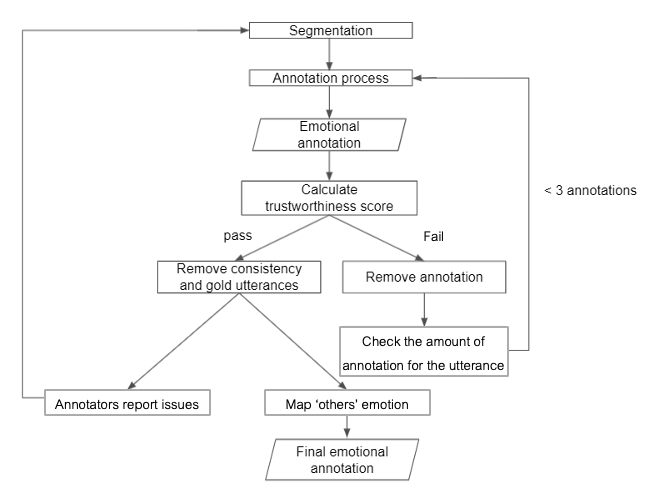}
    \caption{\label{cleaning_process}Overview of the annotation process. Utterances are segmented and distributed for annotation. Each utterance requires at least three trustworthy annotations. ‘other’ emotion are also handled manually by the researchers by either fixing the segmentation or re-annotation.}
\end{figure}

\autoref{cleaning_process} showed the whole annotation quality control loop.
The first three boxes were described in previous sections (\textit{segmentation} in \autoref{subsec:alignment}, \textit{Annotation process} and \textit{Emotional annotation} in \autoref{sec:pretest}, \autoref{subsec:annotation_task}).
This section highlights the process after the aforementioned step on how we iterate over multiple annotations, ensure annotation trustworthiness, and deal with ``others'' emotion (see \autoref{wang_task}).

After annotators completed their annotation tasks, we evaluated the trustworthiness of the annotator which was calculated from the results of annotated \textit{gold utterances} and \textit{consistency utterance} existed in the completed tasks.
\textit{Gold utterances} were used to calculate ``confidence score'' which is the accuracy of the inserted gold utterances.
\textit{Consistency utterances} were also used to calculate the ``consistency score'' which was calculated by matching the annotation of the consistency utterances in each task.
The annotators must score more than 50 \% on both indicators in order to be considered trustworthy. 
In the event that the annotators had no trustworthiness (i.e., they failed at least one indicator), the platform would automatically remove the annotations of that annotator and update the utterance number of annotators accordingly.
If the number of annotations in any utterance had been lower than three, those utterances were re-annotated until they achieved at least three annotations.
\autoref{tab:annotator trustworthiness} presents the statistics of all annotators in terms of trustworthiness.

Upon investigating the annotators that have no trustworthiness, we found that most of the annotators failed the confidence indicator.
Over 40\% of the non-trustworthiness annotators failed the confidence indicator alone, 35\% failed the consistency indicator, and 25\% of them failed both metrics.

\input{tables/annotation/trustworthiness}

After obtaining the annotation, we removed the consistency utterances and gold utterances from the set. 
Any utterances that were reported due to incomplete segmentation were resegmented and then sent to the whole annotation process again.
Then, we manually checked the annotations marked with `other' and manually assigned all unique `other' emotions to the closest available emotion as shown in \autoref{tab:emotional_label_mapped}.
The `other' emotion that can't be mapped back to the five main emotions were unchanged, annotated as `other' in the dataset.

\input{tables/annotation/emotion_mapped}

The final statistics of our data is summarized in \autoref{tab:final_stats}.

\input{tables/annotation/final_stats}

%% file: tables/annotation/crowdsource_qc.tex
\begin{table}[!ht]
\begin{adjustbox}{width=\linewidth,center}
{
    \centering
    \begin{tabular}{cclc}
    \toprule
    \addlinespace[0.2em]
    \textbf{Method} & \textbf{References} & \textbf{Detail} & \textbf{\begin{tabular}[c]{@{}c@{}}Applied in \\ this corpus\end{tabular}} \\
    \addlinespace[0.2em]
    \midrule
    \addlinespace[0.2em]
    \begin{tabular}[c]{@{}c@{}}\textbf{Majority Decision,}\\ \textbf{Data Redundancy,}\\ \textbf{Plurality Answer Agreement}\end{tabular} 
    & 
    \begin{tabular}[c]{@{}c@{}} \cite{cs_majority_1},\\ \cite{cs_majority_2},\\ \cite{cs_majority_3},\\ \cite{cs_majority_4} \end{tabular} 
    &
    \begin{tabular}[c]{@{}l@{}}
        Each data point is annotated by \\
        multiple annotators, and the results \\
        are determined by a majority consensus.\end{tabular} 
    & 
    \checkmark \\
    
    \addlinespace[0.2em]
    \midrule
    \addlinespace[0.2em]
    
    \begin{tabular}[c]{@{}c@{}}\textbf{Control Questions or}\\ \textbf{Gold Standard Questions}\end{tabular} 
    & 
    \begin{tabular}[c]{@{}c@{}} \cite{cs_majority_4},\\ \cite{cs_gold_1},\\ \cite{cs_gold_2},\\ \cite{cs_gold_3},\\ \cite{cs_gold_4},\\ \cite{cs_gold_5} \end{tabular} 
    &
    \begin{tabular}[c]{@{}l@{}}
        A dataset containing known answers \\
        (gold/control standards) is interleaved \\
        with other data. Annotators who fail \\
        to meet the standard have their results \\
        rejected.\end{tabular} 
    & 
    \checkmark \\
    
    \addlinespace[0.2em]
    \midrule
    \addlinespace[0.2em]
    
    \begin{tabular}[c]{@{}c@{}}\textbf{Intrinsic Matrix,}\\ \textbf{Transitivity Rule}\end{tabular} 
    & 
    \begin{tabular}[c]{@{}c@{}} \cite{cs_gold_1},\\ \cite{cs_gold_5},\\ \cite{cs_intrinsic} \end{tabular} 
    &
    \begin{tabular}[c]{@{}l@{}}
        This method applies the transitive property:\\
        if “a equals b” and “b equals c”, then \\
        “a equals c”. It is applicable to tasks \\
        involving equality, preference, or comparison.\\ 
        Annotators whose responses violate this rule \\
        are rejected.\end{tabular} 
    & 
    \checkmark \\
    
    \addlinespace[0.2em]
    \midrule
    \addlinespace[0.2em]
    
    \begin{tabular}[c]{@{}c@{}}\textbf{Iterative Approaches}\end{tabular} 
    & 
    \begin{tabular}[c]{@{}c@{}} \cite{cs_iterative} \end{tabular} 
    &
    \begin{tabular}[c]{@{}l@{}}
        An iterative process enhances data collection \\
        quality, particularly for tasks such as writing \\
        and brainstorming. Annotators build on one \\
        another’s contributions in successive rounds.
        \end{tabular} 
    & 
    {} \\
    
    \addlinespace[0.2em]
    \midrule
    \addlinespace[0.2em]
    
    \begin{tabular}[c]{@{}c@{}}\textbf{Behavior Matrix,}\\ \textbf{Additional Information Rule}\end{tabular} 
    & 
    \begin{tabular}[c]{@{}c@{}} \cite{cs_majority_4},\\ \cite{cs_gold_1},\\ \cite{cs_gold_5},\\ \cite{cs_behavior_1},\\ \cite{cs_behavior_2},\\ \cite{cs_behavior_3},\\ \cite{cs_behavior_4} \end{tabular} 
    &
    \begin{tabular}[c]{@{}l@{}}
        Additional contextual information is used to \\
        verify annotation quality. For example, \\
        timestamps may be used to reject responses \\
        completed in less time than the audio duration.
        \end{tabular} 
    & 
    {} \\
    
    \addlinespace[0.2em]
    \midrule
    \addlinespace[0.2em]
    
    \begin{tabular}[c]{@{}c@{}}\textbf{Control Group,}\\ \textbf{Professional Judge}\end{tabular} 
    & 
    \begin{tabular}[c]{@{}c@{}} \cite{cs_majority_1} \end{tabular} 
    &
    \begin{tabular}[c]{@{}l@{}}
        Following an initial round, annotations are \\
        re-checked by a separate group. The reviewers \\
        may be either an independent set of annotators \\
        or professional judges specially selected for \\
        quality control.
        \end{tabular} 
    & 
    {} \\
    \addlinespace[0.2em]
    \bottomrule
    \end{tabular}
}
\end{adjustbox}
\caption{Existing crowdsourcing quality control methods}
\label{tab:annotation_method}
\end{table}

%% file: tables/annotation/trustworthiness.tex
\begin{table}[!ht]
\begin{adjustbox}{width=\linewidth,center}
{
    \centering
    \begin{tabular}{ccccccc}
    
        \toprule

        \multirow{5}{*}{\textbf{\begin{tabular}[c]{@{}c@{}}Batch\end{tabular}}}
            & \multirow{5}{*}{\textbf{\begin{tabular}[c]{@{}c@{}}Annotators\end{tabular}}}
            & \multirow{5}{*}{\textbf{\begin{tabular}[c]{@{}c@{}}Trust Worthy\end{tabular}}}
            &  \multicolumn{4}{c}{\textbf{Not trustworthy}}  \\ 
            \cline{4-7} 
            &
            &
            & \multicolumn{1}{c}{\textbf{Total}} 
            & \textbf{\begin{tabular}[c]{@{}c@{}}
                Fail both \\ 
                confidence \\ 
                and consistency
                \end{tabular}} 
            & \textbf{\begin{tabular}[c]{@{}c@{}}
                Fail only \\ 
                confidence \\ 
                check
                \end{tabular}} 
            & \textbf{\begin{tabular}[c]{@{}c@{}}
                Fail only \\ 
                consistency \\
                check
                \end{tabular}} \\
                
        \midrule
        
        1   &   334 &   284 (85.03\%) & \multicolumn{1}{c}{50}   &   8   &   27  &   15  \\
        2   & 137   &   126 (91.97\%) & \multicolumn{1}{c}{11}  &   3  &    4   &   4\\
        3   & 129  & 116 (89.92\%)    & \multicolumn{1}{c}{13}   & 7     & 2    & 4  \\
        4   & 128    & 113 (88.28\%) & \multicolumn{1}{c}{15}    & 3     & 8    & 4  \\
        5   & 145     & 107 (73.79\%)  & \multicolumn{1}{c}{38}   & 23   & 9   & 6  \\
        6   & 191    & 175 (91.62\%)   & \multicolumn{1}{c}{16}    & 4     & 5  & 7  \\ 
        \midrule
    \end{tabular}
}
\end{adjustbox}
\caption{\label{tab:annotator trustworthiness}Distribution of annotators trustworthiness across different fail type.}
\end{table}

%% file: tables/annotation/emotion_mapped.tex
\begin{table}[!ht]
\small
\centering
\begin{tabular}{cc}
\toprule
\textbf{List of `other' emotions that can be mapped} & \textbf{Mapped emotion} \\ \midrule
calm, sympathy, encourage, peaceful, optimistic, persuasion    & neutral            \\
mad, aggressive, hateful                                       & angry                 \\
surprise, amazed, excited, joyful, proud, fun                  & happy             \\
guilty, despair, depressed                                     & sad               \\
bored, irritated, infuriated, dissatisfied, sarcastic, annoyed & frustrated           \\
\midrule
\end{tabular}
\caption{\label{tab:emotional_label_mapped}Emotional label mapped. If the annotators put in any of the listed emotion, they are mapped manually to the five main emotions.}
\end{table}

%% file: tables/annotation/final_stats.tex
\begin{table}[!ht]

\begin{adjustbox}{width=\linewidth,center}
{ 
    \centering
    \begin{tabular}{ccccccccccc}
    
        & \multicolumn{4}{c}{\textbf{Studios}}
        & \multicolumn{4}{c}{\textbf{Zoom}}
        & \multicolumn{2}{c}{}
        \\ 
        
        \cline{2-9}
        
        & \multicolumn{2}{c}{\textit{scripted}} 
        & \multicolumn{2}{c}{\textit{improvised}} 
        & \multicolumn{2}{c}{\textit{scripted}}
        & \multicolumn{2}{c}{\textit{improvised}}
        & \multicolumn{2}{c}{\multirow{-2}{*}{\textbf{Total}}} 
        \\ 
        
        \cline{2-11} 
        
        \multirow{-3}{*}{\textbf{Emotion}} 
        & Hours
        & \%
        & Hours
        & \%
        & Hours
        & \%
        & Hours
        & \%
        & Hours
        & \%
        \\ 
        
        \midrule
        
        Neutral
        & 3.79
        & 27.66\%
        & 3.81
        & 21.16\%
        & 0.97
        & 24.18\%
        & 1.12
        & 19.11\%
        & 9.70
        & 23.30\%
        \\

        Angry
        & 1.04
        & 7.57\%
        & 2.12
        & 11.78\%
        & 0.35
        & 8.71\%
        & 0.67
        & 11.42\%
        & 4.18
        & 10.05\%
        \\
        
        Happy
        & 1.68
        & 12.27\%
        & 2.47
        & 13.69\%
        & 0.46
        & 11.45\%
        & 0.94
        & 15.90\%
        & 5.54
        & 13.32\%
        \\
        
        Sad
        & 2.28
        & 16.68\%
        & 2.06
        & 11.42\%
        & 0.63
        & 15.57\%
        & 0.58
        & 9.84\%
        & 5.55
        & 13.33\%
        \\
        
        Frustrated
        & 3.51
        & 25.64\%
        & 5.98
        & 33.20\%
        & 1.17 
        & 29.13\% 
        & 2.04
        & 34.59\%
        & 12.69
        & 30.52\%
        \\
        
        Others
        & 0.00
        & 0.00\%
        & 0.00
        & 0.02\%
        & 0.00
        & 0.00\%  
        & 0.00
        & 0.03\%
        & 0.01
        & 0.01\%
        \\

        None
        & 1.39
        & 10.17\%
        & 1.57
        & 8.73\%
        & 0.44
        & 10.96\% 
        & 0.54
        & 9.11\%
        & 3.94
        & 9.48\%
        \\ 
        
        \midrule
        
        \textbf{Total}
        & 13.69
        &
        & 18.01
        &
        & 4.02
        &
        & 5.89
        &
        & \textbf{41.61}
        & 
        \\
        
        \bottomrule
    
    \end{tabular}
}
\end{adjustbox}

\caption{\label{tab:final_stats}Final statistics of the whole corpus. \textit{Others} denotes the emotion that annotators can't reach consensus.}

\end{table}

%% file: 4-data-evaluation.tex
\section{Data Evaluation}
\label{sec:evaluation}

To ensure the reliability and correctness of our dataset, we evaluate annotation quality using three complementary metrics: majority agreement, inter-annotator reliability (IAR), and human recognition accuracy (HRA).
Based on these metrics, we identify and filter out low-agreement samples to improve the overall statistical reliability of the corpus—measured via Krippendorff’s alpha.
Finally, we apply these metrics to analyze how annotation quality varies with actor demographics (e.g., gender and age), evaluate recognition accuracy before and after filtering, and assess the effectiveness of pre-screening procedures during pilot annotations.

\subsection{Evaluation Metrics}
\label{subsec:eval_metrics}

We evaluate our dataset reliability using three metrics: majority agreement, inter-annotator reliability, and human recognition accuracy.

\subsubsection{Majority Agreement}
\label{sec:majority-agreement}

Given an utterance $x_i$ with $N$ annotators selecting $K$ available emotions, the annotation $\textbf{y}\in \{ 0,1 \}^{N\times K}$ is a binary matrix where $y_{nk}$ is equal to 1 if \textit{annotator $i$ selects an emotion $j$} and is 0 otherwise.
The majority agreement score represents the \textit{consensus score of the majority voted emotion}.
Specifically, the majority agreement of an utterance $x_i$ can be calculated as follows:

\begin{equation}
    \label{majority_equation}
    \text{agreement}(x_i) =
    \begin{cases}
        \displaystyle \max\limits_k \frac{1}{N} \sum\limits_{n=1}^N \frac{\textbf{y}_{nk}}{\Sigma_{k=1}^K \textbf{y}_{nk}}, & \text{if consensus can be reached} \\
        0, & \text{otherwise}
    \end{cases}
\end{equation}

In other words, majority agreement was set to a default of 0 if consensus can't be reached among annotators, and the emotion label will be set to `None'.
If consensus can be reached, the agreement measures the averaged \textit{adjusted} score of annotators to the majority emotion.

\begin{figure}[!ht]
    \centering
    \includegraphics[width=1.0\textwidth]{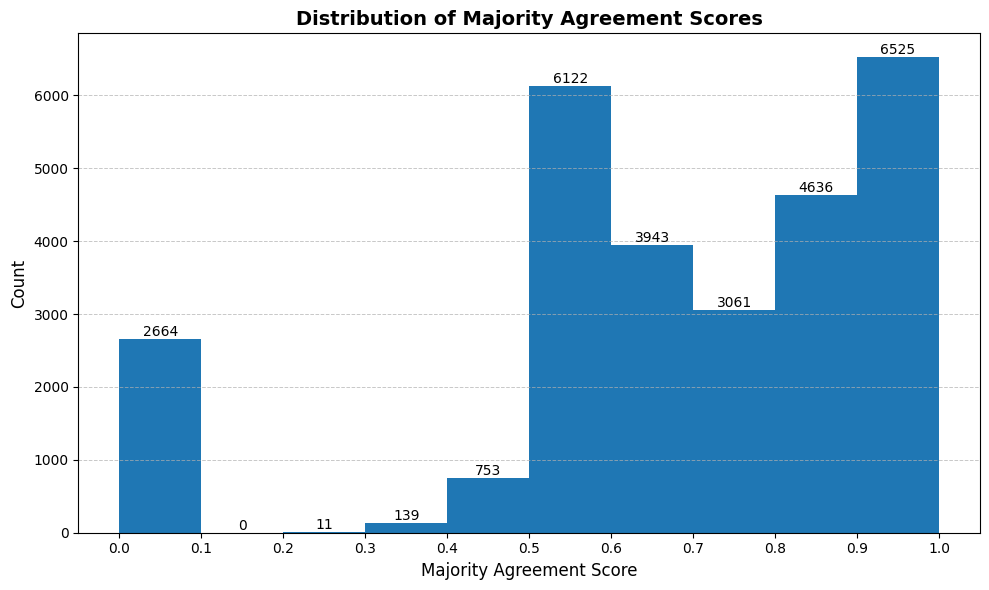}
    \caption{\label{agreement_histrogram}Histogram of the majority agreement score.}
\end{figure}

\autoref{agreement_histrogram} shows the distribution of majority score in the full THAI-SER corpus.

\subsubsection{Inter-Annotator Reliability}
\label{sec:inter-annotator}

In addition to evaluating the consensus among annotators \textit{for each utterance}, we also evaluate the consensus of annotators \textit{of the whole corpus}.
Inter-Annotator Reliability (IAR) measures the consistency of the annotations of the whole dataset.
A suitable IAR metric is crucial, especially given the nature of our data collection, where the number of annotators per utterance varied, ranging between 3 and 8.
\footnote{Initial annotation rounds involved up to 8 annotators per item to refine the guidelines and assess the annotation process thoroughly. Subsequently, most utterances were annotated by 3 individuals.}
Furthermore, our task allowed annotators to select multiple emotion labels for a single utterance, resulting in set-valued annotations.
This makes some IAR scores, such as the Kappa score \citep{kappascore}, which assumed that the number of annotators must be fixed, inapplicable.

Considering these factors, variable number of annotators per item, and set-valued annotation data, Krippendorff's alpha ($\alpha$) \citep{Krippendorff-2004, Krippendorff-2011} is an appropriate choice. 
Krippenorff's alpha is a versatile IAR coefficient that handles multiple annotators, accommodates missing data (implicitly managing the uneven number of annotations per item), and can be adapted for different levels of measurement, including set-valued data when paired with a suitable distance metric.

For our set-valued emotion labels, we employed the Measuring Agreement on Set-valued Items (MASI) distance metric \citep{MASI} within the Krippendorff's alpha calculation. 
The general formula for Krippendorff's alpha is

\begin{equation}
\alpha = 1 - \frac{D_o}{D_e}
\label{eq:alpha}
\end{equation}

where $D_o$ represents the observed disagreement among the annotations provided, and $D_e$ represents the disagreement expected purely by chance.

The observed and expected disagreements are calculated based on the pairwise comparisons of annotations assigned to each item, weighted by the distance between those annotations. 
Using the MASI metric, these are computed as

\begin{subequations}
\begin{align}
    D_{o} &= \sum_{v, v' \in V} o_{vv'} \cdot \delta_{MASI}(v, v') \label{eq:Do} \\
    D_{e} &= \sum_{v, v' \in V} e_{vv'} \cdot \delta_{MASI}(v, v') \label{eq:De}
\end{align}
\end{subequations}

In these equations:
\begin{itemize}
    
    \item $V$ is the set of all unique annotation sets (i.e., combinations of emotion labels, including single labels) observed across all utterances in the dataset.
    
    \item $o_{vv'}$ represents the observed proportion of pairable assignments where annotation set $v$ is paired with annotation set $v'$ within the same utterance. 
    This is derived from a coincidence matrix summarizing the agreement and disagreement patterns in the actual annotation data across all items and annotator pairs.

    \item $e_{vv'}$ represents the expected proportion of pairings between set $v$ and set $v'$ under the null hypothesis of chance assignment. 
    This is calculated based on the marginal distributions of the observed annotation sets $v$ and $v'$.

    \item $\delta_{MASI}(v, v') = 1 - MASI(v, v')$ is the distance function used.
    Since MASI measures \textit{agreement} (similarity) between sets $v$ and $v'$, we use $1 - MASI$ to quantify their \textit{disagreement} or distance, as required by the alpha formula.

\end{itemize}

The MASI metric \citep{MASI} itself quantifies the similarity between two sets, $v$ and $v'$, specifically designed for tasks like multi-label annotation. 
It ranges from 0 (completely disjoint sets) to 1 (identical sets) and is calculated as the product of the Jaccard index ($J$) and a monotonicity measure ($M$):

\begin{equation}
MASI(v, v') = J(v, v') \times M(v, v')
\label{eq:masi}
\end{equation}

The Jaccard index ($J$) measures the overlap between the two sets:

\begin{equation}
J(v, v') = \frac{\lvert v \cap v'\rvert}{\lvert v \cup v'\rvert}
\label{eq:jaccard}
\end{equation}

where $\lvert v \cap v' \rvert$ is the number of elements common to both sets, and $\lvert v \cup v'\rvert$ is the total number of unique elements across both sets. $J=1$ if $v=v'$, and $J=0$ if the sets are disjoint ($v \cap v' = \emptyset$).

The monotonicity term ($M$) refines the agreement score based on the structural relationship between the sets:

\begin{itemize}

    \item $M = 1$, if the sets are identical ($v = v'$).

    \item $M = 2/3$, if one set is a proper subset of the other ($v \subset v'$ or $v' \subset v$). 
    This assigns partial agreement when one annotation is a more specific version of the other.

    \item $M = 1/3$, if the sets intersect but neither is a subset of the other ($v \cap v' \neq \emptyset$, $v \not\subseteq v'$, and $v' \not\subseteq v$). 
    This reflects a lower level of partial agreement.

    \item $M = 0$, if the sets are disjoint ($v \cap v' = \emptyset$). 
    Note that in this case, $J$ is also 0, resulting in $MASI = 0$.

\end{itemize}

Using $1-MASI$ as the distance $\delta_{MASI}$ means identical sets have a distance of 0, disjoint sets have a distance of 1, and partially overlapping or subset-related sets have intermediate distances between 0 and 1.

Krippendorff's alpha values range from $\alpha \le 0$ (indicating disagreement is at or exceeds chance levels) to $\alpha = 1$ (representing perfect reliability). 
Values greater than 0 signify agreement above chance. 
While interpretation can depend on the context and stakes of the research, \citet{Krippendorff-2004} suggests that $\alpha \ge 0.667$ is a minimum acceptable value of a reliable corpus.

\subsubsection{Human recognition accuracy}
\label{sec:human-recognition-accuracy}

Human recognition accuracy (HRA) is a metric that generally describes how accurate the annotators perceive emotions compared to the assigned emotions, which is defined as follows:

\begin{equation}
\text{HRA}(x_i, \textbf{y}_i) = \frac1N\sum_{i=1}^N \textbf{1}(\text{maj}(\textbf{y}_i) =\text{assigned}(x_i))
\end{equation}

where $\text{maj}(\textbf{y}_i)$ denotes the majority voted emotion of annotation matrix $\textbf{y}_i$ and $\text{assigned}(x_i)$ denotes the assigned emotion of the i-th utterance.

\subsection{Results}
\label{sec:result}

\subsubsection{Determining the Optimal Agreement Threshold}
\label{sec:optinal_agreement}

\begin{figure}[!ht]
    \centering
    \includegraphics[width=0.6\textwidth]{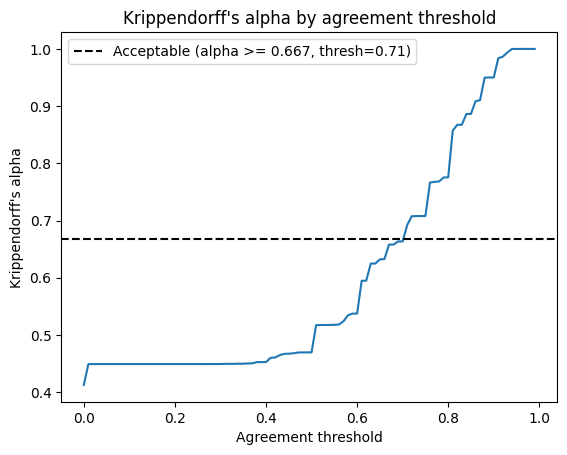}
    \caption{\label{fig:alpha_by_thresh}Krippendorff's alpha scores of the corpus under different agreement threshold filtering conditions.}
\end{figure}

Since emotion annotations can sometimes be ambiguous—especially in our setup, where annotators are not allowed to access the actor's facial expressions—the majority agreement can vary across different annotators, as shown in \autoref{agreement_histrogram}.
As a result, the raw THAI-SER corpus achieves a relatively low Krippendorff's alpha score of 0.413, which falls below the commonly recommended threshold of 0.667 \citep{Krippendorff-2004}.
To address this issue, we filter out utterances with low agreement scores, which in turn increases the Krippendorff's alpha of the corpus.
\autoref{fig:alpha_by_thresh} shows the Krippendorff's alpha values obtained when varying the agreement score filtering threshold.
The optimal threshold that yields an alpha score of $\geq 0.667$ is 0.71.
\textbf{Thus, we strongly recommend that any researcher using the THAI-SER corpus remove utterances with an agreement score lower than 0.71 to ensure higher corpus reliability.}

\subsubsection{Effect of Actor Gender, Age, and Recording Type}

Our corpus includes utterances performed by a diverse demographic of actors, spanning various genders and age groups. 
To assess whether these factors influence acting performance, we analyzed inter-annotator reliability and human recognition accuracy, grouped by actor gender and age.
We further examined the impact of recording type by comparing improvised sessions with scripted sessions of varying emotional intensity (low, high, and all combined).
All analyses were conducted under two conditions: before and after filtering samples using an agreement threshold of 0.71 (see \autoref{sec:optinal_agreement}), to examine how excluding low-agreement utterances affects reliability and recognition.

As shown in \autoref{tab:session-alpha-accuracy}, data filtered to retain only samples above the agreement threshold yield a human recognition accuracy of 77.2\%.
Among the scripted sessions, those with high emotional intensity result in the highest inter-annotator reliability and recognition accuracy, whereas low-intensity sessions show a notable drop in both metrics.
This trend suggests that stronger emotional expressiveness facilitates consensus among annotators and improves recognition outcomes.
\textbf{These findings underscore the importance of emotional intensity in the design of acted corpora, as reduced intensity significantly diminishes annotation agreement, even though lower-intensity expressions may better reflect real-world emotional distributions where strong emotions are rare.}

Interestingly, improvised sessions achieve higher inter-annotator reliability than scripted sessions overall, indicating that naturally elicited emotions may lead to clearer consensus among annotators.
However, despite the higher agreement, human recognition accuracy for improvised utterances is lower than for scripted sessions.
We caution against drawing a strong conclusion that improvised speech is inherently more difficult to recognize; rather, the disparity in expressiveness across scripted intensities (e.g., 0.690 for low-intensity vs. 0.883 for high-intensity) likely dominates this outcome.
Hence, we conclude that the difficulty of emotion recognition in improvised speech lies between that of low- and high-intensity scripted expressions.
\textbf{These findings further highlight the importance of incorporating elicited (improvised) speech in emotion corpora, as it represents a distinct and more consistently annotated acting style compared to scripted speech, which can vary substantially in reliability depending on emotional intensity.}

\input{tables/data_evaluation/session_alpha_accuracy}
\input{tables/data_evaluation/sex_age_alpha_accuracy}

\subsubsection{Emotion-Level Agreement and Recognition Patterns}

\begin{figure}[!ht]
    \centering
    \includegraphics[width=1\textwidth]{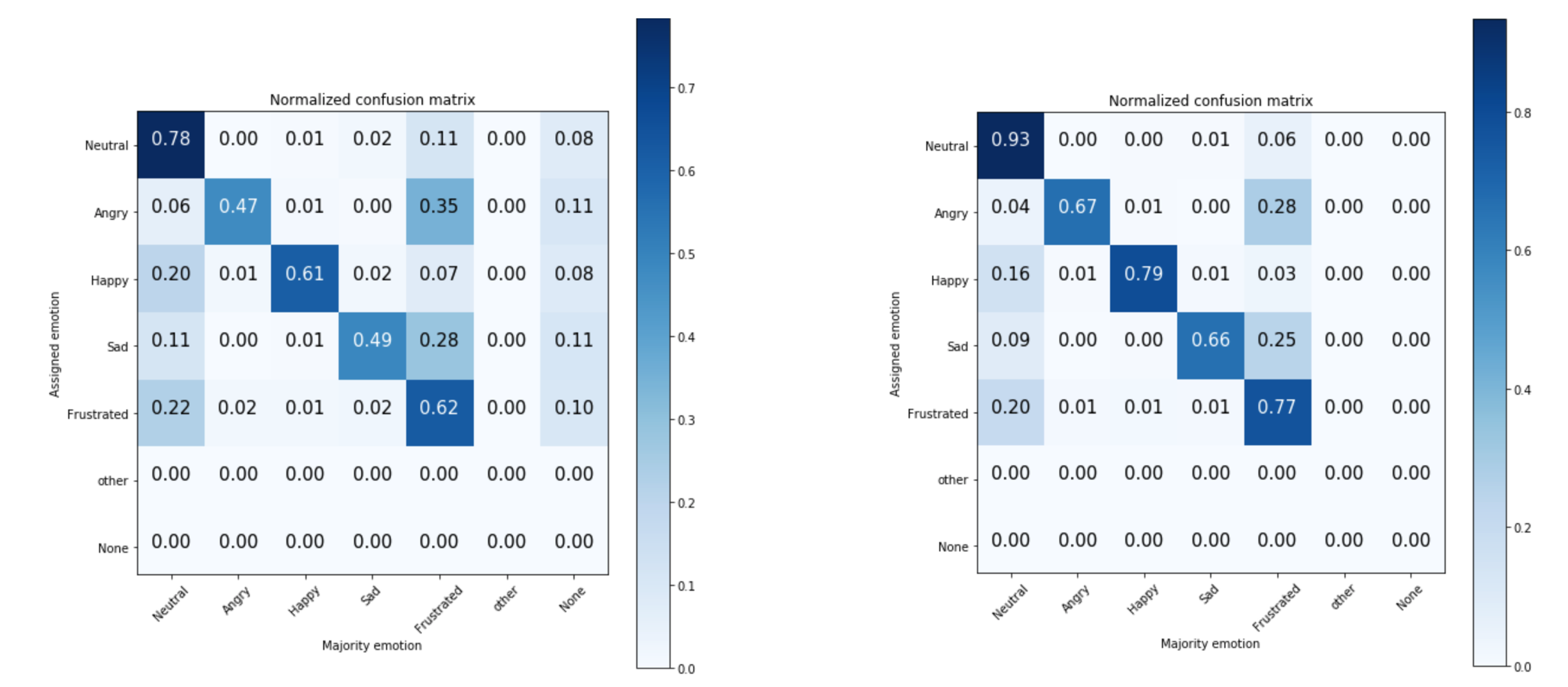}
    \caption{\label{cm}Confusion matrix between the assigned emotion and the majority annotation
            (Left: all data with a total of 27,854, and right: data that have a majority $\geq$ 0.71 with a total of 14,182)
            }
\end{figure}

We also inspect the confusion matrix comparing actors' assigned emotions and the majority-voted annotations to highlight the impact of majority agreement thresholding.
The confusion matrix is presented in \autoref{cm}.

Neutral was the most accurately recognized emotion, with recognition rates of 78\% and 93\% under the two agreement settings.
Emotions such as happy and frustrated showed lower recognition rates, approximately 61\% and 62\% across the entire corpus.
After filtering the data to include only samples with a majority agreement threshold of $\geq$ 0.71, recognition rates for happy and frustrated improved by approximately 18\% and 15\%, respectively.
However, there was notable confusion among annotators: frustrated was often mistaken for angry or sad, and also frequently confused with neutral.

These results provide useful guidelines for corpus development and downstream modeling. 
Emotions such as \textit{frustrated} present a high rate of false negatives due to confusion with semantically adjacent categories like \textit{angry} and \textit{sad}, highlighting the need for clearer annotation definitions.
On the other hand, \textit{neutral}, despite being the most accurately recognized overall, also exhibits a high rate of false positives, often being confused with low-arousal emotions.
\textbf{We recommend treating such ambiguous categories with special care through clearer annotation guidelines to improve both annotation reliability and classification robustness.}

\subsubsection{Effectiveness of Pre-screening in Annotation}

Before launching the full-scale annotation, we conducted two pilot trials to identify potential issues in the annotation pipeline and evaluate the effectiveness of our pre-screening procedures.
For each pilot, we assessed three key metrics: (i) (i) the percentage of annotators passing the pretest (see \autoref{sec:pretest}), (ii) inter-annotator reliability, and (iii) human recognition accuracy. 
The results of the pilot trials and the final annotation setup are shown in \autoref{tab:annotation_prescreen}.

\input{tables/data_evaluation/annotation_prescreen}

In the first pilot (Pilot 1), only 40\% of annotators passed the pretest. The resulting inter-annotator reliability was 37.9\%, and human recognition accuracy was 54.9\%. These results suggested that the pretest may not have been effective at filtering annotators who could reliably perform the task.
To improve quality control, the second pilot (Pilot 2) incorporated three additional pretest items that used a comparative task format: annotators were shown two utterances and asked to choose the one that better expressed a given emotion.
To increase overall pass rates, we also simplified the original pretest items by selecting utterances with more intense emotional expressions.

However, despite these changes, the passing rate dropped to 23.5\%. We hypothesize that the additional comparative questions introduced greater subjectivity, leading to inconsistent judgments and lower scores. 
Although inter-annotator reliability and recognition accuracy showed slight improvements, the gains were marginal compared to the drop in participant retention.

As a result, we excluded the additional comparative questions in the final annotation setup. 
In the actual run, we retained the simplified pretest items from Pilot 2 but reverted to the original question format. 
This setup achieved a higher passing rate of 55.9\%, a modest increase in inter-annotator reliability (41.3\%), and slightly lower human recognition accuracy (59.2\%).

\textbf{These results suggest that while stricter pretests can marginally improve data quality, overly complex screening tasks may reduce annotator throughput without substantial gains in annotation reliability.}

%% file: tables/data_evaluation/session_alpha_accuracy.tex
\begin{table}[!ht]
\begin{adjustbox}{width=\linewidth,center}{
    \centering
    \begin{tabular}{@{}cccccc@{}}
    \toprule
    \multicolumn{1}{c}{} &
    \multicolumn{1}{c}{} &
    \multicolumn{2}{c}{\textbf{Inter-annotator reliability}} &
    \multicolumn{2}{c}{\textbf{Human recognition accuracy}}\\             \cmidrule(lr){3-4} \cmidrule(lr){5-6}
    \textbf{Recording type} & 
    \textbf{Intensity} &
    \begin{tabular}[c]{@{}c@{}}\textbf{Agreement $\geq$ 0} \\ (27,854)\end{tabular}  &
    \begin{tabular}[c]{@{}c@{}}\textbf{Agreement $\geq$ 0.71} \\ (14,182)\end{tabular}  &
    \begin{tabular}[c]{@{}c@{}}\textbf{Agreement $\geq$ 0} \\ (27,854)\end{tabular}  &
    \begin{tabular}[c]{@{}c@{}}\textbf{Agreement $\geq$ 0.71} \\ (14,182)\end{tabular}  \\ \midrule
    All & - & 0.413 & 0.692 & 0.592 & 0.772 \\ 
    Improvised & - & 0.426 & 0.697  & 0.589 & 0.753 \\  
    Scripted & All & 0.392 & 0.680  & 0.596 & 0.798 \\      
    Scripted & Low & 0.315 & 0.622  & 0.501 & 0.690 \\       
    Scripted & High & 0.454 & 0.712  & 0.691 & 0.883 \\ \midrule
    \end{tabular}
}\end{adjustbox}
\caption{Inter-annotator reliability and human recognition accuracy of each session at 0 and 0.71 agreement.}
\label{tab:session-alpha-accuracy}
\end{table}

%% file: tables/data_evaluation/sex_age_alpha_accuracy.tex
\begin{table}[!ht]
\begin{adjustbox}{width=\linewidth,center}{
    \centering
    \begin{tabular}{@{}cccccc@{}}
    \toprule
    & & \multicolumn{2}{c}{\textbf{Inter-annotator reliability}}                                                                 
    & \multicolumn{2}{c}{\textbf{Human recognition accuracy}}                                                           
    \\  \cmidrule(lr){3-4} \cmidrule(lr){5-6}
    \begin{tabular}[c]{@{}c@{}}\textbf{Actor's} \\ \textbf{Gender} \\ \textbf{or Age} \end{tabular}
    &\textbf{Group}                    & \begin{tabular}[c]{@{}c@{}}\textbf{Agreement $\geq$ 0} \\ (27,854)\end{tabular}  &
    \begin{tabular}[c]{@{}c@{}}\textbf{Agreement $\geq$ 0.71} \\ (14,182)\end{tabular}  &
    \begin{tabular}[c]{@{}c@{}}\textbf{Agreement $\geq$ 0} \\ (27,854)\end{tabular}  &
    \begin{tabular}[c]{@{}c@{}}\textbf{Agreement $\geq$ 0.71} \\ (14,182)\end{tabular}  \\ \midrule
    Gender & Male (88)               & 0.402                                & \multicolumn{1}{c}{0.680}                                   & 0.573                                         & 0.748                                            \\ \vspace{1mm}
    & Female (112)             & 0.421                                & \multicolumn{1}{c}{0.701}                                   & 0.608                                         & 0.791                                           \\ \midrule
    Age & $\leq$ 25 (70)  & 0.393                                & \multicolumn{1}{c}{0.677}                                   & 0.567                                         & 0.746                                            \\
    & \textgreater 25 (130) & 0.423                                & \multicolumn{1}{c}{0.699}                                   & 0.606                                         & 0.785                                            \\ 
    \midrule
    \end{tabular}
}\end{adjustbox}
\caption{Inter-annotator reliability and human recognition accuracy by gender and age of actors at 0 and 0.71 agreement.}
\label{tab:sex-age-alpha-accuracy}
\end{table}

%% file: tables/data_evaluation/annotation_prescreen.tex
\begin{table}[!ht]
\begin{adjustbox}{width=\linewidth,center}{
    \centering
    \begin{tabular}{@{}cccccc@{}}

    \toprule
        \begin{tabular}[c]{@{}c@{}}
            \textbf{Demonstration} \\ 
            \textbf{(utterances)} 
        \end{tabular}
        & \multicolumn{2}{c}{\textbf{Pre-screen detail}}
        & \textbf{\begin{tabular}[c]{@{}c@{}}
            Pass  \\ 
            pretest
            \end{tabular}}
        & \textbf{\begin{tabular}[c]{@{}c@{}}
            Inter-\\
            annotator  \\ 
            reliability\end{tabular}} 
        & \textbf{\begin{tabular}[c]{@{}c@{}}
            Human \\
            recognition  \\ 
            accuracy\end{tabular}}               
        \\  \cmidrule(lr){2-3} 
        & \textbf{Tutorial video} &  
        \textbf{Pretest}  & &  &   
        \\ \midrule
        
    Pilot 1 (350)    
    & \begin{tabular}[c]{@{}c@{}} only text description  \\  trick Q\&A in text\end{tabular} 
    & \begin{tabular}[c]{@{}c@{}} 10 normal  \\  + 1 trick Q\&A\end{tabular}       
    & 0.401 & 0.379     & 0.549                                
    \\ \vspace{1mm}
    Pilot 2 (350) 
    & \begin{tabular}[c]{@{}c@{}c@{}} text description \\ + emotional sound example  \\  trick Q\&A in text\end{tabular}          
    & \begin{tabular}[c]{@{}c@{}c@{}c@{}c@{}} \\ 10 normal \\ (with a set of more\\  intense expressions) \\  + 1 trick Q\&A\\ + 3 additional\end{tabular}                                 
    & 0.235 & 0.397 & 0.600     
    \\ 
    \begin{tabular}[c]{@{}c@{}}Actual  \\ (27,854)\end{tabular}
    & \begin{tabular}[c]{@{}c@{}c@{}} text description \\ + emotional sound example  \\  trick Q\&A in text and sound\end{tabular}  
    & \begin{tabular}[c]{@{}c@{}c@{}c@{}} \\ 10 normal \\ (same as Pilot 2) \\  + 1 trick Q\&A\end{tabular}           
    & 0.559 & 0.413 & 0.592                                       
    \\
    \midrule
    \end{tabular}
}
\end{adjustbox}
\caption{Effect of pre-screening on annotation performance.}
\label{tab:annotation_prescreen}
\end{table}

%% file: 6-experiments.tex
\section{Downstream Experiments}
\label{sec:experiments}

This section describes the experimental setup for benchmarking the THAI-SER dataset in the speech emotion recognition task.
We also include some of our experimental results regarding the effect of training with different recording environments and the experiments on cross-corpus settings compared to IEMOCAP.

\subsection{Experimental Setups}

\paragraph{Data Preparation} 
Our experiments follow the speaker-independent k-fold cross-validation method adapted from \citep{ser_kfold:journals/corr/abs-1802-05630}. 
The goal of speaker-independent splitting is to ensure no speaker overlap between each split, while the k-fold settings ensure model consistency between different training/testing data. 
As stated in Section \ref{sec:design}, the corpus consists of 100 sessions, where 80 sessions were recorded in a studio environment and 20 sessions were recorded via Zoom.

This baseline experiment will exclude the 20 Zoom recordings because of two main reasons.
\begin{enumerate}
    \item The distribution of Zoom and studio recordings is totally different. Including Zoom recordings can cause data distribution to be mismatched and would cause k-fold cross-validation results to be misleading when the Zoom recording was chosen as a validation or test set. 
    \item As Zoom recording is considered out-of-domain (due to Zoom's audio encoding and compression scheme as well as robust and uncontrolled environment), we would like to leave this recording set as a challenge set, which we will discuss later in section \ref{subsec:challenges}. 
\end{enumerate}

This would leave us with 80 sessions. 
The 80 sessions were then split into 8 different folds, each with 10 distinct sessions. 
For example, the first fold contains sessions 1 to 10, the second fold contains sessions 11 to 20, and so forth. 
For the fold that was selected as a validation and test set, the first five sessions were chosen as the validation set, and the rest were used as the test set. 
With this setting, there will be no speakers' overlap between the training, validation, and test sets. 
For the input features, we use 64 mel-scale filterbank coefficients with a frame length of 25 ms and a frame shift of 10 ms. 
The features were extracted using Kaldi \citep{Povey11thekaldi}. 
The audio was split into 3-second intervals, while the remaining intervals were padded by repeating the audio until the duration of 3 seconds was reached. 
We also apply Vocal Tract Length Perturbation (VTLP) \citep{Jaitly2013VocalTL} to all of the training set samples. 
Each input feature was normalized via Cepstral Mean Variance Normalization (CMVN).

\paragraph{Agreement Score} 
Since the corpus was annotated by multiple annotators, each sample will have a different number of agreement scores, and it is important for the training data to have a consistent label. 
Hence, if our training data contains a sample with a low agreement score, it could potentially confuse the model. 
However, if we filtered the corpus with too high of an agreement score, there would be a lower number of samples in the corpus.
To select the optimal value of the agreement score, we use the result obtained from Section \ref{sec:result} where the corpus with an agreement score lower than 0.71 was filtered out. 
This left us with a corpus that was 7 hours and 43 minutes long.

\paragraph{Evaluation Metrics}
We chose weighted accuracy and unweighted accuracy as the evaluation metrics. 
These two metrics have been used in various literature (\cite{satt17_interspeech}, \cite{soft-label1}, and \cite{ser_kfold:journals/corr/abs-1802-05630}). 
The weighted accuracy is computed by dividing the number of correctly classified samples by the total number of samples.
On the other hand, unweighted accuracy is calculated by taking an average of per-class accuracy. 

For each fold, we train five models with different random seed initialization.
We report the average weighted and unweighted accuracy as well as the standard deviation of these five runs.

\paragraph{Model Architecture} 
In all of our experiments, we use a simple CNN+LSTM architecture as described in \cite{ser_kfold:journals/corr/abs-1802-05630}. 
The architecture consists of 4 shallow convolution blocks, each block consists of 1D-convolution with kernel sizes of (5, 3, 3, 3) and (64, 64, 128, 128) channels, LeakyReLU Activation, layer normalization \citep{ba2016layer}, and a max pooling layer of size 2. 
The feature maps are then passed to the bidirectional LSTM \citep{lstm} with 128 hidden units, and the last hidden state was passed to a softmax classifier. 
The model was optimized using the Adam Optimizer \citep{kingma2017adam} with a learning rate of 0.0001 for 40 epochs, and the batch size was 32.

\subsection{Experimental Results}
\label{subsec:expresults}

\paragraph{Baseline} 
Our work provides two baseline models for this corpus. 
The first is the baseline that was evaluated on all emotions, while the second baseline was evaluated on four basic emotions: neutral, angry, happy, and sad. 
We conducted an experiment using an 8-fold cross validation setting, neglecting the Zoom recording session when training. 
Each fold was separated for each of the 10 studio sessions, and the fold that was used as a test set was split in half for validation and test set. 
Table \ref{tab:baseline} shows the split configuration as well as the baseline results. 
Unsurprisingly, the performance of the model trained with all emotions has a worse performance compared to using only four basic emotions.
As shown in figure \ref{cm}, the \textit{frustrated} emotion are easily confused with \textit{sad} and \textit{angry} by humans, so it is reasonable that the model will be confused with these emotions as well, causing a moderate performance degradation.

\input{tables/experiments/baseline}

\subsubsection{Effects of different acting sessions}
\label{sec:acting}
Neumann et al. found that the model trained on IEMOCAP dataset using only \textit{improvised} session yield a better performance compared to using improvised+scripted or scripted sessions alone \citep{neumann2017attentive}. 
Thus, we would like to investigate whether our corpus will have a similar effect or not. 
To do so, we trained the model using three separate session settings: scripted only, improvised only, and using both scripted and improvised (\textit{all}). 
The results are illustrated in table \ref{tab:turntype}. 
The results stated that the model trained on \textit{scripted} sessions outperformed that when trained on \textit{improvised} or \textit{all} sessions. 
These findings contradicted what was found in the IEMOCAP. 
The hypothesis for this incident is that the data collection methodologies as well as the languages were different. 
Our approach asked actors to repeat a similar sentence while for IEMOCAP, the scripts were prepared as a dialogue. 
This could potentially indicate that collecting the scripted data by using fixed sentences could allow more consistency in corpus annotation. 
Nevertheless, one should note that despite having more consistent annotations, it can also be more prone to overfitting as the variance of the sentence combinations is decreased. 

Additionally, we compared models trained with all emotions (including \textit{frustration}) using a different weight initialization scheme: randomly initialization and finetuned from a model that was trained with 4 basic emotions. 
We found that there is no improvement compared to randomly initialized weights.

\input{tables/experiments/turntype}

\subsubsection{Cross-corpus settings} 
\label{sec:cross}
We investigate the results of training and evaluating the model using the THAI-SER and IEMOCAP corpora where two additional corpora (Emo-DB and EMOVO) were also added for evaluation. 
The cross corpus results are shown in Table \ref{tab:crosscorpus}. 

\input{tables/experiments/crosscorpus}

When training with IEMOCAP, it is clear that training with improvised sessions yields the best results for almost all corpora apart from Emo-DB. 
these findings align with the findings that \cite{neumann2017attentive} found that the IEMOCAP corpus tends to perform best when trained only with improvised data. 
On the other hand, when training with the THAI-SER, the model tends to perform best when training with both improvised and scripted sessions. 
This indicates that our scripted corpus contains more consistent emotional information compared to IEMOCAP, as it does not degrade the model performance when mixed with the improvised portion. 
Nevertheless, one should note that the performance of the improvised only model performs better than the scripted only model in most cases. 
Additionally, the performance of the model trained with THAI-SER also performs better emotion recognition on both the Emo-DB and EMOVO corpus.

Although Table \ref{tab:crosscorpus} stated that the overall results of THAI-SER are better as a whole, these comparisons are still unfair due to the difference in number of speakers and hours in both corpora. 
Therefore, we also conducted another experiment where both corpora were pruned to have similar settings such as the number of speakers and the number of hours for each fold. 

We prune both THAI-SER and IEMOCAP according to the following procedure:
\begin{enumerate}
    \item We select the studio session from THAI-SER that have the closest gender distribution, portion of improvised and scripted, and emotion class distribution with each IEMOCAP fold session.
    \item For each fold on IEMOCAP and it's corresponded studio session, we prune the data to make sure that both pruned fold have the equal number of hours.
\end{enumerate}

The pruned corpus statistics are showed in \autoref{tab:prunedcorpus_stats}, and
the results of the pruned corpus are shown in Table \ref{tab:prunedcrosscorpus}.
Despite being pruned, the model trained on THAI-SER still shows a dominant performance when evaluated on both EMO-DB and EMOVO.

\input{tables/experiments/pruned_stats}
\input{tables/experiments/pruned_crosscorpus}

%% file: tables/experiments/baseline.tex
\begin{table}[!ht]
\begin{adjustbox}{width=\linewidth,center}{
    \centering
    \begin{tabular}{@{}ccccccc@{}}
    \toprule
                           &                                      &                                    & \multicolumn{2}{c}{\textbf{All Emotions}} & \multicolumn{2}{c}{\textbf{4 Basic Emotions}} \\ \cmidrule(l){4-7} 
    \multirow{-2}{*}{Fold} & \multirow{-2}{*}{Validation Studios} & \multirow{-2}{*}{Test Studios}     & Weighted Accuracy                  & Unweighted Accuracy                  & Weighted Accuracy                    & Unweighted Accuracy                    \\ \midrule
    0                      & Studio 71-75                         & Studio 76-80                       & 61.27 ± 1.44\%      & 60.90 ± 1.50\%      & 69.75 ± 2.02\%        & 66.80 ± 1.31\%        \\
    1                      & Studio 61-65                         & Studio 66-70                       & 58.22 ± 2.46\%      & 50.86 ± 3.62\%      & 65.24 ± 2.56\%        & 57.99 ± 1.24\%        \\
    2                      & Studio 51-55                         & Studio 56-60                       & 58.27 ± 1.56\%      & 55.44 ± 2.28\%      & 68.37 ± 0.16\%        & 58.79 ± 1.60\%        \\
    3                      & Studio 41-45                         & Studio 46-50                       & 60.91 ± 2.98\%      & 59.21 ± 4.02\%      & 69.27 ± 0.96\%        & 64.59 ± 1.12\%        \\
    4                      & Studio 31-35                         & Studio 36-40                       & 62.53 ± 1.20\%      & 60.55 ± 2.24\%      & 69.71 ± 1.09\%        & 65.06 ± 1.20\%        \\
    5                      & Studio 21-25                         & Studio 26-30                       & 61.38 ± 2.43\%      & 60.13±3.35\%        & 67.03 ± 1.31\%        & 62.23 ± 1.39\%        \\
    6                      & Studio 11-15                         & Studio 16-20                       & 59.97 ± 1.40\%      & 58.33±1.83\%        & 67.81 ± 1.42\%        & 64.37 ± 1.19\%        \\
    7                      & {\color[HTML]{333333} Studio 1-5}    & { Studio 6-10} & 55.85 ± 2.28\%      & 57.03±2.31\%        & 61.57 ± 1.07\%        & 61.08 ± 0.81\%        \\ \midrule
    \multicolumn{3}{c}{\textbf{Avg}}                                                                   & 59.80 ± 2.91\%      & 57.81 ± 4.20\%      & 67.34 ± 3.05\%        & 62.61 ± 3.19\%        \\ \midrule
    \multicolumn{3}{c}{\textbf{Zoom Evaluation}}                                                       & 46.64 ± 2.93\%      & 46.57 ± 2.09\%      & 56.05 ± 2.37\%        & 53.85 ± 1.48\%        \\ \bottomrule
    \end{tabular}
}\end{adjustbox}
\caption{\label{tab:baseline} The baseline results when trained with all emotions and only 4 basic emotions.}
\end{table}

%% file: tables/experiments/turntype.tex
\begin{table}[!ht]
    \centering
    \begin{tabular}{@{}lcc@{}}
    \toprule
    \textbf{Training Data} & \textbf{Weighted Accuracy} & \textbf{Unweighted Accuracy} \\ \midrule
    Improvised Only & 61.80 ± 3.24\% & 57.03 ± 4.84\% \\
    Scripted Only   & 73.99 ± 4.60\% & 65.93 ± 4.40\% \\
    All             & 67.34 ± 3.05\% & 62.61 ± 3.19\% \\
    \bottomrule[0.15em]
    \end{tabular}
\caption{\label{tab:turntype} The effect of training data type on model performance using only the 4 basic emotions.}
\end{table}

%% file: tables/experiments/crosscorpus.tex
\begin{table}[!ht]
\begin{adjustbox}{width=\linewidth,center}{
    \centering
    \begin{tabular}{@{}cllllrr@{}}
    \toprule
    \multicolumn{2}{l}{} &
    \multicolumn{3}{c}{\textbf{\begin{tabular}[c]{@{}c@{}}Testing Data (WA/UA)\\(THAI SER)\end{tabular}}} &
    \textbf{Emo-DB} &
    \textbf{EMOVO} \\
    \cmidrule[0.15em](l){3-5}
    \multicolumn{2}{l}{\multirow{-2}{*}{}} 
    & \textbf{Improvised} 
    & \textbf{Scripted} 
    & \textbf{All} 
    & 
    &  \\ 
    \midrule
    \multicolumn{1}{c}{\multirow{3}{*}{\textbf{\begin{tabular}[c]{@{}c@{}}Training\\Data\\(IEMOCAP)\end{tabular}}}} 
    & \textbf{Improvised} 
      & \textbf{45.56} / \textbf{37.22} 
      & \textbf{45.91} / \textbf{39.58} 
      & \textbf{45.71} / \textbf{38.09} 
      & 39.97 / 34.88 
      & \textbf{34.32}\\
    
    \multicolumn{1}{c}{} 
    & \textbf{Scripted} 
      & 29.20 / 31.20 
      & 20.54 / 31.70 
      & 25.54 / 31.38 
      & 42.32 / 31.67 
      & 31.00 \\
    
    \multicolumn{1}{c}{} 
    & \textbf{All} 
      & 40.90 / 37.39 
      & 34.87 / 37.82 
      & 38.36 / 37.53 
      & \textbf{45.09} / \textbf{36.61} 
      & 33.80 \\
    \midrule
    
    \multicolumn{2}{l}{} &
    \multicolumn{3}{c}{\textbf{\begin{tabular}[c]{@{}c@{}}Testing Data (WA/UA)\\(IEMOCAP)\end{tabular}}} &
    \textbf{Emo-DB} &
    \textbf{EMOVO} \\
    \cmidrule(l){3-5}
    \multicolumn{2}{l}{\multirow{-2}{*}{}} 
    & \textbf{Improvised} 
    & \textbf{Scripted} 
    & \textbf{All} 
    & 
    &  \\
    \midrule
    \multicolumn{1}{c}{\multirow{3}{*}{\textbf{\begin{tabular}[c]{@{}c@{}}Training\\Data\\(THAI-SER)\end{tabular}}}} 
    & \textbf{Improvised} 
      & 45.44 / 40.16 
      & 31.26 / 32.15 
      & 38.46 / 36.40 
      & 50.79 / 49.84 
      & 40.94\\
    
    \multicolumn{1}{c}{} 
    & \textbf{Scripted} 
      & 44.89 / 39.97 
      & 27.84 / 30.89 
      & 36.5 / 35.80  
      & 41.55 / 46.19 
      & 41.12\\
    
    \multicolumn{1}{c}{} 
    & \textbf{All} 
      & \textbf{46.88} / \textbf{42.22} 
      & \textbf{31.86} / \textbf{32.84} 
      & \textbf{39.49} / \textbf{37.87} 
      & \textbf{51.15} / \textbf{52.02} 
      & \textbf{43.10}\\
    \bottomrule
    \end{tabular}
}\end{adjustbox}
\caption{\label{tab:crosscorpus}The results of the cross-corpus settings. The numbers are showed in weighted accuracy / unweighted accuracy format. Note that for the EMOVO corpus, the value of weighted and unweighted accuracy are equal as the number of samples for each class are equal. Thus, we present only a single value for EMOVO corpus.}
\end{table}

%% file: tables/experiments/pruned_stats.tex
\begin{table}[!ht]
\centering
\begin{tabular}{cccc}
\midrule
                     & \multicolumn{2}{c}{\textbf{IEMOCAP}}                    \\ \midrule
                     & \textbf{Full}         & \textbf{Pruned}    & \textbf{Pruned \%}             \\ \midrule
\textbf{Improvised}  & 2 hr 46 min 32.20 sec & 1 hr 24 min 25.49 sec & 50.69\% \\
\textbf{Scripted}    & 2 hr 49 min 51.32 sec & 1 hr 07 min 32.38 sec & 39.76\% \\
\textbf{Total}       & 5 hr 36 min 23.52 sec & 2 hr 31 min 57.87 sec & 45.17\% \\ \midrule
\multicolumn{1}{l}{} & \multicolumn{2}{c}{\textbf{THAI-SER}}                    \\ \midrule
\multicolumn{1}{l}{} & \textbf{Full}         & \textbf{Pruned}  & \textbf{Pruned \%}               \\ \midrule
\textbf{Improvised}  & 4 hr 14 min 53.58 sec & 1 hr 22 min 49.85 sec  & 32.50\% \\
\textbf{Scripted}    & 3 hr 28 min 23.27 sec & 1 hr 05 min 58.80 sec  & 31.66\% \\
\textbf{Total}       & 7 hr 43 min 16.85 sec & 2 hr 28 min 48.65 sec  & 32.12\% \\
\midrule
\end{tabular}
\caption{\label{tab:prunedcorpus_stats}The statistics of the pruned IEMOCAP and THAI-SER corpus.}
\end{table}

%% file: tables/experiments/pruned_crosscorpus.tex
\begin{table}[!ht]
\begin{adjustbox}{width=\linewidth,center}{
    \centering
    \begin{tabular}{@{}clllllr@{}}
    \toprule
    \multicolumn{2}{l}{}                                                                                                                                                     & \multicolumn{5}{c}{\textbf{\begin{tabular}[c]{@{}c@{}}Testing Data\\ (THAI SER - pruned)\end{tabular}}}                                                                                            \\ \cmidrule(l){3-7} 
    \multicolumn{2}{l}{\multirow{-2}{*}{}}                                                                                                                                   & \textbf{Improvised}                   & \textbf{Scripted}                     & \textbf{All}                          & \textbf{Emo-DB}                       & \multicolumn{1}{l}{\textbf{EMOVO}} \\ \midrule
    \multicolumn{1}{c}{}                                                                                                         & \textbf{Improvised}                       & 40.05 / 36.57                         & \textbf{38.99} / \textbf{41.46}                         & \textbf{39.61} / 38.13                         & 37.35 / 33.36                         & 33.33                              \\
    \multicolumn{1}{c}{}                                                                                                         & \textbf{Scripted} & 39.09 / 31.99 & 22.15 / 32.61 & 31.98 / 32.11 & 43.42 / 33.28 & 31.98      \\
    \multicolumn{1}{c}{\multirow{-3}{*}{\textbf{\begin{tabular}[c]{@{}c@{}}Training Data\\ (IEMOCAP \\- pruned)\end{tabular}}}}    & \textbf{All}                              & \textbf{44.82} / \textbf{39.03}                         & 32.04 / 38.94                         & 39.46 / \textbf{38.58}                         & \textbf{43.92} / \textbf{35.78}                         & \textbf{33.55}                              \\ \midrule
    \multicolumn{2}{l}{}                                                                                                                                                     & \multicolumn{5}{c}{\textbf{\begin{tabular}[c]{@{}c@{}}Testing Data\\ (IEMOCAP - pruned)\end{tabular}}}                                                                                             \\ \cmidrule(l){3-7} 
    \multicolumn{2}{l}{\multirow{-2}{*}{}}                                                                                                                                   & \textbf{Improvised}                   & \textbf{Scripted}                     & \textbf{All}                          & \textbf{Emo-DB}                       & \multicolumn{1}{l}{\textbf{EMOVO}} \\ \midrule
    \multicolumn{1}{c}{}                                                                                                         & \textbf{Improvised}                       & 42.87 / 35.81                         & \textbf{28.85} / 29.38                         & 36.82 / 33.61                         & \textbf{49.54} / 49.43                         & 42.74                              \\
    \multicolumn{1}{c}{}                                                                                                         & \textbf{Scripted} & 41.05 / \textbf{37.88} & 26.31 / 29.48 & 34.69 / 34.51 & 42.28 / 45.48 & 37.54      \\
    \multicolumn{1}{c}{\multirow{-3}{*}{\textbf{\begin{tabular}[c]{@{}c@{}}Training Data\\ (THAI-SER\\  - pruned)\end{tabular}}}} & \textbf{All}                              & \textbf{43.31} / 37.70                         & 28.29 / \textbf{29.64}                         & \textbf{36.83} / \textbf{34.65}                         & 49.40 / \textbf{50.39}                          & \textbf{42.44}                              \\ \bottomrule
    \end{tabular}
}\end{adjustbox}
\caption{\label{tab:prunedcrosscorpus}The results of the cross-corpus settings using pruned corpus. Both corpora are pruned to have equal numbers of speaker and training hours.}
\end{table}

%% file: 7-discussion.tex
\section{Discussion}
\label{sec:discussion}

THAI-SER was designed to create a valuable corpus for worldwide affective computing research, not only for Thailand. 
This corpus is challenging and needs knowledge in diverse areas. 
Therefore, we want to discuss problems that we inevitably experienced during the process of creating the corpus and discuss possible solutions through each section below.

\subsection{Recordings}

Although we conducted the pilot recording multiple times in both recording environments (Studio and Zoom) to get the best recording procedure, we still encountered many problems that affected the quality of the data. 
Here, we will discuss the problems of Studio and Zoom recordings independently.
 
\subsubsection{Studio recording}

\paragraph{Reducing Noises and Audio Calibration}
Sound is sensitive to all components of the studio.
When actors perform some movements, we encounter unwanted noise from the chairs and floor.
Therefore, we decided to purchase the new chairs and cover the floor with a mat.
Importantly, we raise this issue as it directly affects the audio quality. 
As a result, we leave this as a reminder for future work that environmental noise should be a concern.

Another point to note about the recording process is that the actors' voice levels can vary depending on their characteristics or the emotions they portray. 
For example, consider situation 15 from \autoref{tab:situations_improvisation} (angry versus sad). 
The actors typically use significantly different voice levels to express their assigned emotions. 
The volume of the angry actor is usually very high, whereas the sad actor's volume is much lower.

This disparity can sometimes lead to issues, such as the audio from the angry actor being picked up by the sad actor's microphone, which is undesirable. 
Additionally, moving the microphone stand closer to the quieter actor is not a viable solution, as it would obstruct the camera's view.

To address this, we first ask both actors to speak louder than usual and use that loudness to calibrate input signal gain such that both actors loudness level are similar.  
However, this approach inevitably introduces noise, which reduces the overall audio quality.

\paragraph{Maintaining Consistent Recordings Quality}
It is also important to mark the locations of devices (cameras, microphones, chairs, etc.) to maintain full control of the recording environment. 
Uncontrolled device positions can lead to minor inconsistencies in audio quality and recording frames, which, in turn, affect the overall dataset quality. 
For example, it is essential to maintain a consistent microphone-to-face distance for all actors and ensure that the camera frame fully captures the actor's face and mouth.

While performing, the director must also monitor whether an actor's hand gestures block the camera view; ideally, such obstructions should be avoided. 
Additionally, a common issue arises when actors focus on the opposing actor instead of the camera during their performance. 
To address this, the director should carefully monitor the footage during recording to ensure that the actors maintain proper eye contact with the camera.

All of the aforementioned issues are more frequently observed during improvised sessions, as actors tend to express their movements more dynamically compared to scripted sessions.

\paragraph{Device Malfunctions}
Lastly, devices can sometimes malfunction unpredictably during recording (e.g., a camera overheating, accidental shutdowns, or electrical interference from nearby devices introducing noise into the microphone). 
Therefore, having backup files and devices is essential to prevent the loss of valuable assets and materials.

The large output files from all devices must be recorded in the same sequence to simplify post-processing. 
Additionally, since we were working with a team of three staff members, proper management skills were crucial to ensure smooth operation between recording sessions.
The project owner must also oversee the experimental process and maintain rigorous quality control throughout the production.

\subsubsection{Zoom recording}

Since Zoom recordings were conducted online, the challenges differed significantly from those of studio recordings. 

\paragraph{Connectivity Issues}
During our Zoom sessions, we often encountered internet connectivity issues. 
To mitigate this, we conducted pre-recording sessions with all actors and staff, checking the internet connection quality, device functionality, and other tools before starting the actual recordings.

Despite these precautions, we still experienced connection problems, such as low latency and signal loss, which impacted data quality and recording time. 
For example, during one Zoom recording session, the process extended to six hours due to delays in rendering audio files with the external audio recording service, Zencastr, which was affected by internet latency. 
This prolonged session led to lower-quality recordings as both the actors and staff became exhausted, directly impacting the actors' performance quality.

As noted in section \ref{subsec:recordingprocess}, we typically aim to complete each recording session within three hours to maintain consistency in the actors' performances. 
Consequently, we decided to postpone that session to another day. 
This incident not only wasted time but also incurred additional costs, as we had to pay the actors for an extra session. 
Following this experience, we agreed to promptly postpone any recording session if similar issues arise in the future to prevent further disruptions.

\subsection{Directing}

Directing plays an important roles in controlling actor's acting quality.
When directing actors' performances, it is crucial to establish clear emotional definitions that both actors and directors agree upon. 
This ensures that actors can express their emotions accurately and consistently.
To further elaborate the challenges in directing actors' performance, we highlight the challenges as outlined below.

\subsubsection{Dealing with Ambiguity in Frustrated Emotion}
As shown in \autoref{cm}, annotators often confused frustration with anger. 
This confusion may stem from the fact that frustration can be viewed as a subset of anger. 
To mitigate this issue, we have chosen to restrict the actor's expression of anger directly. 
We observed that the key difference between acting frustrated and acting angry lies in the intensity of the emotion; frustration is expressed with lower intensity than anger. 
It is also unsurprising that \autoref{cm} shows confusion between frustration and sadness, as both emotions are expressed with similarly low intensity. 
However, some actors may naturally appear aggressive or negative due to their inherent style. 
In such cases, the director often disregards the intensity to prioritize the naturalness of the performance. 
These factors contribute to the challenges actors, annotators, and SER models face in accurately recognizing frustration, particularly in short sentences.

\subsubsection{Scripted session}

When recording the scripted session, we provided actors with the utterances in order as shown in \autoref{fig:scriptorder}. 
In some cases some actors are not familiar with the given emotional orders, leading to multiple failures due to wrong word usage. 
To tackle this problem, the director can decide to rearrange the utterance order during acting. 
This was done to reduce the chances of mistakes and carefully guide the actor through each take.

\subsubsection{Improvised session}

In some improvised situations, it is challenging to maintain the assigned emotion for over three minutes under a specific scenario. 
To ensure the actors' performances feel natural, the director sometimes allows slight adjustments to the scenarios, enabling the actors to sustain the assigned emotion throughout the three-minute session. 
Additionally, during the breaks between each improvised session, actors are given time to reset and discard any lingering emotions from the previous scenario. 
The directors play an essential role in assisting the actors with this process, ensuring they can begin the new scenario with a fresh emotional state, free from any potential disruptions caused by residual emotions from the prior session.

\subsection{Post-recording}

In addition to recordings and directing, we highlight the issues we found during postprocessing audio files as well as during annotation phases.

\subsubsection{Segmentation} 
Since an improvised session is recorded as a continuous dialogue, segmenting the recordings into multiple segments is necessary for further annotation. 
During this process, the most challenging part is dealing with overlapping speech. 
In an improvised session, it is not always possible to ask actors to refrain from speaking while another actor is speaking, especially when the actor is expressing intense emotions. 
This often results in overlapping speech or multiple speakers being present in a single utterance. 
To avoid multiple speakers in an utterance, the editor may choose to remove the overlapping parts. 
After segmentation, the segmented utterances are checked twice to prevent segmentation errors by the editor or studio staff. 
If a segmentation error is found, the file is either sent back for re-segmentation or deleted from the database.

\subsubsection{Emotion annotation} 

We explored three annotation approaches, as detailed in \autoref{tab:annotation_method}, with the initial plan to include a behavior matrix as a tool for filtering poor or careless annotations. 
However, due to the limitation of the annotation platform providers, we didn't include behavior matrix filtering in our annotation process.
Thus, it is always important to check up with annotation platform provider limitation during planing phase to ensure that all of the filtering approach can be implemented properly.

\autoref{tab:annotator trustworthiness} shows that Annotation Batch 1 had the highest number of annotators.
This is because, in the first batch, we opted to use two independent crowdsourcing platforms (\textit{Wang.in.th} and \textit{Hope Data Annotation}). 
Following this initial phase, we chose to proceed solely with Wang.in.th due to its larger and more active user base, which provided greater diversity among annotators. 
Such diversity is vital for reliable crowdsourced annotation, as it helps maintain consistent data quality and accuracy.

Building on these observations, we established best practices for emotion annotation at two key stages as follows: 

\paragraph{1. A Clear Annotation Guideline with Pretest Assessment}
During the pre-screening phase, we recommend that the annotation platform offer: 

\begin{itemize} 
    \item Clear and concise emotional descriptors 
    \item A tutorial video with straightforward instructions, accompanying emotion samples, and at least one “trick question” 
    \item A short pretest assessment featuring a “trick question” from the tutorial material 
\end{itemize} 

These “trick questions” effectively filter out participants who are inattentive or fail to follow instructions.

\paragraph{2. A Robust Quality Control Measures on Annotators}
During the annotation, we prioritized an easy-to-use platform equipped with various quality-control measures. 
We also suggest incorporating additional methods beyond those shown in \autoref{tab:annotation_method}, as resources permit. 
In our own setup, we discreetly inserted “gold” (previously validated) and “consistent” (repeated) utterances to regularly assess annotator performance. 
Furthermore, we limited the number of tasks assigned to each annotator in every phase. 
This practice prevents fatigue and reduces bias, ensuring that annotators remain engaged and accurate when identifying emotional states.

\subsection{Challenges and Open Questions}
\label{subsec:challenges}

The THAI-SER corpus can play an important role in the research area of speech emotion recognition. 
This section address several challenges in this corpus that can be further explored in future literature.

\paragraph{Hard label \& soft label} 
The THAI-SER is annotated by multiple users per sample. 
This means that each sample can be represented as a probability distribution of each emotion's weighting by the scores voted by each annotator. 
In other words, one can use soft labels instead of hard labels to train the model. 
There is a literature that investigates the results of leveraging soft label instead of hard label in the IEMOCAP, such as \cite{soft-label1}, \cite{soft-label2}, \cite{soft-label3}, and \cite{soft-label4}. 
Unlike the IEMOCAP, which fixed the number of annotators to 3, the THAI-SER corpus contains up to 8 annotators per sample. 
This means our soft labels will contain richer information compared to those of IEMOCAP. 
We leave this as a future research questions for future researches.

\textbf{Leveraging samples with low agreement values:}
The baseline results that shown in section \ref{subsec:expresults} are trained using the samples where the agreement value is higher or equal to 0.71, as mentioned in section \ref{sec:optinal_agreement}. 
This leaves quite a number of ambiguous samples untouched during the training process. 
This raises the question of whether it’s possible to leverage this unused data to improve the model's performance. 
On the IEMOCAP, \citep{curriculum} leveraged low agreement samples by generating a curriculum based on the inter-evaluation agreement. 
This curriculum can be generated on the THAI-SER corpus as well. 
Therefore, we leave this as an open question for future work to explore this area.

\paragraph{Environmental Robustness:} 
This corpus also enables evaluation of the generalization of the recording environments. 
Since the THAI-SER corpus contains a Zoom recording session, one may leave these zoom recordings out as a special test set, as done in section \ref{subsec:expresults}. 
Several studies have been conducted to study the robustness of the SER engine in different environments \citep{envrobust}. 
However, previous work requires the reconstruction of an old corpus with additional setups. 
Therefore, we leave these Zoom recording sessions as a challenge for future work. 
Researchers can come up with different methodologies that would increase the environmental robustness of the model by benchmarking the Zoom session as a test corpus.

\paragraph{Cross corpus generalization:} 
It is also worth mentioning that in section \ref{subsec:expresults}, the performance of the cross-corpus can be further improved. 
There are several studies regrading cross-corpus generalization that have been done on the IEMOCAP, such as \cite{crosscorpus1}, \cite{crosscorpus2}, \cite{crosscorpus3}, and \cite{crosscorpus4}. 
These literatures can also be applied to THAI-SER to further improve the baseline results mentioned in section \ref{subsec:expresults}. 
However, it would be beyond the scope of this work, so we leave this as an open challenge for researchers in the speech emotion recognition area.

\paragraph{Enhancing Multimodal LLMs Performance}
While this study primarily focuses on CNN-based emotion recognition, the THAI SER corpus also provides valuable resources for enhancing multimodal large language models (LLMs). 
Models integrating speech inputs, such as Qwen-Audio \citep{chu2023qwenaudioadvancinguniversalaudio} and AudioPaLM \citep{rubenstein2023audiopalmlargelanguagemodel}, can significantly benefit from THAI SER by improving their emotional comprehension capabilities.
Specifically, training or fine-tuning LLMs on this dataset can facilitate more accurate emotional speech-to-text and expressive speech-to-speech generation, enabling these models to respond with greater emotional awareness and realism.

\paragraph{Facilitating Emotionally-Conditioned TTS Generation}
The THAI SER corpus is also well-suited to advance emotionally-conditioned text-to-speech (TTS) technologies. 
Contemporary neural TTS architectures, such as VALL-E \citep{wang2023neuralcodeclanguagemodels}, can utilize THAI SER for fine-grained emotional control, allowing synthesis systems to produce realistic, emotionally expressive speech \citep{zhou2025emotionaldimensioncontrollanguage}.

%% file: 8-conclusion.tex
\section{Conclusion}

This paper proposes THAI-SER, the largest-scale speech emotion corpus in Thai. 
THAI-SER contains 27,854 utterances (41.61 hours) recorded by 200 actors with five categorical emotions (neutrality, anger, happiness, sadness, and frustration), two recording sessions (scripted and improvised), and two environments (Zoom and studio). 
We collect the annotations from many annotators using crowdsourcing. 
The annotations are calculated using the majority agreement, and we also evaluate the annotation quality using inter-annotator reliability and human recognition accuracy. 
This work also addresses the benchmarking standard as well as its performance under different settings. 
The results show that our corpus performs generally well in cross-corpus settings and still leave some open challenges for SER. 
Finally, we believe that THAI-SER is an essential corpus for affective computing and language processing research, supporting extensive research on the real-world application. 
The available corpus can be downloaded under a Creative Commons BY-SA 4.0 license at \href{https://github.com/vistec-AI/dataset-releases/releases/tag/v1}{\texttt{https://github.com/vistec-AI/dataset-releases/releases/tag/v1}}.
The codes for the experiments can be found at \href{https://github.com/tann9949/thaiser-experiments}{\texttt{https://github.com/tann9949/thaiser-experiments}}.

\section*{Acknowledgement}
This corpus is supported by Advanced Info Services Public Company Limited and cooperating (AIS), Digital Economy Promotion Agency (DEPA) Thailand, and Siam Commercial Bank (SCB) together with the cooporation between Department of Computer Engineering - Faculty of Engineering, and Department of Dramatic Arts - Faculty of Arts, Chulalongkorn University. The authors would like to thank all the professors (Kiattipoom Nantanukul), directors (Tanyathorn Khunapinya, Peangdao Jariyapun, and Arpassorn Patitanon), staff/interns (Tin Charoenrungutai, Thanapat Trachu, Ornida Yuktanandana, and Pannarong Thongtangsai), and actors for their contributions and support.

\section*{Funding}
The THAI-SER dataset proposed in this study was funded by Advanced Info Services Public Company Limited (AIS) and the Digital Economy Promotion Agency (DEPA).

%% file: 0-main.bbl
\begin{thebibliography}{}
\providecommand{\doi}[1]{\url{https://doi.org/#1}}
\bibcommenthead

\bibitem [\protect \citeauthoryear {%
Abrilian%
, Devillers%
, Buisine%
\BCBL {}\ \BBA {} Martin%
}{%
Abrilian%
\ \protect \BOthers {.}}{%
{\protect \APACyear {2005}}%
}]{%
Abrilan_4}
\APACinsertmetastar {%
Abrilan_4}%
\begin{APACrefauthors}%
Abrilian, S.%
, Devillers, L.%
, Buisine, S.%
\BCBL {} Martin, J\BHBI C.%
\end{APACrefauthors}%
\unskip\
\newblock
\APACrefYearMonthDay{2005}{}{}.
\newblock
{\BBOQ}\APACrefatitle {{EmoTV1}: Annotation of real-life emotions for the
  specification of multimodal affective interfaces} {{EmoTV1}: Annotation of
  real-life emotions for the specification of multimodal affective
  interfaces}.{\BBCQ}
\newblock
 \APACrefbtitle {HCI International} {Hci international}\ (\BVOL~401,
  \BPG~407-408).
\PrintBackRefs{\CurrentBib}

\bibitem [\protect \citeauthoryear {%
Akçay%
\ \BBA {} Oğuz%
}{%
Akçay%
\ \BBA {} Oğuz%
}{%
{\protect \APACyear {2020}}%
}]{%
Mehmet_2}
\APACinsertmetastar {%
Mehmet_2}%
\begin{APACrefauthors}%
Akçay, M.B.%
\BCBT {}\ \BBA {} Oğuz, K.%
\end{APACrefauthors}%
\unskip\
\newblock
\APACrefYearMonthDay{2020}{}{}.
\newblock
{\BBOQ}\APACrefatitle {Speech emotion recognition: Emotional models, databases,
  features, preprocessing methods, supporting modalities, and classifiers}
  {Speech emotion recognition: Emotional models, databases, features,
  preprocessing methods, supporting modalities, and classifiers}.{\BBCQ}
\newblock
\APACjournalVolNumPages{Speech Communication}{116}{}{56-76}.
\newblock

\newblock

\PrintBackRefs{\CurrentBib}

\bibitem [\protect \citeauthoryear {%
Ando%
\ \protect \BOthers {.}}{%
Ando%
\ \protect \BOthers {.}}{%
{\protect \APACyear {2018}}%
}]{%
soft-label1}
\APACinsertmetastar {%
soft-label1}%
\begin{APACrefauthors}%
Ando, A.%
, Kobashikawa, S.%
, Kamiyama, H.%
, Masumura, R.%
, Ijima, Y.%
\BCBL {} Aono, Y.%
\end{APACrefauthors}%
\unskip\
\newblock
\APACrefYearMonthDay{2018}{}{}.
\newblock
{\BBOQ}\APACrefatitle {Soft-Target Training with Ambiguous Emotional Utterances
  for DNN-Based Speech Emotion Classification} {Soft-target training with
  ambiguous emotional utterances for dnn-based speech emotion
  classification}.{\BBCQ}
\newblock
 \APACrefbtitle {2018 IEEE International Conference on Acoustics, Speech and
  Signal Processing (ICASSP)} {2018 ieee international conference on acoustics,
  speech and signal processing (icassp)}\ (\BPG~4964-4968).
\newblock
\begin{APACrefDOI} \doi{10.1109/ICASSP.2018.8461299} \end{APACrefDOI}
\PrintBackRefs{\CurrentBib}

\bibitem [\protect \citeauthoryear {%
Ando%
, Masumura%
, Kamiyama%
, Kobashikawa%
\BCBL {}\ \BBA {} Aono%
}{%
Ando%
\ \protect \BOthers {.}}{%
{\protect \APACyear {2019}}%
}]{%
soft-label3}
\APACinsertmetastar {%
soft-label3}%
\begin{APACrefauthors}%
Ando, A.%
, Masumura, R.%
, Kamiyama, H.%
, Kobashikawa, S.%
\BCBL {} Aono, Y.%
\end{APACrefauthors}%
\unskip\
\newblock
\APACrefYearMonthDay{2019}{09}{}.
\newblock
{\BBOQ}\APACrefatitle {Speech Emotion Recognition Based on Multi-Label Emotion
  Existence Model} {Speech emotion recognition based on multi-label emotion
  existence model}.{\BBCQ}
\newblock
 (\BPG~2818-2822).
\newblock
\begin{APACrefDOI} \doi{10.21437/Interspeech.2019-2524} \end{APACrefDOI}
\PrintBackRefs{\CurrentBib}

\bibitem [\protect \citeauthoryear {%
Anolli%
, Wang%
, Mantovani%
\BCBL {}\ \BBA {} Toni%
}{%
Anolli%
\ \protect \BOthers {.}}{%
{\protect \APACyear {2008}}%
}]{%
doi:10.1177/0022022108321178}
\APACinsertmetastar {%
doi:10.1177/0022022108321178}%
\begin{APACrefauthors}%
Anolli, L.%
, Wang, L.%
, Mantovani, F.%
\BCBL {} Toni, A.D.%
\end{APACrefauthors}%
\unskip\
\newblock
\APACrefYearMonthDay{2008}{}{}.
\newblock
{\BBOQ}\APACrefatitle {The Voice of Emotion in Chinese and Italian Young
  Adults} {The voice of emotion in chinese and italian young adults}.{\BBCQ}
\newblock
\APACjournalVolNumPages{Journal of Cross-Cultural Psychology}{39}{5}{565-598}.
\newblock
\begin{APACrefURL} {https://doi.org/10.1177/0022022108321178} \end{APACrefURL}
\newblock
{\href{https://arxiv.org/abs/https://doi.org/10.1177/0022022108321178}{{https://doi.org/10.1177/0022022108321178}}}
\newblock

\newblock
\begin{APACrefDOI} \doi{10.1177/0022022108321178} \end{APACrefDOI}
\PrintBackRefs{\CurrentBib}

\bibitem [\protect \citeauthoryear {%
Ba%
, Kiros%
\BCBL {}\ \BBA {} Hinton%
}{%
Ba%
\ \protect \BOthers {.}}{%
{\protect \APACyear {2016}}%
}]{%
ba2016layer}
\APACinsertmetastar {%
ba2016layer}%
\begin{APACrefauthors}%
Ba, J.L.%
, Kiros, J.R.%
\BCBL {} Hinton, G.E.%
\end{APACrefauthors}%
\unskip\
\newblock
\APACrefYearMonthDay{2016}{}{}.
\newblock
\APACrefbtitle {Layer Normalization.} {Layer normalization.}
\PrintBackRefs{\CurrentBib}

\bibitem [\protect \citeauthoryear {%
Buchholz%
\ \BBA {} Latorre%
}{%
Buchholz%
\ \BBA {} Latorre%
}{%
{\protect \APACyear {2011}}%
}]{%
cs_gold_1}
\APACinsertmetastar {%
cs_gold_1}%
\begin{APACrefauthors}%
Buchholz, S.%
\BCBT {}\ \BBA {} Latorre, J.%
\end{APACrefauthors}%
\unskip\
\newblock
\APACrefYearMonthDay{2011}{01}{}.
\newblock
{\BBOQ}\APACrefatitle {Crowdsourcing Preference Tests, and How to Detect
  Cheating.} {Crowdsourcing preference tests, and how to detect
  cheating.}{\BBCQ}
\newblock
 (\BPG~3053-3056).
\PrintBackRefs{\CurrentBib}

\bibitem [\protect \citeauthoryear {%
Burkhardt%
, Paeschke%
, Rolfes%
, Sendlmeier%
\BCBL {}\ \BBA {} Weiss%
}{%
Burkhardt%
\ \protect \BOthers {.}}{%
{\protect \APACyear {2005}}%
}]{%
Emo-DB}
\APACinsertmetastar {%
Emo-DB}%
\begin{APACrefauthors}%
Burkhardt, F.%
, Paeschke, A.%
, Rolfes, M.%
, Sendlmeier, W.F.%
\BCBL {} Weiss, B.%
\end{APACrefauthors}%
\unskip\
\newblock
\APACrefYearMonthDay{2005}{}{}.
\newblock
{\BBOQ}\APACrefatitle {A database of German emotional speech} {A database of
  german emotional speech}.{\BBCQ}
\newblock
 \APACrefbtitle {Ninth European Conference on Speech Communication and
  Technology.} {Ninth european conference on speech communication and
  technology.}
\PrintBackRefs{\CurrentBib}

\bibitem [\protect \citeauthoryear {%
Busso%
\ \protect \BOthers {.}}{%
Busso%
\ \protect \BOthers {.}}{%
{\protect \APACyear {2008}}%
}]{%
IEMOCAP}
\APACinsertmetastar {%
IEMOCAP}%
\begin{APACrefauthors}%
Busso, C.%
, Bulut, M.%
, Lee, C\BHBI C.%
, Kazemzadeh, A.%
, Mower, E.%
, Kim, S.%
\BDBL {}Narayanan, S.S.%
\end{APACrefauthors}%
\unskip\
\newblock
\APACrefYearMonthDay{2008}{}{}.
\newblock
{\BBOQ}\APACrefatitle {{IEMOCAP}: Interactive emotional dyadic motion capture
  database} {{IEMOCAP}: Interactive emotional dyadic motion capture
  database}.{\BBCQ}
\newblock
\APACjournalVolNumPages{Language resources and evaluation}{42}{4}{335-359}.
\newblock

\newblock

\PrintBackRefs{\CurrentBib}

\bibitem [\protect \citeauthoryear {%
Busso%
\ \BBA {} Narayanan%
}{%
Busso%
\ \BBA {} Narayanan%
}{%
{\protect \APACyear {2008}}%
}]{%
CarlosBusso}
\APACinsertmetastar {%
CarlosBusso}%
\begin{APACrefauthors}%
Busso, C.%
\BCBT {}\ \BBA {} Narayanan, S.%
\end{APACrefauthors}%
\unskip\
\newblock
\APACrefYearMonthDay{2008}{}{}.
\newblock
{\BBOQ}\APACrefatitle {Scripted dialogs versus improvisation: Lessons learned
  about emotional elicitation techniques from the IEMOCAP database} {Scripted
  dialogs versus improvisation: Lessons learned about emotional elicitation
  techniques from the iemocap database}.{\BBCQ}
\newblock
 (\BPG~1670-1673).
\PrintBackRefs{\CurrentBib}

\bibitem [\protect \citeauthoryear {%
{Busso}%
\ \protect \BOthers {.}}{%
{Busso}%
\ \protect \BOthers {.}}{%
{\protect \APACyear {2017}}%
}]{%
MSP-IMPROV}
\APACinsertmetastar {%
MSP-IMPROV}%
\begin{APACrefauthors}%
{Busso}, C.%
, {Parthasarathy}, S.%
, {Burmania}, A.%
, {AbdelWahab}, M.%
, {Sadoughi}, N.%
\BCBL {} {Provost}, E.M.%
\end{APACrefauthors}%
\unskip\
\newblock
\APACrefYearMonthDay{2017}{}{}.
\newblock
{\BBOQ}\APACrefatitle {{MSP-IMPROV}: An Acted Corpus of Dyadic Interactions to
  Study Emotion Perception} {{MSP-IMPROV}: An acted corpus of dyadic
  interactions to study emotion perception}.{\BBCQ}
\newblock
\APACjournalVolNumPages{IEEE Transactions on Affective Computing}{8}{1}{67-80}.
\newblock

\newblock

\PrintBackRefs{\CurrentBib}

\bibitem [\protect \citeauthoryear {%
{Cao}%
\ \protect \BOthers {.}}{%
{Cao}%
\ \protect \BOthers {.}}{%
{\protect \APACyear {2014}}%
}]{%
CREMA-D}
\APACinsertmetastar {%
CREMA-D}%
\begin{APACrefauthors}%
{Cao}, H.%
, {Cooper}, D.G.%
, {Keutmann}, M.K.%
, {Gur}, R.C.%
, {Nenkova}, A.%
\BCBL {} {Verma}, R.%
\end{APACrefauthors}%
\unskip\
\newblock
\APACrefYearMonthDay{2014}{}{}.
\newblock
{\BBOQ}\APACrefatitle {{CREMA-D}: Crowd-Sourced Emotional Multimodal Actors
  Dataset} {{CREMA-D}: Crowd-sourced emotional multimodal actors
  dataset}.{\BBCQ}
\newblock
\APACjournalVolNumPages{IEEE Transactions on Affective
  Computing}{5}{4}{377-390}.
\newblock

\newblock

\PrintBackRefs{\CurrentBib}

\bibitem [\protect \citeauthoryear {%
Carvalho%
, Lease%
\BCBL {}\ \BBA {} Yilmaz%
}{%
Carvalho%
\ \protect \BOthers {.}}{%
{\protect \APACyear {2011}}%
}]{%
cs_gold_2}
\APACinsertmetastar {%
cs_gold_2}%
\begin{APACrefauthors}%
Carvalho, V.%
, Lease, M.%
\BCBL {} Yilmaz, E.%
\end{APACrefauthors}%
\unskip\
\newblock
\APACrefYearMonthDay{2011}{05}{}.
\newblock
{\BBOQ}\APACrefatitle {Crowdsourcing for search and data mining} {Crowdsourcing
  for search and data mining}.{\BBCQ}
\newblock
 (\BVOL~45, \BPG~5-6).
\newblock
\begin{APACrefDOI} \doi{10.1145/1988852.1988856} \end{APACrefDOI}
\PrintBackRefs{\CurrentBib}

\bibitem [\protect \citeauthoryear {%
Cen%
, Wu%
, Yu%
\BCBL {}\ \BBA {} Hu%
}{%
Cen%
\ \protect \BOthers {.}}{%
{\protect \APACyear {2016}}%
}]{%
onlinelearning}
\APACinsertmetastar {%
onlinelearning}%
\begin{APACrefauthors}%
Cen, L.%
, Wu, F.%
, Yu, Z.L.%
\BCBL {} Hu, F.%
\end{APACrefauthors}%
\unskip\
\newblock
\APACrefYearMonthDay{2016}{}{}.
\newblock
{\BBOQ}\APACrefatitle {Chapter 2 - A Real-Time Speech Emotion Recognition
  System and its Application in Online Learning} {Chapter 2 - a real-time
  speech emotion recognition system and its application in online
  learning}.{\BBCQ}
\newblock
 S.Y.~Tettegah\ \BBA {} M.~Gartmeier\ (\BEDS), \APACrefbtitle {Emotions,
  Technology, Design, and Learning} {Emotions, technology, design, and
  learning}\ (\BPG~27-46).
\newblock
\APACaddressPublisher{San Diego}{Academic Press}.
\newblock
\begin{APACrefURL}
  {https://www.sciencedirect.com/science/article/pii/B9780128018569000025}
  \end{APACrefURL}
\newblock
\begin{APACrefDOI} \doi{https://doi.org/10.1016/B978-0-12-801856-9.00002-5}
  \end{APACrefDOI}
\PrintBackRefs{\CurrentBib}

\bibitem [\protect \citeauthoryear {%
Chantarapratin%
}{%
Chantarapratin%
}{%
{\protect \APACyear {2020}}%
}]{%
hopewebsite}
\APACinsertmetastar {%
hopewebsite}%
\begin{APACrefauthors}%
Chantarapratin, N.%
\end{APACrefauthors}%
\unskip\
\newblock
\APACrefYearMonthDay{2020}{}{}.
\newblock
\APACrefbtitle {Hope: Data Annotations.} {Hope: Data annotations.}
\newblock
\APAChowpublished {\url{https://www.hopedata.org/}}.
\newblock
\APACrefnote{Online; accessed between 15 June 2020 to 15 July 2020}
\PrintBackRefs{\CurrentBib}

\bibitem [\protect \citeauthoryear {%
Chen%
, Wu%
, Chang%
\BCBL {}\ \BBA {} Lei%
}{%
Chen%
\ \protect \BOthers {.}}{%
{\protect \APACyear {2009}}%
}]{%
cs_intrinsic}
\APACinsertmetastar {%
cs_intrinsic}%
\begin{APACrefauthors}%
Chen, K\BHBI T.%
, Wu, C\BHBI C.%
, Chang, Y\BHBI C.%
\BCBL {} Lei, C\BHBI L.%
\end{APACrefauthors}%
\unskip\
\newblock
\APACrefYearMonthDay{2009}{}{}.
\newblock
{\BBOQ}\APACrefatitle {A Crowdsourceable QoE Evaluation Framework for
  Multimedia Content} {A crowdsourceable qoe evaluation framework for
  multimedia content}.{\BBCQ}
\newblock
\APACaddressPublisher{New York, NY, USA}{Association for Computing Machinery}.
\newblock
\begin{APACrefURL} {https://doi.org/10.1145/1631272.1631339} \end{APACrefURL}
\newblock
\begin{APACrefDOI} \doi{10.1145/1631272.1631339} \end{APACrefDOI}
\PrintBackRefs{\CurrentBib}

\bibitem [\protect \citeauthoryear {%
Chong%
, Kim%
\BCBL {}\ \BBA {} Davis%
}{%
Chong%
\ \protect \BOthers {.}}{%
{\protect \APACyear {2015}}%
}]{%
chong2015exploring}
\APACinsertmetastar {%
chong2015exploring}%
\begin{APACrefauthors}%
Chong, C.S.%
, Kim, J.%
\BCBL {} Davis, C.%
\end{APACrefauthors}%
\unskip\
\newblock
\APACrefYearMonthDay{2015}{}{}.
\newblock
{\BBOQ}\APACrefatitle {Exploring acoustic differences between Cantonese (tonal)
  and English (non-tonal) spoken expressions of emotions} {Exploring acoustic
  differences between cantonese (tonal) and english (non-tonal) spoken
  expressions of emotions}.{\BBCQ}
\newblock
 \APACrefbtitle {Sixteenth Annual Conference of the International Speech
  Communication Association.} {Sixteenth annual conference of the international
  speech communication association.}
\PrintBackRefs{\CurrentBib}

\bibitem [\protect \citeauthoryear {%
Chopra%
, Mathur%
, Sawhney%
\BCBL {}\ \BBA {} Shah%
}{%
Chopra%
\ \protect \BOthers {.}}{%
{\protect \APACyear {2021}}%
}]{%
crosscorpus3}
\APACinsertmetastar {%
crosscorpus3}%
\begin{APACrefauthors}%
Chopra, S.%
, Mathur, P.%
, Sawhney, R.%
\BCBL {} Shah, R.R.%
\end{APACrefauthors}%
\unskip\
\newblock
\APACrefYearMonthDay{2021}{}{}.
\newblock
{\BBOQ}\APACrefatitle {Meta-Learning for Low-Resource Speech Emotion
  Recognition} {Meta-learning for low-resource speech emotion
  recognition}.{\BBCQ}
\newblock
 \APACrefbtitle {ICASSP 2021 - 2021 IEEE International Conference on Acoustics,
  Speech and Signal Processing (ICASSP)} {Icassp 2021 - 2021 ieee international
  conference on acoustics, speech and signal processing (icassp)}\
  (\BPG~6259-6263).
\newblock
\begin{APACrefDOI} \doi{10.1109/ICASSP39728.2021.9414373} \end{APACrefDOI}
\PrintBackRefs{\CurrentBib}

\bibitem [\protect \citeauthoryear {%
Chou%
\ \BBA {} Lee%
}{%
Chou%
\ \BBA {} Lee%
}{%
{\protect \APACyear {2019}}%
}]{%
soft-label4}
\APACinsertmetastar {%
soft-label4}%
\begin{APACrefauthors}%
Chou, H\BHBI C.%
\BCBT {}\ \BBA {} Lee, C\BHBI C.%
\end{APACrefauthors}%
\unskip\
\newblock
\APACrefYearMonthDay{2019}{}{}.
\newblock
{\BBOQ}\APACrefatitle {Every Rating Matters: Joint Learning of Subjective
  Labels and Individual Annotators for Speech Emotion Classification} {Every
  rating matters: Joint learning of subjective labels and individual annotators
  for speech emotion classification}.{\BBCQ}
\newblock
 \APACrefbtitle {ICASSP 2019 - 2019 IEEE International Conference on Acoustics,
  Speech and Signal Processing (ICASSP)} {Icassp 2019 - 2019 ieee international
  conference on acoustics, speech and signal processing (icassp)}\
  (\BPG~5886-5890).
\newblock
\begin{APACrefDOI} \doi{10.1109/ICASSP.2019.8682170} \end{APACrefDOI}
\PrintBackRefs{\CurrentBib}

\bibitem [\protect \citeauthoryear {%
Chu%
\ \protect \BOthers {.}}{%
Chu%
\ \protect \BOthers {.}}{%
{\protect \APACyear {2023}}%
}]{%
chu2023qwenaudioadvancinguniversalaudio}
\APACinsertmetastar {%
chu2023qwenaudioadvancinguniversalaudio}%
\begin{APACrefauthors}%
Chu, Y.%
, Xu, J.%
, Zhou, X.%
, Yang, Q.%
, Zhang, S.%
, Yan, Z.%
\BDBL {}Zhou, J.%
\end{APACrefauthors}%
\unskip\
\newblock
\APACrefYearMonthDay{2023}{}{}.
\newblock
\APACrefbtitle {Qwen-Audio: Advancing Universal Audio Understanding via Unified
  Large-Scale Audio-Language Models.} {Qwen-audio: Advancing universal audio
  understanding via unified large-scale audio-language models.}
\newblock
\begin{APACrefURL} {https://arxiv.org/abs/2311.07919} \end{APACrefURL}
\PrintBackRefs{\CurrentBib}

\bibitem [\protect \citeauthoryear {%
Cohen%
}{%
Cohen%
}{%
{\protect \APACyear {1960}}%
}]{%
kappascore}
\APACinsertmetastar {%
kappascore}%
\begin{APACrefauthors}%
Cohen, J.%
\end{APACrefauthors}%
\unskip\
\newblock
\APACrefYearMonthDay{1960}{}{}.
\newblock
{\BBOQ}\APACrefatitle {A Coefficient of Agreement for Nominal Scales} {A
  coefficient of agreement for nominal scales}.{\BBCQ}
\newblock
\APACjournalVolNumPages{Educational and Psychological
  Measurement}{20}{1}{37-46}.
\newblock
\begin{APACrefURL} {https://doi.org/10.1177/001316446002000104}
  \end{APACrefURL}
\newblock
{\href{https://arxiv.org/abs/https://doi.org/10.1177/001316446002000104}{{https://doi.org/10.1177/001316446002000104}}}
\newblock

\newblock
\begin{APACrefDOI} \doi{10.1177/001316446002000104} \end{APACrefDOI}
\PrintBackRefs{\CurrentBib}

\bibitem [\protect \citeauthoryear {%
Costantini%
, Iaderola%
, Paoloni%
\BCBL {}\ \BBA {} Todisco%
}{%
Costantini%
\ \protect \BOthers {.}}{%
{\protect \APACyear {2014}}%
}]{%
EMOVO}
\APACinsertmetastar {%
EMOVO}%
\begin{APACrefauthors}%
Costantini, G.%
, Iaderola, I.%
, Paoloni, A.%
\BCBL {} Todisco, M.%
\end{APACrefauthors}%
\unskip\
\newblock
\APACrefYearMonthDay{2014}{}{}.
\newblock
{\BBOQ}\APACrefatitle {{EMOVO} corpus: an Italian emotional speech database}
  {{EMOVO} corpus: an italian emotional speech database}.{\BBCQ}
\newblock
 \APACrefbtitle {International Conference on Language Resources and Evaluation
  (LREC 2014)} {International conference on language resources and evaluation
  (lrec 2014)}\ (\BPG~3501-3504).
\PrintBackRefs{\CurrentBib}

\bibitem [\protect \citeauthoryear {%
Crowston%
}{%
Crowston%
}{%
{\protect \APACyear {2012}}%
}]{%
amazonmechanicalturk}
\APACinsertmetastar {%
amazonmechanicalturk}%
\begin{APACrefauthors}%
Crowston, K.%
\end{APACrefauthors}%
\unskip\
\newblock
\APACrefYearMonthDay{2012}{}{}.
\newblock
{\BBOQ}\APACrefatitle {Amazon Mechanical Turk: A Research Tool for
  Organizations and Information Systems Scholars} {Amazon mechanical turk: A
  research tool for organizations and information systems scholars}.{\BBCQ}
\newblock
 A.~Bhattacherjee\ \BBA {} B.~Fitzgerald\ (\BEDS), \APACrefbtitle {Shaping the
  Future of ICT Research. Methods and Approaches} {Shaping the future of ict
  research. methods and approaches}\ (\BPGS\ 210--221).
\newblock
\APACaddressPublisher{Berlin, Heidelberg}{Springer Berlin Heidelberg}.
\PrintBackRefs{\CurrentBib}

\bibitem [\protect \citeauthoryear {%
Etienne%
, Fidanza%
, Petrovskii%
, Devillers%
\BCBL {}\ \BBA {} Schmauch%
}{%
Etienne%
\ \protect \BOthers {.}}{%
{\protect \APACyear {2018}}%
}]{%
ser_kfold:journals/corr/abs-1802-05630}
\APACinsertmetastar {%
ser_kfold:journals/corr/abs-1802-05630}%
\begin{APACrefauthors}%
Etienne, C.%
, Fidanza, G.%
, Petrovskii, A.%
, Devillers, L.%
\BCBL {} Schmauch, B.%
\end{APACrefauthors}%
\unskip\
\newblock
\APACrefYearMonthDay{2018}{}{}.
\newblock
{\BBOQ}\APACrefatitle {Speech Emotion Recognition with Data Augmentation and
  Layer-wise Learning Rate Adjustment} {Speech emotion recognition with data
  augmentation and layer-wise learning rate adjustment}.{\BBCQ}
\newblock
\APACjournalVolNumPages{CoRR}{abs/1802.05630}{}{}.
\newblock
\begin{APACrefURL} {http://arxiv.org/abs/1802.05630} \end{APACrefURL}
\newblock
{\href{https://arxiv.org/abs/1802.05630}{{1802.05630}}}
\newblock

\PrintBackRefs{\CurrentBib}

\bibitem [\protect \citeauthoryear {%
Fan%
, Xu%
, Xing%
, Chen%
\BCBL {}\ \BBA {} Huang%
}{%
Fan%
\ \protect \BOthers {.}}{%
{\protect \APACyear {2021}}%
}]{%
lssed}
\APACinsertmetastar {%
lssed}%
\begin{APACrefauthors}%
Fan, W.%
, Xu, X.%
, Xing, X.%
, Chen, W.%
\BCBL {} Huang, D.%
\end{APACrefauthors}%
\unskip\
\newblock
\APACrefYearMonthDay{2021}{}{}.
\newblock
{\BBOQ}\APACrefatitle {LSSED: A Large-Scale Dataset and Benchmark for Speech
  Emotion Recognition} {Lssed: A large-scale dataset and benchmark for speech
  emotion recognition}.{\BBCQ}
\newblock
 \APACrefbtitle {ICASSP 2021 - 2021 IEEE International Conference on Acoustics,
  Speech and Signal Processing (ICASSP)} {Icassp 2021 - 2021 ieee international
  conference on acoustics, speech and signal processing (icassp)}\
  (\BPG~641-645).
\newblock
\begin{APACrefDOI} \doi{10.1109/ICASSP39728.2021.9414542} \end{APACrefDOI}
\PrintBackRefs{\CurrentBib}

\bibitem [\protect \citeauthoryear {%
Gangwanpongpun%
}{%
Gangwanpongpun%
}{%
{\protect \APACyear {2020}}%
}]{%
wangwebsite}
\APACinsertmetastar {%
wangwebsite}%
\begin{APACrefauthors}%
Gangwanpongpun, K.%
\end{APACrefauthors}%
\unskip\
\newblock
\APACrefYearMonthDay{2020}{}{}.
\newblock
\APACrefbtitle {Wang: Data Market.} {Wang: Data market.}
\newblock
\APAChowpublished {\url{https://www.wang.in.th/}}.
\newblock
\APACrefnote{Online; accessed between 1 August 2020 to 15 March 2021}
\PrintBackRefs{\CurrentBib}

\bibitem [\protect \citeauthoryear {%
Goel%
\ \BBA {} Beigi%
}{%
Goel%
\ \BBA {} Beigi%
}{%
{\protect \APACyear {2020}}%
}]{%
crosscorpus2}
\APACinsertmetastar {%
crosscorpus2}%
\begin{APACrefauthors}%
Goel, S.%
\BCBT {}\ \BBA {} Beigi, H.%
\end{APACrefauthors}%
\unskip\
\newblock
\APACrefYearMonthDay{2020}{}{}.
\newblock
\APACrefbtitle {Cross Lingual Cross Corpus Speech Emotion Recognition.} {Cross
  lingual cross corpus speech emotion recognition.}
\PrintBackRefs{\CurrentBib}

\bibitem [\protect \citeauthoryear {%
{Grimm}%
, {Kroschel}%
\BCBL {}\ \BBA {} {Narayanan}%
}{%
{Grimm}%
\ \protect \BOthers {.}}{%
{\protect \APACyear {2008}}%
}]{%
Grimm_3}
\APACinsertmetastar {%
Grimm_3}%
\begin{APACrefauthors}%
{Grimm}, M.%
, {Kroschel}, K.%
\BCBL {} {Narayanan}, S.%
\end{APACrefauthors}%
\unskip\
\newblock
\APACrefYearMonthDay{2008}{}{}.
\newblock
{\BBOQ}\APACrefatitle {The {Vera} am {Mittag} German audio-visual emotional
  speech database} {The {Vera} am {Mittag} german audio-visual emotional speech
  database}.{\BBCQ}
\newblock
 \APACrefbtitle {2008 IEEE International Conference on Multimedia and Expo}
  {2008 ieee international conference on multimedia and expo}\ (\BPG~865-868).
\PrintBackRefs{\CurrentBib}

\bibitem [\protect \citeauthoryear {%
Hirth%
, Hoßfeld%
\BCBL {}\ \BBA {} Tran-Gia%
}{%
Hirth%
\ \protect \BOthers {.}}{%
{\protect \APACyear {2011}}%
}]{%
cs_majority_1}
\APACinsertmetastar {%
cs_majority_1}%
\begin{APACrefauthors}%
Hirth, M.%
, Hoßfeld, T.%
\BCBL {} Tran-Gia, P.%
\end{APACrefauthors}%
\unskip\
\newblock
\APACrefYearMonthDay{2011}{}{}.
\newblock
{\BBOQ}\APACrefatitle {Cost-Optimal Validation Mechanisms and Cheat-Detection
  for Crowdsourcing Platforms} {Cost-optimal validation mechanisms and
  cheat-detection for crowdsourcing platforms}.{\BBCQ}
\newblock
 \APACrefbtitle {2011 Fifth International Conference on Innovative Mobile and
  Internet Services in Ubiquitous Computing} {2011 fifth international
  conference on innovative mobile and internet services in ubiquitous
  computing}\ (\BPG~316-321).
\newblock
\begin{APACrefDOI} \doi{10.1109/IMIS.2011.91} \end{APACrefDOI}
\PrintBackRefs{\CurrentBib}

\bibitem [\protect \citeauthoryear {%
Hirth%
, Scheuring%
, Hossfeld%
, Schwartz%
\BCBL {}\ \BBA {} Tran-Gia%
}{%
Hirth%
\ \protect \BOthers {.}}{%
{\protect \APACyear {2014}}%
}]{%
cs_behavior_2}
\APACinsertmetastar {%
cs_behavior_2}%
\begin{APACrefauthors}%
Hirth, M.%
, Scheuring, S.%
, Hossfeld, T.%
, Schwartz, C.%
\BCBL {} Tran-Gia, P.%
\end{APACrefauthors}%
\unskip\
\newblock
\APACrefYearMonthDay{2014}{07}{}.
\newblock
{\BBOQ}\APACrefatitle {Predicting Result Quality in Crowdsourcing Using
  Application Layer Monitoring} {Predicting result quality in crowdsourcing
  using application layer monitoring}.{\BBCQ}.
\newblock
\begin{APACrefDOI} \doi{10.1109/CCE.2014.6916756} \end{APACrefDOI}
\PrintBackRefs{\CurrentBib}

\bibitem [\protect \citeauthoryear {%
Hochreiter%
\ \BBA {} Schmidhuber%
}{%
Hochreiter%
\ \BBA {} Schmidhuber%
}{%
{\protect \APACyear {1997}}%
}]{%
lstm}
\APACinsertmetastar {%
lstm}%
\begin{APACrefauthors}%
Hochreiter, S.%
\BCBT {}\ \BBA {} Schmidhuber, J.%
\end{APACrefauthors}%
\unskip\
\newblock
\APACrefYearMonthDay{1997}{12}{}.
\newblock
{\BBOQ}\APACrefatitle {Long Short-term Memory} {Long short-term memory}.{\BBCQ}
\newblock
\APACjournalVolNumPages{Neural computation}{9}{}{1735-80}.
\newblock

\newblock

\newblock
\begin{APACrefDOI} \doi{10.1162/neco.1997.9.8.1735} \end{APACrefDOI}
\PrintBackRefs{\CurrentBib}

\bibitem [\protect \citeauthoryear {%
Huahu%
, Jue%
\BCBL {}\ \BBA {} Jian%
}{%
Huahu%
\ \protect \BOthers {.}}{%
{\protect \APACyear {2010}}%
}]{%
robot}
\APACinsertmetastar {%
robot}%
\begin{APACrefauthors}%
Huahu, X.%
, Jue, G.%
\BCBL {} Jian, Y.%
\end{APACrefauthors}%
\unskip\
\newblock
\APACrefYearMonthDay{2010}{}{}.
\newblock
{\BBOQ}\APACrefatitle {Application of Speech Emotion Recognition in Intelligent
  Household Robot} {Application of speech emotion recognition in intelligent
  household robot}.{\BBCQ}
\newblock
 \APACrefbtitle {2010 International Conference on Artificial Intelligence and
  Computational Intelligence} {2010 international conference on artificial
  intelligence and computational intelligence}\ (\BVOL~1, \BPG~537-541).
\newblock
\begin{APACrefDOI} \doi{10.1109/AICI.2010.118} \end{APACrefDOI}
\PrintBackRefs{\CurrentBib}

\bibitem [\protect \citeauthoryear {%
Ipeirotis%
, Provost%
\BCBL {}\ \BBA {} Wang%
}{%
Ipeirotis%
\ \protect \BOthers {.}}{%
{\protect \APACyear {2010}}%
}]{%
cs_gold_5}
\APACinsertmetastar {%
cs_gold_5}%
\begin{APACrefauthors}%
Ipeirotis, P.G.%
, Provost, F.%
\BCBL {} Wang, J.%
\end{APACrefauthors}%
\unskip\
\newblock
\APACrefYearMonthDay{2010}{}{}.
\newblock
{\BBOQ}\APACrefatitle {Quality Management on Amazon Mechanical Turk} {Quality
  management on amazon mechanical turk}.{\BBCQ}
\newblock
 (\BPG~64–67).
\newblock
\APACaddressPublisher{New York, NY, USA}{Association for Computing Machinery}.
\newblock
\begin{APACrefURL} {https://doi.org/10.1145/1837885.1837906} \end{APACrefURL}
\newblock
\begin{APACrefDOI} \doi{10.1145/1837885.1837906} \end{APACrefDOI}
\PrintBackRefs{\CurrentBib}

\bibitem [\protect \citeauthoryear {%
Jaitly%
\ \BBA {} Hinton%
}{%
Jaitly%
\ \BBA {} Hinton%
}{%
{\protect \APACyear {2013}}%
}]{%
Jaitly2013VocalTL}
\APACinsertmetastar {%
Jaitly2013VocalTL}%
\begin{APACrefauthors}%
Jaitly, N.%
\BCBT {}\ \BBA {} Hinton, E.%
\end{APACrefauthors}%
\unskip\
\newblock
\APACrefYearMonthDay{2013}{}{}.
\newblock
{\BBOQ}\APACrefatitle {Vocal Tract Length Perturbation (VTLP) improves speech
  recognition} {Vocal tract length perturbation (vtlp) improves speech
  recognition}.{\BBCQ}.
\PrintBackRefs{\CurrentBib}

\bibitem [\protect \citeauthoryear {%
Kasuriya%
, Sornlertlamvanich%
, Cotsomrong%
, Kanokphara%
\BCBL {}\ \BBA {} Thatphithakkul%
}{%
Kasuriya%
\ \protect \BOthers {.}}{%
{\protect \APACyear {2004}}%
}]{%
Kasuriya_1}
\APACinsertmetastar {%
Kasuriya_1}%
\begin{APACrefauthors}%
Kasuriya, S.%
, Sornlertlamvanich, V.%
, Cotsomrong, P.%
, Kanokphara, S.%
\BCBL {} Thatphithakkul, N.%
\end{APACrefauthors}%
\unskip\
\newblock
\APACrefYearMonthDay{2004}{}{}.
\newblock
{\BBOQ}\APACrefatitle {Thai Speech Recognition Corpora} {Thai speech
  recognition corpora}.{\BBCQ}
\newblock
\APACjournalVolNumPages{Journal of Chinese Language and Computing}{14}{4}{}.
\newblock

\newblock

\PrintBackRefs{\CurrentBib}

\bibitem [\protect \citeauthoryear {%
Kazai%
\ \BBA {} Zitouni%
}{%
Kazai%
\ \BBA {} Zitouni%
}{%
{\protect \APACyear {2016}}%
}]{%
cs_behavior_4}
\APACinsertmetastar {%
cs_behavior_4}%
\begin{APACrefauthors}%
Kazai, G.%
\BCBT {}\ \BBA {} Zitouni, I.%
\end{APACrefauthors}%
\unskip\
\newblock
\APACrefYearMonthDay{2016}{}{}.
\newblock
{\BBOQ}\APACrefatitle {Quality Management in Crowdsourcing Using Gold Judges
  Behavior} {Quality management in crowdsourcing using gold judges
  behavior}.{\BBCQ}
\newblock
 (\BPG~267–276).
\newblock
\APACaddressPublisher{New York, NY, USA}{Association for Computing Machinery}.
\newblock
\begin{APACrefURL} {https://doi.org/10.1145/2835776.2835835} \end{APACrefURL}
\newblock
\begin{APACrefDOI} \doi{10.1145/2835776.2835835} \end{APACrefDOI}
\PrintBackRefs{\CurrentBib}

\bibitem [\protect \citeauthoryear {%
Kingma%
\ \BBA {} Ba%
}{%
Kingma%
\ \BBA {} Ba%
}{%
{\protect \APACyear {2017}}%
}]{%
kingma2017adam}
\APACinsertmetastar {%
kingma2017adam}%
\begin{APACrefauthors}%
Kingma, D.P.%
\BCBT {}\ \BBA {} Ba, J.%
\end{APACrefauthors}%
\unskip\
\newblock
\APACrefYearMonthDay{2017}{}{}.
\newblock
\APACrefbtitle {Adam: A Method for Stochastic Optimization.} {Adam: A method
  for stochastic optimization.}
\PrintBackRefs{\CurrentBib}

\bibitem [\protect \citeauthoryear {%
{Kossaifi}%
\ \protect \BOthers {.}}{%
{Kossaifi}%
\ \protect \BOthers {.}}{%
{\protect \APACyear {2021}}%
}]{%
SEWA}
\APACinsertmetastar {%
SEWA}%
\begin{APACrefauthors}%
{Kossaifi}, J.%
, {Walecki}, R.%
, {Panagakis}, Y.%
, {Shen}, J.%
, {Schmitt}, M.%
, {Ringeval}, F.%
\BDBL {}{Pantic}, M.%
\end{APACrefauthors}%
\unskip\
\newblock
\APACrefYearMonthDay{2021}{}{}.
\newblock
{\BBOQ}\APACrefatitle {SEWA DB: A Rich Database for Audio-Visual Emotion and
  Sentiment Research in the Wild} {Sewa db: A rich database for audio-visual
  emotion and sentiment research in the wild}.{\BBCQ}
\newblock
\APACjournalVolNumPages{IEEE Transactions on Pattern Analysis and Machine
  Intelligence}{43}{3}{1022-1040}.
\newblock

\newblock

\PrintBackRefs{\CurrentBib}

\bibitem [\protect \citeauthoryear {%
Krippendorff%
}{%
Krippendorff%
}{%
{\protect \APACyear {2004}}%
}]{%
Krippendorff-2004}
\APACinsertmetastar {%
Krippendorff-2004}%
\begin{APACrefauthors}%
Krippendorff, K.%
\end{APACrefauthors}%
\unskip\
\newblock
\APACrefYear{2004}.
\newblock
\APACrefbtitle {Content Analysis: An Introduction to Its Methodology (second
  edition)} {Content analysis: An introduction to its methodology (second
  edition)}.
\newblock
\APACaddressPublisher{}{Sage Publications}.
\PrintBackRefs{\CurrentBib}

\bibitem [\protect \citeauthoryear {%
Krippendorff%
}{%
Krippendorff%
}{%
{\protect \APACyear {2011}}%
}]{%
Krippendorff-2011}
\APACinsertmetastar {%
Krippendorff-2011}%
\begin{APACrefauthors}%
Krippendorff, K.%
\end{APACrefauthors}%
\unskip\
\newblock
\APACrefYearMonthDay{2011}{}{}.
\newblock
{\BBOQ}\APACrefatitle {Computing {Krippendorff's} alpha-reliability} {Computing
  {Krippendorff's} alpha-reliability}.{\BBCQ}
\newblock

\newblock

\newblock

\PrintBackRefs{\CurrentBib}

\bibitem [\protect \citeauthoryear {%
Lakomkin%
, Zamani%
, Weber%
, Magg%
\BCBL {}\ \BBA {} Wermter%
}{%
Lakomkin%
\ \protect \BOthers {.}}{%
{\protect \APACyear {2018}}%
}]{%
envrobust}
\APACinsertmetastar {%
envrobust}%
\begin{APACrefauthors}%
Lakomkin, E.%
, Zamani, M.A.%
, Weber, C.%
, Magg, S.%
\BCBL {} Wermter, S.%
\end{APACrefauthors}%
\unskip\
\newblock
\APACrefYearMonthDay{2018}{}{}.
\newblock
{\BBOQ}\APACrefatitle {On the Robustness of Speech Emotion Recognition for
  Human-Robot Interaction with Deep Neural Networks} {On the robustness of
  speech emotion recognition for human-robot interaction with deep neural
  networks}.{\BBCQ}
\newblock
 \APACrefbtitle {2018 IEEE/RSJ International Conference on Intelligent Robots
  and Systems (IROS)} {2018 ieee/rsj international conference on intelligent
  robots and systems (iros)}\ (\BPG~854-860).
\newblock
\begin{APACrefDOI} \doi{10.1109/IROS.2018.8593571} \end{APACrefDOI}
\PrintBackRefs{\CurrentBib}

\bibitem [\protect \citeauthoryear {%
Lee%
}{%
Lee%
}{%
{\protect \APACyear {2021}}%
}]{%
crosscorpus1}
\APACinsertmetastar {%
crosscorpus1}%
\begin{APACrefauthors}%
Lee, S\BHBI w.%
\end{APACrefauthors}%
\unskip\
\newblock
\APACrefYearMonthDay{2021}{}{}.
\newblock
{\BBOQ}\APACrefatitle {Domain Generalization with Triplet Network for
  Cross-Corpus Speech Emotion Recognition} {Domain generalization with triplet
  network for cross-corpus speech emotion recognition}.{\BBCQ}
\newblock
 \APACrefbtitle {2021 IEEE Spoken Language Technology Workshop (SLT)} {2021
  ieee spoken language technology workshop (slt)}\ (\BPG~389-396).
\newblock
\begin{APACrefDOI} \doi{10.1109/SLT48900.2021.9383534} \end{APACrefDOI}
\PrintBackRefs{\CurrentBib}

\bibitem [\protect \citeauthoryear {%
Li%
, Tao%
, Chao%
, Bao%
\BCBL {}\ \BBA {} Liu%
}{%
Li%
\ \protect \BOthers {.}}{%
{\protect \APACyear {2017}}%
}]{%
Li_5}
\APACinsertmetastar {%
Li_5}%
\begin{APACrefauthors}%
Li, Y.%
, Tao, J.%
, Chao, L.%
, Bao, W.%
\BCBL {} Liu, Y.%
\end{APACrefauthors}%
\unskip\
\newblock
\APACrefYearMonthDay{2017}{}{}.
\newblock
{\BBOQ}\APACrefatitle {{CHEAVD}: a Chinese natural emotional audio-visual
  database} {{CHEAVD}: a chinese natural emotional audio-visual
  database}.{\BBCQ}
\newblock
\APACjournalVolNumPages{Journal of Ambient Intelligence and Humanized
  Computing}{8}{6}{913-924}.
\newblock

\newblock

\PrintBackRefs{\CurrentBib}

\bibitem [\protect \citeauthoryear {%
Little%
, Chilton%
, Goldman%
\BCBL {}\ \BBA {} Miller%
}{%
Little%
\ \protect \BOthers {.}}{%
{\protect \APACyear {2010}}%
}]{%
cs_iterative}
\APACinsertmetastar {%
cs_iterative}%
\begin{APACrefauthors}%
Little, G.%
, Chilton, L.B.%
, Goldman, M.%
\BCBL {} Miller, R.%
\end{APACrefauthors}%
\unskip\
\newblock
\APACrefYearMonthDay{2010}{}{}.
\newblock
{\BBOQ}\APACrefatitle {Exploring iterative and parallel human computation
  processes} {Exploring iterative and parallel human computation
  processes}.{\BBCQ}
\newblock
 \APACrefbtitle {HCOMP '10.} {Hcomp '10.}
\PrintBackRefs{\CurrentBib}

\bibitem [\protect \citeauthoryear {%
Livingstone%
\ \BBA {} Russo%
}{%
Livingstone%
\ \BBA {} Russo%
}{%
{\protect \APACyear {2018}}%
}]{%
RAVDESS}
\APACinsertmetastar {%
RAVDESS}%
\begin{APACrefauthors}%
Livingstone, S.R.%
\BCBT {}\ \BBA {} Russo, F.A.%
\end{APACrefauthors}%
\unskip\
\newblock
\APACrefYearMonthDay{2018}{}{}.
\newblock
{\BBOQ}\APACrefatitle {The {Ryerson} Audio-Visual Database of Emotional Speech
  and Song ({RAVDESS}): A dynamic, multimodal set of facial and vocal
  expressions in North American English} {The {Ryerson} audio-visual database
  of emotional speech and song ({RAVDESS}): A dynamic, multimodal set of facial
  and vocal expressions in north american english}.{\BBCQ}
\newblock
\APACjournalVolNumPages{PLOS ONE}{13}{5}{1-35}.
\newblock

\newblock

\PrintBackRefs{\CurrentBib}

\bibitem [\protect \citeauthoryear {%
Lotfian%
\ \BBA {} Busso%
}{%
Lotfian%
\ \BBA {} Busso%
}{%
{\protect \APACyear {2017}}%
}]{%
soft-label2}
\APACinsertmetastar {%
soft-label2}%
\begin{APACrefauthors}%
Lotfian, R.%
\BCBT {}\ \BBA {} Busso, C.%
\end{APACrefauthors}%
\unskip\
\newblock
\APACrefYearMonthDay{2017}{}{}.
\newblock
{\BBOQ}\APACrefatitle {Formulating emotion perception as a probabilistic model
  with application to categorical emotion classification} {Formulating emotion
  perception as a probabilistic model with application to categorical emotion
  classification}.{\BBCQ}
\newblock
 \APACrefbtitle {2017 Seventh International Conference on Affective Computing
  and Intelligent Interaction (ACII)} {2017 seventh international conference on
  affective computing and intelligent interaction (acii)}\ (\BPG~415-420).
\newblock
\begin{APACrefDOI} \doi{10.1109/ACII.2017.8273633} \end{APACrefDOI}
\PrintBackRefs{\CurrentBib}

\bibitem [\protect \citeauthoryear {%
Lotfian%
\ \BBA {} Busso%
}{%
Lotfian%
\ \BBA {} Busso%
}{%
{\protect \APACyear {2019}}%
{\protect \APACexlab {{\protect \BCnt {1}}}}}]{%
MSP-PODCAST}
\APACinsertmetastar {%
MSP-PODCAST}%
\begin{APACrefauthors}%
Lotfian, R.%
\BCBT {}\ \BBA {} Busso, C.%
\end{APACrefauthors}%
\unskip\
\newblock
\APACrefYearMonthDay{2019{\protect \BCnt {1}}}{October-December}{}.
\newblock
{\BBOQ}\APACrefatitle {Building Naturalistic Emotionally Balanced Speech Corpus
  by Retrieving Emotional Speech From Existing Podcast Recordings} {Building
  naturalistic emotionally balanced speech corpus by retrieving emotional
  speech from existing podcast recordings}.{\BBCQ}
\newblock
\APACjournalVolNumPages{IEEE Transactions on Affective
  Computing}{10}{4}{471-483}.
\newblock

\newblock

\newblock
\begin{APACrefDOI} \doi{10.1109/TAFFC.2017.2736999} \end{APACrefDOI}
\PrintBackRefs{\CurrentBib}

\bibitem [\protect \citeauthoryear {%
Lotfian%
\ \BBA {} Busso%
}{%
Lotfian%
\ \BBA {} Busso%
}{%
{\protect \APACyear {2019}}%
{\protect \APACexlab {{\protect \BCnt {2}}}}}]{%
curriculum}
\APACinsertmetastar {%
curriculum}%
\begin{APACrefauthors}%
Lotfian, R.%
\BCBT {}\ \BBA {} Busso, C.%
\end{APACrefauthors}%
\unskip\
\newblock
\APACrefYearMonthDay{2019{\protect \BCnt {2}}}{}{}.
\newblock
{\BBOQ}\APACrefatitle {Curriculum Learning for Speech Emotion Recognition From
  Crowdsourced Labels} {Curriculum learning for speech emotion recognition from
  crowdsourced labels}.{\BBCQ}
\newblock
\APACjournalVolNumPages{IEEE/ACM Transactions on Audio, Speech, and Language
  Processing}{27}{4}{815-826}.
\newblock

\newblock

\newblock
\begin{APACrefDOI} \doi{10.1109/TASLP.2019.2898816} \end{APACrefDOI}
\PrintBackRefs{\CurrentBib}

\bibitem [\protect \citeauthoryear {%
Mok%
, Chang%
\BCBL {}\ \BBA {} Li%
}{%
Mok%
\ \protect \BOthers {.}}{%
{\protect \APACyear {2017}}%
}]{%
cs_behavior_3}
\APACinsertmetastar {%
cs_behavior_3}%
\begin{APACrefauthors}%
Mok, R.K.P.%
, Chang, R.K.C.%
\BCBL {} Li, W.%
\end{APACrefauthors}%
\unskip\
\newblock
\APACrefYearMonthDay{2017}{}{}.
\newblock
{\BBOQ}\APACrefatitle {Detecting Low-Quality Workers in QoE Crowdtesting: A
  Worker Behavior-Based Approach} {Detecting low-quality workers in qoe
  crowdtesting: A worker behavior-based approach}.{\BBCQ}
\newblock
\APACjournalVolNumPages{IEEE Transactions on Multimedia}{19}{3}{530-543}.
\newblock

\newblock

\newblock
\begin{APACrefDOI} \doi{10.1109/TMM.2016.2619901} \end{APACrefDOI}
\PrintBackRefs{\CurrentBib}

\bibitem [\protect \citeauthoryear {%
Neumann%
\ \BBA {} Vu%
}{%
Neumann%
\ \BBA {} Vu%
}{%
{\protect \APACyear {2017}}%
}]{%
neumann2017attentive}
\APACinsertmetastar {%
neumann2017attentive}%
\begin{APACrefauthors}%
Neumann, M.%
\BCBT {}\ \BBA {} Vu, N.T.%
\end{APACrefauthors}%
\unskip\
\newblock
\APACrefYearMonthDay{2017}{}{}.
\newblock
\APACrefbtitle {Attentive Convolutional Neural Network based Speech Emotion
  Recognition: A Study on the Impact of Input Features, Signal Length, and
  Acted Speech.} {Attentive convolutional neural network based speech emotion
  recognition: A study on the impact of input features, signal length, and
  acted speech.}
\PrintBackRefs{\CurrentBib}

\bibitem [\protect \citeauthoryear {%
Oleson%
\ \protect \BOthers {.}}{%
Oleson%
\ \protect \BOthers {.}}{%
{\protect \APACyear {2011}}%
}]{%
cs_gold_4}
\APACinsertmetastar {%
cs_gold_4}%
\begin{APACrefauthors}%
Oleson, D.%
, Sorokin, A.%
, Laughlin, G.%
, Hester, V.%
, Le, J.%
\BCBL {} Biewald, L.%
\end{APACrefauthors}%
\unskip\
\newblock
\APACrefYearMonthDay{2011}{}{}.
\newblock
{\BBOQ}\APACrefatitle {Programmatic Gold: Targeted and Scalable Quality
  Assurance in Crowdsourcing} {Programmatic gold: Targeted and scalable quality
  assurance in crowdsourcing}.{\BBCQ}
\newblock
 \APACrefbtitle {Proceedings of the 11th AAAI Conference on Human Computation}
  {Proceedings of the 11th aaai conference on human computation}\
  (\BPG~43–48).
\newblock
\APACaddressPublisher{}{AAAI Press}.
\PrintBackRefs{\CurrentBib}

\bibitem [\protect \citeauthoryear {%
Parada-Cabaleiro%
, Costantini%
, Batliner%
, Schmitt%
\BCBL {}\ \BBA {} Schuller%
}{%
Parada-Cabaleiro%
\ \protect \BOthers {.}}{%
{\protect \APACyear {2019}}%
}]{%
DEMoS}
\APACinsertmetastar {%
DEMoS}%
\begin{APACrefauthors}%
Parada-Cabaleiro, E.%
, Costantini, G.%
, Batliner, A.%
, Schmitt, M.%
\BCBL {} Schuller, B.W.%
\end{APACrefauthors}%
\unskip\
\newblock
\APACrefYearMonthDay{2019}{}{}.
\newblock
{\BBOQ}\APACrefatitle {{DEMoS}: An Italian emotional speech corpus} {{DEMoS}:
  An italian emotional speech corpus}.{\BBCQ}
\newblock
\APACjournalVolNumPages{Language Resources and Evaluation}{}{}{1-43}.
\newblock

\newblock

\PrintBackRefs{\CurrentBib}

\bibitem [\protect \citeauthoryear {%
Parry%
\ \protect \BOthers {.}}{%
Parry%
\ \protect \BOthers {.}}{%
{\protect \APACyear {2019}}%
}]{%
crosscorpus4}
\APACinsertmetastar {%
crosscorpus4}%
\begin{APACrefauthors}%
Parry, J.%
, Palaz, D.%
, Clarke, G.%
, Lecomte, P.%
, Mead, R.%
, Berger, M.%
\BCBL {} Hofer, G.%
\end{APACrefauthors}%
\unskip\
\newblock
\APACrefYearMonthDay{2019}{09}{}.
\newblock
{\BBOQ}\APACrefatitle {Analysis of Deep Learning Architectures for Cross-Corpus
  Speech Emotion Recognition} {Analysis of deep learning architectures for
  cross-corpus speech emotion recognition}.{\BBCQ}
\newblock
 (\BPG~1656-1660).
\newblock
\begin{APACrefDOI} \doi{10.21437/Interspeech.2019-2753} \end{APACrefDOI}
\PrintBackRefs{\CurrentBib}

\bibitem [\protect \citeauthoryear {%
Passonneau%
}{%
Passonneau%
}{%
{\protect \APACyear {2006}}%
}]{%
MASI}
\APACinsertmetastar {%
MASI}%
\begin{APACrefauthors}%
Passonneau, R.%
\end{APACrefauthors}%
\unskip\
\newblock
\APACrefYearMonthDay{2006}{}{}.
\newblock
{\BBOQ}\APACrefatitle {Measuring Agreement on Set-valued Items ({MASI}) for
  Semantic and Pragmatic Annotation} {Measuring agreement on set-valued items
  ({MASI}) for semantic and pragmatic annotation}.{\BBCQ}
\newblock
 \APACrefbtitle {Proceedings of the Fifth International Conference on Language
  Resources and Evaluation ({LREC}{'}06).} {Proceedings of the fifth
  international conference on language resources and evaluation ({LREC}{'}06).}
\PrintBackRefs{\CurrentBib}

\bibitem [\protect \citeauthoryear {%
Pell%
, Monetta%
, Paulmann%
\BCBL {}\ \BBA {} Kotz%
}{%
Pell%
\ \protect \BOthers {.}}{%
{\protect \APACyear {2009}}%
}]{%
langdiif}
\APACinsertmetastar {%
langdiif}%
\begin{APACrefauthors}%
Pell, M.%
, Monetta, L.%
, Paulmann, S.%
\BCBL {} Kotz, S.%
\end{APACrefauthors}%
\unskip\
\newblock
\APACrefYearMonthDay{2009}{06}{}.
\newblock
{\BBOQ}\APACrefatitle {Recognizing Emotions in a Foreign Language} {Recognizing
  emotions in a foreign language}.{\BBCQ}
\newblock
\APACjournalVolNumPages{Journal of Nonverbal Behavior}{33}{}{107-120}.
\newblock

\newblock

\newblock
\begin{APACrefDOI} \doi{10.1007/s10919-008-0065-7} \end{APACrefDOI}
\PrintBackRefs{\CurrentBib}

\bibitem [\protect \citeauthoryear {%
Petrushin%
}{%
Petrushin%
}{%
{\protect \APACyear {2000}}%
}]{%
callcenter}
\APACinsertmetastar {%
callcenter}%
\begin{APACrefauthors}%
Petrushin, V.%
\end{APACrefauthors}%
\unskip\
\newblock
\APACrefYearMonthDay{2000}{01}{}.
\newblock
{\BBOQ}\APACrefatitle {Emotion in Speech: Recognition and Application to Call
  Centers} {Emotion in speech: Recognition and application to call
  centers}.{\BBCQ}
\newblock
\APACjournalVolNumPages{Proceedings of Artificial Neural Networks in
  Engineering}{}{}{}.
\newblock

\newblock

\PrintBackRefs{\CurrentBib}

\bibitem [\protect \citeauthoryear {%
Pichora-Fuller%
\ \BBA {} Dupuis%
}{%
Pichora-Fuller%
\ \BBA {} Dupuis%
}{%
{\protect \APACyear {2020}}%
}]{%
TESS}
\APACinsertmetastar {%
TESS}%
\begin{APACrefauthors}%
Pichora-Fuller, M.K.%
\BCBT {}\ \BBA {} Dupuis, K.%
\end{APACrefauthors}%
\unskip\
\newblock
\APACrefYearMonthDay{2020}{}{}.
\newblock
\APACrefbtitle {{Toronto emotional speech set (TESS)}.} {{Toronto emotional
  speech set (TESS)}.}
\newblock
\APACaddressPublisher{}{Borealis}.
\newblock
\begin{APACrefURL} {https://doi.org/10.5683/SP2/E8H2MF} \end{APACrefURL}
\newblock
\begin{APACrefDOI} \doi{10.5683/SP2/E8H2MF} \end{APACrefDOI}
\PrintBackRefs{\CurrentBib}

\bibitem [\protect \citeauthoryear {%
Povey%
\ \protect \BOthers {.}}{%
Povey%
\ \protect \BOthers {.}}{%
{\protect \APACyear {2011}}%
}]{%
Povey11thekaldi}
\APACinsertmetastar {%
Povey11thekaldi}%
\begin{APACrefauthors}%
Povey, D.%
, Ghoshal, A.%
, Boulianne, G.%
, Goel, N.%
, Hannemann, M.%
, Qian, Y.%
\BDBL {}Stemmer, G.%
\end{APACrefauthors}%
\unskip\
\newblock
\APACrefYearMonthDay{2011}{}{}.
\newblock
{\BBOQ}\APACrefatitle {The kaldi speech recognition toolkit} {The kaldi speech
  recognition toolkit}.{\BBCQ}
\newblock
 \APACrefbtitle {In IEEE 2011 workshop.} {In ieee 2011 workshop.}
\PrintBackRefs{\CurrentBib}

\bibitem [\protect \citeauthoryear {%
Ribeiro%
, Florêncio%
, Zhang%
\BCBL {}\ \BBA {} Seltzer%
}{%
Ribeiro%
\ \protect \BOthers {.}}{%
{\protect \APACyear {2011}}%
}]{%
cs_behavior_1}
\APACinsertmetastar {%
cs_behavior_1}%
\begin{APACrefauthors}%
Ribeiro, F.%
, Florêncio, D.%
, Zhang, C.%
\BCBL {} Seltzer, M.%
\end{APACrefauthors}%
\unskip\
\newblock
\APACrefYearMonthDay{2011}{}{}.
\newblock
{\BBOQ}\APACrefatitle {CROWDMOS: An approach for crowdsourcing mean opinion
  score studies} {Crowdmos: An approach for crowdsourcing mean opinion score
  studies}.{\BBCQ}
\newblock
 \APACrefbtitle {2011 IEEE International Conference on Acoustics, Speech and
  Signal Processing (ICASSP)} {2011 ieee international conference on acoustics,
  speech and signal processing (icassp)}\ (\BPG~2416-2419).
\newblock
\begin{APACrefDOI} \doi{10.1109/ICASSP.2011.5946971} \end{APACrefDOI}
\PrintBackRefs{\CurrentBib}

\bibitem [\protect \citeauthoryear {%
Rubenstein%
\ \protect \BOthers {.}}{%
Rubenstein%
\ \protect \BOthers {.}}{%
{\protect \APACyear {2023}}%
}]{%
rubenstein2023audiopalmlargelanguagemodel}
\APACinsertmetastar {%
rubenstein2023audiopalmlargelanguagemodel}%
\begin{APACrefauthors}%
Rubenstein, P.K.%
, Asawaroengchai, C.%
, Nguyen, D.D.%
, Bapna, A.%
, Borsos, Z.%
, de Chaumont~Quitry, F.%
\BDBL {}Frank, C.%
\end{APACrefauthors}%
\unskip\
\newblock
\APACrefYearMonthDay{2023}{}{}.
\newblock
\APACrefbtitle {AudioPaLM: A Large Language Model That Can Speak and Listen.}
  {Audiopalm: A large language model that can speak and listen.}
\newblock
\begin{APACrefURL} {https://arxiv.org/abs/2306.12925} \end{APACrefURL}
\PrintBackRefs{\CurrentBib}

\bibitem [\protect \citeauthoryear {%
Satt%
, Rozenberg%
\BCBL {}\ \BBA {} Hoory%
}{%
Satt%
\ \protect \BOthers {.}}{%
{\protect \APACyear {2017}}%
}]{%
satt17_interspeech}
\APACinsertmetastar {%
satt17_interspeech}%
\begin{APACrefauthors}%
Satt, A.%
, Rozenberg, S.%
\BCBL {} Hoory, R.%
\end{APACrefauthors}%
\unskip\
\newblock
\APACrefYearMonthDay{2017}{}{}.
\newblock
{\BBOQ}\APACrefatitle {{Efficient Emotion Recognition from Speech Using Deep
  Learning on Spectrograms}} {{Efficient Emotion Recognition from Speech Using
  Deep Learning on Spectrograms}}.{\BBCQ}
\newblock
 \APACrefbtitle {Proc. Interspeech 2017} {Proc. interspeech 2017}\ (\BPGS\
  1089--1093).
\newblock
\begin{APACrefDOI} \doi{10.21437/Interspeech.2017-200} \end{APACrefDOI}
\PrintBackRefs{\CurrentBib}

\bibitem [\protect \citeauthoryear {%
Sheng%
, Provost%
\BCBL {}\ \BBA {} Ipeirotis%
}{%
Sheng%
\ \protect \BOthers {.}}{%
{\protect \APACyear {2008}}%
}]{%
cs_majority_2}
\APACinsertmetastar {%
cs_majority_2}%
\begin{APACrefauthors}%
Sheng, V.%
, Provost, F.%
\BCBL {} Ipeirotis, P.%
\end{APACrefauthors}%
\unskip\
\newblock
\APACrefYearMonthDay{2008}{08}{}.
\newblock
{\BBOQ}\APACrefatitle {Get Another Label? Improving Data Quality and Data
  Mining Using Multiple, Noisy Labelers} {Get another label? improving data
  quality and data mining using multiple, noisy labelers}.{\BBCQ}
\newblock
 (\BPG~614-622).
\newblock
\begin{APACrefDOI} \doi{10.1145/1401890.1401965} \end{APACrefDOI}
\PrintBackRefs{\CurrentBib}

\bibitem [\protect \citeauthoryear {%
Venetis%
\ \BBA {} Garcia-Molina%
}{%
Venetis%
\ \BBA {} Garcia-Molina%
}{%
{\protect \APACyear {2012}}%
}]{%
cs_majority_4}
\APACinsertmetastar {%
cs_majority_4}%
\begin{APACrefauthors}%
Venetis, P.%
\BCBT {}\ \BBA {} Garcia-Molina, H.%
\end{APACrefauthors}%
\unskip\
\newblock
\APACrefYearMonthDay{2012}{August}{}.
\newblock
{\BBOQ}\APACrefatitle {Quality Control for Comparison Microtasks} {Quality
  control for comparison microtasks}.{\BBCQ}
\newblock
 \APACrefbtitle {CrowdKDD 2012.} {Crowdkdd 2012.}
\newblock
\APACaddressPublisher{}{Stanford InfoLab}.
\newblock
\begin{APACrefURL} {http://ilpubs.stanford.edu:8090/1044/} \end{APACrefURL}
\PrintBackRefs{\CurrentBib}

\bibitem [\protect \citeauthoryear {%
Wang%
\ \protect \BOthers {.}}{%
Wang%
\ \protect \BOthers {.}}{%
{\protect \APACyear {2023}}%
}]{%
wang2023neuralcodeclanguagemodels}
\APACinsertmetastar {%
wang2023neuralcodeclanguagemodels}%
\begin{APACrefauthors}%
Wang, C.%
, Chen, S.%
, Wu, Y.%
, Zhang, Z.%
, Zhou, L.%
, Liu, S.%
\BDBL {}Wei, F.%
\end{APACrefauthors}%
\unskip\
\newblock
\APACrefYearMonthDay{2023}{}{}.
\newblock
\APACrefbtitle {Neural Codec Language Models are Zero-Shot Text to Speech
  Synthesizers.} {Neural codec language models are zero-shot text to speech
  synthesizers.}
\newblock
\begin{APACrefURL} {https://arxiv.org/abs/2301.02111} \end{APACrefURL}
\PrintBackRefs{\CurrentBib}

\bibitem [\protect \citeauthoryear {%
Yoon%
, Cho%
\BCBL {}\ \BBA {} Park%
}{%
Yoon%
\ \protect \BOthers {.}}{%
{\protect \APACyear {2007}}%
}]{%
mobileservice}
\APACinsertmetastar {%
mobileservice}%
\begin{APACrefauthors}%
Yoon, W\BHBI J.%
, Cho, Y\BHBI H.%
\BCBL {} Park, K\BHBI S.%
\end{APACrefauthors}%
\unskip\
\newblock
\APACrefYearMonthDay{2007}{}{}.
\newblock
{\BBOQ}\APACrefatitle {A Study of Speech Emotion Recognition and Its
  Application to Mobile Services} {A study of speech emotion recognition and
  its application to mobile services}.{\BBCQ}
\newblock
 \APACrefbtitle {Proceedings of the 4th International Conference on Ubiquitous
  Intelligence and Computing} {Proceedings of the 4th international conference
  on ubiquitous intelligence and computing}\ (\BPG~758–766).
\newblock
\APACaddressPublisher{Berlin, Heidelberg}{Springer-Verlag}.
\PrintBackRefs{\CurrentBib}

\bibitem [\protect \citeauthoryear {%
Zaidan%
\ \BBA {} Callison-Burch%
}{%
Zaidan%
\ \BBA {} Callison-Burch%
}{%
{\protect \APACyear {2011}}%
}]{%
cs_majority_3}
\APACinsertmetastar {%
cs_majority_3}%
\begin{APACrefauthors}%
Zaidan, O.%
\BCBT {}\ \BBA {} Callison-Burch, C.%
\end{APACrefauthors}%
\unskip\
\newblock
\APACrefYearMonthDay{2011}{01}{}.
\newblock
{\BBOQ}\APACrefatitle {Crowdsourcing Translation: Professional Quality from
  Non-Professionals} {Crowdsourcing translation: Professional quality from
  non-professionals}.{\BBCQ}
\newblock
 (\BPG~1220-1229).
\PrintBackRefs{\CurrentBib}

\bibitem [\protect \citeauthoryear {%
Zhou%
\ \protect \BOthers {.}}{%
Zhou%
\ \protect \BOthers {.}}{%
{\protect \APACyear {2025}}%
}]{%
zhou2025emotionaldimensioncontrollanguage}
\APACinsertmetastar {%
zhou2025emotionaldimensioncontrollanguage}%
\begin{APACrefauthors}%
Zhou, K.%
, Zhang, Y.%
, Zhao, S.%
, Wang, H.%
, Pan, Z.%
, Ng, D.%
\BDBL {}Ma, B.%
\end{APACrefauthors}%
\unskip\
\newblock
\APACrefYearMonthDay{2025}{}{}.
\newblock
\APACrefbtitle {Emotional Dimension Control in Language Model-Based
  Text-to-Speech: Spanning a Broad Spectrum of Human Emotions.} {Emotional
  dimension control in language model-based text-to-speech: Spanning a broad
  spectrum of human emotions.}
\newblock
\begin{APACrefURL} {https://arxiv.org/abs/2409.16681} \end{APACrefURL}
\PrintBackRefs{\CurrentBib}

\bibitem [\protect \citeauthoryear {%
Zhu%
\ \BBA {} Carterette%
}{%
Zhu%
\ \BBA {} Carterette%
}{%
{\protect \APACyear {2010}}%
}]{%
cs_gold_3}
\APACinsertmetastar {%
cs_gold_3}%
\begin{APACrefauthors}%
Zhu, D.%
\BCBT {}\ \BBA {} Carterette, B.%
\end{APACrefauthors}%
\unskip\
\newblock
\APACrefYearMonthDay{2010}{07}{}.
\newblock
{\BBOQ}\APACrefatitle {An Analysis of Assessor Behavior in Crowdsourced
  Preference Judgments} {An analysis of assessor behavior in crowdsourced
  preference judgments}.{\BBCQ}.
\PrintBackRefs{\CurrentBib}

\end{thebibliography}
